\documentclass[letterpaper,11pt,fleqn]{article}
\pdfoutput=1
\usepackage{jheppub}

\usepackage{graphicx}
\usepackage[figuresright]{rotating}
\usepackage{bm,amsmath,amssymb}
\usepackage[mathscr]{eucal}
\usepackage{color}
\usepackage{caption}

\long\def\comment#1{ }
\newcommand{\eqn}[1]{Eq.~\eqref{#1}}
\newcommand{\beq}{\begin{equation}}
\newcommand{\eeq}{\end{equation}}
\newcommand{\nn}{\nonumber\\}
\newcommand{\dif}{{\rm d}}
\newcommand{\rmd}{{\rm d}}
\newcommand{\rme}{{\rm e}}
\newcommand{\rmi}{{\rm i}}
\newcommand{\rmP}{{\rm P}}

\newcommand{\tr}{{\rm tr}}

\newcommand{\rmI}{{\rm I}}
\newcommand{\rmJ}{{\rm J}}
\newcommand{\del}{\partial}

\newcommand{\order}[1]{\mcal{O}{(#1)}}
\newcommand{\mcal}{\mathcal}

\newcommand{\bl}{\bm{\ell}}
\newcommand{\bk}{\bm{k}}
\newcommand{\bq}{\bm{q}}
\newcommand{\bp}{\bm{p}}
\newcommand{\bx}{\bm{x}}
\newcommand{\by}{\bm{y}}
\newcommand{\bu}{\bm{u}}

\newcommand{\bz}{\bm{z}}

\newcommand{\br}{\bm{r}}
\newcommand{\bb}{\bm{b}}

\newcommand{\abar}{\bar{\alpha}_s}

\newcommand{\sdla}{{\rm \scriptscriptstyle DLA}}
\newcommand{\ssl}{{\rm \scriptscriptstyle SL}}

\newcommand{\nnlo}{{\rm \scriptscriptstyle NNLO}}
\newcommand{\nlo}{{\rm \scriptscriptstyle NLO}}
\newcommand{\lo}{{\rm \scriptscriptstyle LO}}
\newcommand{\slog}{{\rm \scriptscriptstyle STL}}
\newcommand{\gbw}{{\rm \scriptscriptstyle GBW}}

\newcommand{\mv}{{\rm \scriptscriptstyle MV}}
\newcommand{\CXY}{{\rm \scriptscriptstyle CXY}}
\newcommand{\rcBK}{{\rm \scriptscriptstyle rcBK}}

\newcommand{\collBK}{{\rm \scriptscriptstyle collBK}}
\newcommand{\sT}{{\rm \scriptscriptstyle T}}
\newcommand{\sP}{{\rm \scriptscriptstyle P}}

\newcommand{\Nc}{N_{\rm c}}
\newcommand{\CF}{C_{\rm F}}
\newcommand{\Nf}{N_{\rm f}}
\newcommand{\calS}{\mathcal{S}}
\newcommand{\calN}{\mathcal{N}}

\newcommand{\calT}{\mathcal{T}}
\newcommand{\calK}{\mathcal{K}}
\newcommand{\minus}{\!-\!}

\newcommand{\xm}{x_{\rm m}}

\newcommand{\jcal}{\mathcal{J}}

\newcommand{\pt}{p_\perp} 
\newcommand{\kt}{k_\perp} 
\newcommand{\qt}{q_\perp} 
\newcommand{\lt}{\ell_\perp}

\title{\Large  CGC factorization for forward particle production in proton-nucleus collisions at next-to-leading
order}

\author[a]{E. Iancu,}

\author[b]{A.H. Mueller,}

\author[c]{and D.N.~Triantafyllopoulos\,}

\affiliation[a]{Institut de physique th\'{e}orique, Universit\'{e} Paris Saclay, CNRS, CEA, F-91191 Gif-sur-Yvette, France}

\affiliation[b]{Department of Physics, Columbia University, New York, NY 10027, USA}

\affiliation[c]{European Centre for Theoretical Studies in Nuclear Physics and Related Areas (ECT*)\\and Fondazione Bruno Kessler, Strada delle Tabarelle 286, I-38123 Villazzano (TN), Italy}

\emailAdd{edmond.iancu@cea.fr}
\emailAdd{amh@phys.columbia.edu}
\emailAdd{trianta@ectstar.eu}

\abstract{Within the Color Glass Condensate effective theory, we reconsider the next-to-leading 
order (NLO) calculation of the single inclusive particle production at forward rapidities in
proton-nucleus collisions at high energy. 
Focusing on quark production for definiteness, we establish a new factorization scheme,
perturbatively correct through NLO, in which there is no `rapidity subtraction'.
That is, the NLO correction to the impact factor is not explicitly separated from the high-energy 
evolution. Our construction exploits the skeleton structure of the (NLO) Balitsky-Kovchegov
equation, in which the first step of the evolution is explicitly singled out. The NLO impact factor
is included by computing this first emission with the exact kinematics for the emitted
gluon, rather than by using the eikonal approximation. This particular calculation
has already been presented in the literature \cite{Chirilli:2011km,Chirilli:2012jd}, 
but the reorganization of the perturbation theory that we propose is new. 
As compared to the proposal in \cite{Chirilli:2011km,Chirilli:2012jd}, 
our scheme is free of the fine-tuning inherent in the
rapidity subtraction, which might be the
origin of the negativity of the NLO cross-section observed in previous studies.
}


\begin{document}
\maketitle

\section{Introduction}
\label{sec:intro}

Using perturbative QCD, we would like to study particle production in high-energy  
proton-nucleus ($pA$) collisions in the kinematical regime where the produced 
particle is {\em semi-hard to hard} (meaning that its transverse momenta can be
larger than the nuclear saturation momentum $Q_s$, but not {\em much} larger)
and it propagates at {\em forward rapidity} in the proton fragmentation region
(that is, it makes a very small angle w.r.t. the collision axis)
\cite{Kovchegov:1998bi,Kovchegov:2001sc,Dumitru:2002qt,Albacete:2003iq,Kharzeev:2003wz,Iancu:2004bx,Blaizot:2004wu,Blaizot:2004wv,Dumitru:2005gt,Albacete:2010bs,Tribedy:2011aa,Rezaeian:2012ye,Lappi:2013zma}.

 What is special about this 
kinematics is that the scattering probes the small-$x$ part of the nuclear wavefunction,
but the large-$x$ part of the proton wavefunction, so it acts as a clean probe 
of the nuclear gluon distribution in the interesting regime where one expects
large gluon occupation numbers and strong non-linear phenomena, like gluon saturation. 
This probe is `clean' since the large-$x$ part of the proton wavefunction is
very dilute and hence well described by the standard QCD parton picture
and the associated collinear factorization. Accordingly, the overall process can be
depicted as follows: a collinear parton from the proton undergoes
multiple scattering off the dense gluon system in the nuclear target and hence acquires
some transverse momentum $\kt$, before eventually fragmenting into the hadrons
that are measured in the final state. 

The above physical picture naturally lends itself to a {\em hybrid
factorization} scheme \cite{Dumitru:2002qt,Dumitru:2005gt}  for the calculation of the single-inclusive hadron multiplicity,  
which combines the {\em collinear factorization} for the parton distribution of the
incoming proton and also for the fragmentation of the produced quark or gluon
\cite{Ellis:1991qj},
with the {\em CGC factorization} for the high-energy scattering between
the collinear parton and the nucleus. The `CGC' refers to 
the Color Glass Condensate effective theory, which is the appropriate pQCD framework
to address the problem of high-energy scattering in the presence of high gluon densities
\cite{Iancu:2002xk,Iancu:2003xm,Gelis:2010nm,Kovchegov:2012mbw}.
This is essentially a theory for the gauge-invariant correlations of Wilson lines
and their evolution with increasing energy. A {\em Wilson line} (a unitary matrix in
the color group SU$(N_c)$) is the $S$-matrix of an
energetic parton which undergoes multiple scattering off a strong color field
representing the gluon distribution of the target. 
The CGC factorization\footnote{The CGC factorization can be viewed as the
generalization to high gluon density of the $\kt$-factorization \cite{Catani:1990eg,Catani:1994sq},
which deals with the `unintegrated gluon distribution' and the associated BFKL evolution 
\cite{Kovchegov:2012mbw}. The $\kt$-factorization 
applies so long as the gluon density is moderately low and non-linear effects
like gluon saturation and multiple scattering can be still neglected.}
 for `dilute-dense' scattering associates 
one such a Wilson line to each of the partons partaking in the collision, 
separately in the direct amplitude and the complex conjugate amplitude. Cross-sections
are obtained by averaging over all the configurations of the color fields in the target, a 
procedure which generates the Wilson-line correlators aforementioned.
In the simplest case, that is for single-inclusive particle production at leading order,
this correlator involves the trace of the product of two Wilson lines\footnote{The Wilson
lines are in the fundamental representation of SU$(N_c)$ if the colliding parton is
a quark and in the adjoint representation if this parton is a gluon. Accordingly, the color dipole
is either a quark-antiquark pair, or a pair of two gluons, in an overall
color singlet state.}, which can be identified with the elastic $S$-matrix of a
{\em color dipole} which scatters off the nuclear target.

Whereas this hybrid factorization may indeed look natural, in view of the underlying physical
picture, its foundation in pQCD is not obvious, nor easy to establish. In order to make sense
beyond tree-level, this scheme must be consistent with the QCD radiative corrections
and notably with the collinear and high-energy evolutions.  As we shall shortly explain,
this issue is already non-trivial at leading-order (LO) and it becomes even more so at
next-to-leading order (NLO) and beyond.

The LO version of the hybrid factorization for single-inclusive hadron production 
\cite{Dumitru:2002qt,Dumitru:2005gt}
 includes the LO DGLAP evolution for the parton distribution in the proton and for the parton
fragmentation in the final state. It furthermore
includes the LO B-JIMWLK evolution of the dipole $S$-matrix.  The B-JIMWLK 
(from Balitsky, Jalilian-Marian, Iancu, McLerran, Weigert, Leonidov and Kovner) equations
\cite{Balitsky:1995ub,JalilianMarian:1997jx,JalilianMarian:1997gr,Kovner:2000pt,Iancu:2000hn,Iancu:2001ad,Ferreiro:2001qy} form an infinite hierarchy of coupled equations which describes 
the non-linear evolution of the $n$-point correlations of the Wilson lines.
(Operators with different number of Wilson lines couple under the evolution due to multiple scattering.) 
This hierarchy drastically simplifies in the limit of a large number of colors $N_c\gg 1$, in which
expectation values of gauge-invariant operators factorize from each other. In that limit,
the evolution of the dipole $S$-matrix is governed by a closed non-linear equation, 
known as the Balitsky-Kovchegov (BK) equation  \cite{Balitsky:1995ub,Kovchegov:1999yj}.

A first subtle point, which arises already at LO, refers to the relation between the cross-section for 
parton-nucleus scattering on one hand, and the dipole scattering amplitude on the other hand. 
In general, cross-sections and amplitudes are different quantities
(e.g. they have different analytic properties) and it is only due to the high-energy approximations  
--- notably, due to the fact that the high-energy amplitudes are
purely absorptive --- that such an identification becomes possible in the problem at hand.
Yet, the consistency between this relation and the high-energy evolution
is far from being trivial (see the discussion in \cite{Mueller:2012bn}). 
So far, this has been demonstrated up to next-to-leading order 
\cite{Kovchegov:2001sc,Mueller:2012bn} and there is no obvious reason
why it should remain true in higher orders.

With this in mind, we can address the calculation of single-inclusive hadron production
in $pA$ collisions at NLO. The NLO version of the DGLAP equation is known since long
(see e.g. the textbook \cite{Ellis:1991qj} for a pedagogical discussion). Recently, the BK
equation and the full B-JIMWLK hierarchy have been promoted to NLO accuracy as well
\cite{Balitsky:2008zza,Balitsky:2013fea,Kovner:2013ona}.
By itself, the NLO approximation turns out to be unstable 
\cite{Avsar:2011ds,Lappi:2015fma,Iancu:2015vea}, 
due to the presence of large NLO corrections enhanced by  transverse logarithms.
A similar difficulty was already encountered for the NLO version of the BFKL equation
(the linearized version of the BK equation valid when the scattering is week; see e.g. the textbook
\cite{Kovchegov:2012mbw}). As in that case 
 \cite{Kwiecinski:1997ee,Salam:1998tj,Ciafaloni:1999yw,Altarelli:1999vw,Ciafaloni:2003rd,Vera:2005jt},
resummation schemes have been devised also for the non-linear, BK and B-JIMWLK,
equations \cite{Beuf:2014uia,Iancu:2015vea,Iancu:2015joa,Hatta:2016ujq}, 
to restore the convergence of perturbation theory. In particular, the collinearly-improved BK equation
\cite{Iancu:2015vea,Iancu:2015joa},
which resums to all orders the double-collinear logarithms together with a subset of
the single-collinear logarithms and with the running coupling corrections, appears to be
a convenient tool for the phenomenology \cite{Iancu:2015joa,Albacete:2015xza}. 
Moreover, the full NLO BK equation with collinear
improvement has recently been shown to be stable and tractable via numerical methods
\cite{Lappi:2016fmu}.

Besides the NLO evolution, a calculation of the particle production to NLO accuracy must 
include an equally accurate version of the {\em impact factor}. The `impact factor' refers to
the partonic subprocess and for the present purposes can be simply defined as the
cross-section for parton-nucleus scattering in the absence of any QCD evolution.
For more clarity, from now on, we shall assume that the parton from the proton which participates 
in the collision is a quark. 

At LO, the impact factor is simply the cross-section for the scattering between a bare quark
and a nucleus or, equivalently, the $S$-matrix for a bare dipole.
At NLO, the wavefunction of the incoming quark (or dipole) may contain an additional gluon,
to be referred to as the `primary gluon' in what follows. This 
gluon can be released in the final state (`real correction'), or not (`virtual correction'), but
in any case its emission modifies the cross-section (or the dipole amplitude) w.r.t. to its LO value. 

Since the kinematics of this primary gluon is integrated over, the corresponding correction
to the impact factor is truly a one-loop effect. So far, this correction has been computed
via two different approaches, \cite{Chirilli:2011km,Chirilli:2012jd} and
respectively  \cite{Altinoluk:2011qy,Altinoluk:2014eka}, with results which are quite
difficult to compare with each other, but {\em a priori} look different at NLO accuracy. 
In what follows, we shall mostly refer to the NLO calculation in 
Refs.~\cite{Chirilli:2011km,Chirilli:2012jd}. This is better suited for our
new developments in this paper and this is also the context in which emerged the
problem of the negativity of the cross-section \cite{Stasto:2013cha}, which attracted 
our interest on this topic. But our general philosophy for attacking this problem
is perhaps closer in spirit to that in \cite{Altinoluk:2014eka}, in that it involves
no subtraction for the `rapidity divergence' (see below).

When performing the one-loop integration alluded to above,
one should keep in mind that there are regions in phase-space that have already
been included at LO, at least approximately, via the collinear and the high-energy evolutions.
In the absence of physical cutoffs, these regions would generate logarithmic divergences. 
The physical cutoffs are truly needed when solving the evolution equations 
(they define the boundaries of the corresponding phase-space),
but they can often be avoided when computing the NLO correction to the impact 
factor. Namely, one can directly subtract the would-be 
divergences by using a suitable `renormalization prescription',  tuned to match the 
resummation performed by the LO evolution equations.  This is the strategy followed
by the authors of Refs.~\cite{Chirilli:2011km,Chirilli:2012jd}. 

Specifically, Refs.~\cite{Chirilli:2011km,Chirilli:2012jd} used dimensional
regularization plus minimal subtraction to `remove' the collinear divergences.
This is a rather standard procedure in the context of the collinear factorization
and relies on the fact that the collinear divergences can be factorized from
the transverse integrations, as they refer to the renormalization
of the {\em integrated} parton distributions and fragmentation functions.

Refs.~\cite{Chirilli:2011km,Chirilli:2012jd}
furthermore proposed a `plus' prescription in order to subtract the `rapidity
divergence', i.e. the would-be divergence\footnote{This is also known as the 
`soft divergence', or the `small-$x$ divergence';
Refs.~\cite{Chirilli:2011km,Chirilli:2012jd} used the variable $\xi\equiv 1-x$,
so for them the `rapidity divergence'  appears in the limit $\xi\to 1$.}
 at $x\to 0$, where $x$ is the longitudinal
momentum fraction of the primary gluon w.r.t. the incoming quark.
This prescription is quite common in the context of the $k_\perp$-factorization
at next-to-leading order (see e.g. \cite{Ivanov:2012iv} and references therein)
and its `non-linear' extension to the CGC factorization may look natural.
Recall however that, underlying this prescription, there is
the strong assumption that cross-sections in perturbative QCD at high-energy can
be factorized in rapidity.  This assumption is highly non-trivial (and still unproven 
in the general case and beyond LO accuracy) because the perturbative corrections
are truly non-local in rapidity. Accordingly, it is {\em a priori} not obvious that the small-$x$
divergence can be factorized from the high-energy evolution of the various scattering 
operators ---  the `dipole $S$-matrices' which describe the eikonal scattering between
the quark-gluon projectile and the target. 
This being said, we shall explicitly demonstrate
in this paper that, via an appropriate reorganization of the perturbative corrections, 
one can indeed obtain such a factorized expression, valid to NLO,
which involves the `plus' prescription and agrees with the 
proposal in Refs.~\cite{Chirilli:2011km,Chirilli:2012jd}.
However, we shall also argue that the manipulations associated with this
reorganization --- namely, with the subtraction of the high-energy evolution
from the NLO impact factor --- involve a considerable amount of fine-tuning, 
which may be dangerous in practice.

This fine-tuning might be at the origin of the negativity problem observed in explicit numerical
calculations based on the factorization scheme in \cite{Chirilli:2011km,Chirilli:2012jd}:
the cross-section  for single-inclusive particle production suddenly turns negative 
for transverse momenta
slightly larger than the target saturation momentum $Q_s$ \cite{Stasto:2013cha,Stasto:2014sea}. 
Various proposals to circumvent this problem, by either introducing a cutoff in the
rapidity subtraction scheme \cite{Kang:2014lha,Xiao:2014uba,Ducloue:2016shw}, 
or via a more careful implementation of the kinematics
\cite{Stasto:2014sea,Altinoluk:2014eka,Watanabe:2015tja},
have managed to alleviate the problem (by pushing it to somewhat larger values of the 
transverse momentum), but without offering a fully satisfactory solution,
at either conceptual or practical level (see also the discussion in the recent
review paper \cite{Stasto:2016wrf}). Whereas high-energy approximations are expected 
to become less accurate at sufficiently large transverse momenta $k_\perp\gg Q_s$, 
we find it surprising that they fail already in the transition region towards saturation  
($k_\perp\gtrsim Q_s$) --- a region whose description is in fact the main focus of the CGC
effective theory \cite{Iancu:2002xk,Iancu:2003xm,Gelis:2010nm,Kovchegov:2012mbw}.


This negativity problem encouraged us to reconsider the overall calculation from a new perspective
and thus propose a new factorization scheme for the high-energy aspects of the problem. In presenting
this proposal below, we shall restrict ourselves to quark production, that is, we shall ignore other
partonic channels and also the fragmentation of the quark into hadrons in the final state. Also,
we shall omit all the NLO corrections associated with the collinear resummation, i.e. the finite
terms which remain after subtracting the collinear divergence. These terms can
be unambiguously distinguished from those referring to the high-energy factorization
 \cite{Ducloue:2016shw} and can be simply added to our final results.

Our main observations and new results can be summarized as follows: 

\texttt{(i)} In our opinion, the negativity problem
is most likely related to the severe fine-tuning inherent in the rapidity subtraction. 
The `fine-tuning' refers to the delicate balance between the NLO corrections
which are included in the evolution of the dipole $S$-matrix and those which are subtracted via
the `plus' prescription in order to construct the NLO correction to the impact factor.
As we shall explain in detail in Sect.~\ref{sec:sub}, this subtraction amounts to a 
reorganization of the perturbation theory which exploits the integral 
representation for the solution to the BK equation. Any approximation in solving
this equation, as well as the subsequent approximations which are in practice needed
to derive the `plus' prescription, will lead to an imbalance between the
large, `added' and `subtracted', contributions and thus possibly to unphysical
results.

\texttt{(ii)} The `plus' prescription is actually not needed: the would-be `rapidity divergence' is truly
cut off by physical mechanisms, namely by energy conservation for the `real' corrections and
by probability conservation for the `virtual' ones. The role of the energy conservation in
constraining the longitudinal phase-space for the primary emission is in fact well appreciated
\cite{Stasto:2014sea,Altinoluk:2014eka}. This constraint has been used to cut off the 
soft divergence in Ref.~\cite{Altinoluk:2014eka} 
and to alleviate the negativity problem in the context of the  `plus' prescription in
Ref.~\cite{Watanabe:2015tja}. The corresponding constraint on the `virtual' corrections has not been
discussed to our knowledge, so we shall devote an appendix to an explicit NLO calculation 
which demonstrates this. Specifically, in App.~\ref{app:canc} we show
that the `virtual' corrections with very short lifetimes --- outside the physical range 
for the `real' corrections --- mutually cancel each other.  This result is in fact natural: 
the `real' and `virtual' corrections must have the same support in longitudinal 
phase-space, since they should combine with each other to ensure probability conservation.

\texttt{(iii)} To NLO accuracy, the calculation of the single-inclusive forward particle production in 
dilute-dense collisions can be given a different factorization, cf. \eqn{NLO},
in which the small-$x$ logarithm associated with the primary gluon emission is not included
in the high-energy evolution, but is implicitly kept within the impact factor\footnote{At this
level, we use the notion of `impact factor' in a rather informal way, as a proxy for the
dilute projectile made with the incoming quark and its primary gluon. The more conventional
impact factor which is defined order-by-order in perturbation theory and involves no 
high-energy evolution (i.e. no small-$x$ logarithms),
will be computed to NLO in Sect.~\ref{sec:sub}, after separating
our general result into leading-order plus next-to-leading order contributions.}.
The latter includes the incoming quark and its (not necessarily soft) primary gluon, 
which together scatter off the gluon distribution in the nucleus and thus measure its high-energy
evolution. This is the same picture as for the CGC calculation of 
di-hadron production  \cite{Marquet:2007vb,Albacete:2010pg,Dominguez:2011wm,Iancu:2013dta}, except that the kinematics of the primary gluon is now integrated over
and one must add the `virtual' corrections. 

\comment{This picture was also the starting
point of the NLO calculation of the impact factor in \cite{Chirilli:2011km,Chirilli:2012jd}.
But as compared to that previous work, we show that by carefully specifying the integration
limits in $x$ and also the ($x$-dependent) rapidity variables of the various $S$-matrices,
we can promote that calculation to a new factorization scheme, valid through NLO, in which the
high-energy evolution of the target is factorized from the quark-gluon
impact factor described above.
}

\texttt{(iv)}  In the limit where the
primary gluon is soft and treated in the eikonal approximation, our general formula in
Eq.~\eqref{NLO} reduces to the integral representation
 \eqref{rcBKint} of the solution to rcBK.
This is indeed the correct result for the quark multiplicity at LO.
Vice-versa, our formula may be viewed as a generalization of the LO BK evolution
in which the very first gluon emission (and that emission only) is treated beyond the
eikonal approximation. In view of that, we expect our result for the cross-section to be
positive semi-definite, albeit we have not been able to prove this explicitly.

\texttt{(v)} Our general formula \eqref{NLO} is probably too cumbersome to be
used in practice, due to the complicated structure of the transverse and longitudinal
integrations, which are entangled with each other. Fortunately though, the problem
can be considerably simplified in the interesting regime where the transverse momentum $\kt$
of the produced quark is sufficiently hard, $\kt\gtrsim Q_s$.
In that regime, the primary gluon is relatively hard as well, 
with a transverse momentum $\pt\sim \kt$ (see the discussion in Sect.~\ref{sec:dijet}).
This allows us to replace  $\pt\sim \kt$ within the rapidity variables in \eqn{NLO} and
thus deduce a much simpler result, \eqn{NLOkt}, which is of the same degree of difficulty
as the formulae used within previous numerical simulations 
\cite{Zaslavsky:2014asa,Stasto:2013cha,Stasto:2014sea,Watanabe:2015tja,Ducloue:2016shw}, 
while at the same time avoiding the rapidity subtraction and the associated fine-tuning.

\texttt{(vi)} To ensure the desired NLO accuracy of the overall scheme, the high-energy
evolution of the color dipoles must be computed to NLO as well. In Sect.~\ref{sec:evol}
we complete our factorization scheme by specifying the NLO corrections associated with
the high-energy evolution. As we also explain there, the inclusion of the NLO evolution
in the problem at hand is {\em a priori} problematic, for two reasons:
(a) the evolution of a dense wavefunction, like a nucleus, is not known beyond 
leading-order, and (b) the strict NLO approximation is expected to be unstable, due to large
corrections enhanced by collinear logarithms. 
We provide solutions to these problems by relating the target evolution
to that of the dilute projectile, which is indeed known to
NLO accuracy \cite{Balitsky:2008zza}, including the all-order resummation of the
collinear logarithms \cite{Iancu:2015vea,Iancu:2015joa}.
 This is furthermore discussed in Appendices \ref{sec:proj} and \ref{sec:NLOBK}.

\texttt{(vii)} To make contact with the formalism in 
\cite{Chirilli:2011km,Chirilli:2012jd}, we consider in Sect.~\ref{sec:sub} the decomposition
of our general result \eqref{NLOkt} between NLO dipole evolution and NLO corrections
to the impact factor. This decomposition, shown in schematic notations in \eqn{toyNLO1}, 
relies in an essential way on the fact that the dipole $S$-matrix obeys a specific evolution
equation --- either the LO BK equation with  running-coupling
(rcBK) \cite{Kovchegov:2006wf,Kovchegov:2006vj,Balitsky:2006wa}, or the NLO BK equation
\cite{Balitsky:2008zza} with collinear improvement \cite{Iancu:2015vea,Iancu:2015joa}, 
depending upon the desired accuracy. Indeed, the integral version of
the evolution equation is used to reshuffle the largest contribution to the cross-section, 
that associated with the LO evolution. 

\eqn{toyNLO1} is very similar, but not fully identical, to the factorization scheme
proposed in \cite{Chirilli:2011km,Chirilli:2012jd}, which is schematically shown
in \eqn{toyCXY}.  As explained in Sect.~\ref{sec:sub},
the differences between Eqs.~\eqref{toyNLO1} and  \eqref{toyCXY} are irrelevant
to NLO accuracy, but they can be nevertheless important in practice, as they
introduce an imbalance between the terms included in the dipole evolution and
those subtracted via the `plus' prescription. As already mentioned at point \texttt{(i)},
we believe that this imbalance is responsible for the problem of the negativity
of the cross-section.

To summarize, all the potential difficulties with the subtraction method can be avoided
by computing the cross-section directly from our formula \eqref{NLOkt}, 
which involves no subtraction at all. This formula can be evaluated with a suitable
approximation for the high-energy evolution, like rcBK or the more elaborated approximations
described in Sect.~\ref{sec:evol} and in Appendix \ref{sec:proj}.

This paper includes two major sections, 
devoted to the LO and the NLO calculations respectively, and three appendices.
Sect.~\ref{sec:LO} starts with a discussion of the kinematics and of
the importance of the choice of a Lorentz frame for building a physical picture. Such a picture is
first developed in the target infinite momentum frame (in Sects.~\ref{sec:kin} and \ref{sec:dip}), 
then extended to a mixed frame, where one of the evolution gluons (the `primary gluon') 
is viewed as an emission by the incoming quark, whereas all the subsequent ones are 
included in the evolution of the nuclear target (in Sect.~\ref{sec:TvsP}). In Sect.~\ref{sec:dijet},
we discuss the di-jet configurations which control the final state in the regime where the
produced quark has a large transverse momentum $\kt \gg Q_s$. The first subsection of
Sect.~\ref{sec:NLO} summarizes the result for the NLO impact factor 
obtained in \cite{Chirilli:2011km,Chirilli:2012jd} and also extends that result
by specifying the rapidity
variables for the evolution of the various dipole $S$-matrices. This discussion motivates
our main result in this paper, that is, the NLO factorization displayed in \eqn{NLO}.
This general but rather cumbersome expression is rendered more tractable and also more explicit 
in Sects.~\ref{sec:approx}, \ref{sec:evol} and in Appendix \ref{sec:proj}, 
where we simplify the kinematics via approximations appropriate
at large $\kt\gtrsim Q_s$ and we replace the unknown NLO evolution of the nuclear target by that
of the dilute quark-gluon projectile (with collinear improvement). 
In Sect.~\ref{sec:sub}, we isolate LO from NLO contributions,
as described at point \texttt{(vii)} above. Sect.~\ref{sec:conc} contains our conclusions.
In Appendix \ref{app:canc} we demonstrate the
mutual cancellation of the  `virtual' fluctuations whose lifetime is shorter than the longitudinal
extent of the target. Finally, Appendices \ref{sec:proj} and \ref{sec:NLOBK} give more
details on the NLO evolution of color dipoles.

\section{The leading order calculation}
\label{sec:LO}

In this section, we shall briefly review the leading-order (LO) calculation of 
single-inclusive quark production in high-energy proton-nucleus ($pA$) 
at forward rapidities (i.e. in the proton fragmentation region).
This calculation relies on a hybrid factorization scheme \cite{Dumitru:2005gt} which involves
collinear factorization at the level of the proton wavefunction together with the dipole 
picture for the scattering between a collinear quark from the proton and the nuclear 
gluon distribution.

\subsection{General picture and kinematics}
\label{sec:kin}

To LO in perturbative QCD and in a suitable Lorentz frame,
the forward production of a quark in $pA$ collisions
proceeds via the {\em transverse momentum broadening} of one of the quarks from the incoming 
proton: the quark, which was originally collinear with the proton,
acquires a transverse momentum $\bk$ via scattering off the small-$x$ gluons in the 
nuclear wavefunction and thus emerges at a small angle $\theta \simeq \kt/k^+$
w.r.t. the collision axis. The typical situation
is such that the quark undergoes multiple soft scattering and thus accumulates
a transverse momentum of order $Q_s$ --- the target saturation momentum
at the longitudinal resolution probed by the scattering (see below).
But the $\kt$-distribution of the produced quark also features a power-like
tail at high momenta $\kt\gg Q_s$, which is the result of a single, relatively hard, 
Coulomb scattering off the color sources in the target.

The physical picture actually depends upon the choice of a Lorentz frame.
The picture that we have just described only holds in a `target infinite momentum frame', 
where the nuclear target carries most of the total energy, so the high-energy
evolution via the successive emissions of soft gluons is fully encoded in the nuclear
gluon distribution. On the other hand, the picture would be different in a frame where
the projectile proton carries most of the total energy; in that case, the wavefunction
of the incoming quark is highly evolved, in the sense that it contains many soft gluons,
which can be put on-shell by their scattering  off the (un-evolved) nucleus. 
The final transverse momentum  $\bk$ acquired by the quark is then
the result of the recoil from this induced gluon radiation.

To transform these pictures into actual calculations, we need to better 
specify the kinematics. We work in a Lorentz frame where the proton is a right mover, 
with longitudinal momentum $Q^+$, while the nuclear target is a left mover, with longitudinal 
momentum $P^-$ per nucleon. The high-energy regime corresponds to the situation where the
center-of-mass energy $\sqrt{s}$, with $s=2Q^+P^-$, is much larger than any of the transverse
momentum (or virtuality) scales in the problem; in particular,  $s\gg k_\perp^2$ and
$s\gg Q_s^2$. For our purposes, the longitudinal momentum $k^+$ of the produced quark is
most conveniently parametrized in terms of the boost-invariant ratio $x_p\equiv k^+/Q^+\le 1$ 
(the quark longitudinal momentum fraction w.r.t. the incoming proton). 

Let us start by choosing a {\em target infinite-momentum frame}, where the scattering
involves a bare quark from the proton and the highly-evolved gluon distribution of the nucleus.  
Prior to the collision, the quark has only a  `plus' momentum $q^+_0$. 
After the scattering, which can involve  
one or several gluon exchanges with color sources from the target, 
the quark emerges with the same longitudinal momentum, $k^+= q^+_0$
(since gluons from the target have negligible `plus' momenta), but
it acquires a transverse momentum $\bk$ 
and also a `minus' component $k^-$, which is needed for the produced quark to be 
on-shell:  $k^-=\kt^2/2k^+$. This condition fixes the total longitudinal momentum
fraction carried by the gluons from the target that were involved in the collision\footnote{Clearly,
in the case of multiple scattering, the light-cone energy $q^-$ which is individually carried by 
the exchanged gluons can be smaller than this overall value $k^-=\kt^2/2k^+$, as emphasized
in Ref.~\cite{Altinoluk:2014eka}. However, we disagree with the conclusion there that the 
longitudinal fraction $X_g$ which counts for the target evolution can be {\em
parametrically} different from the estimate \eqref{Xg} (and in particular independent of the
quark transverse momentum $\kt$). Indeed, the relevant value of $X$ is the one 
which controls the energy dependence of the target saturation momentum $Q_s(X)$.
As well known, the latter is fully determined by the condition that the amplitude 
for a {\em single} scattering become of order one \cite{Iancu:2002tr,Mueller:2002zm}.
} :
\beq\label{Xg}
X_g= \,\frac{k^-}{P^-}\,=\,\frac{\kt^2}{2k^+P^-}\,=\,\frac{\kt^2}{x_p s}
\,\equiv\,\frac{\kt^2}{\hat s}\,.\eeq
To relate to the experimental situation, it is customary to express the longitudinal fractions 
$x_p$ and $X_g$ in terms of the 
rapidity $\eta\equiv (1/2)\ln(k^+/k^-)$ of the produced quark in the center-of-mass frame
(where $Q^+=P^-= \sqrt{s/2}$). Using $x_p=k^+/Q^+$ and $k^+=(\kt/\sqrt{2})\rme^\eta$, one finds
\beq
\label{COM}
x_p=\frac{\kt}{\sqrt{s}}\,\rme^\eta\,,\qquad X_g=\frac{\kt}{\sqrt{s}}\,\rme^{-\eta}\,.
\eeq
The {\em forward kinematics} corresponds to the situation where $\eta$ 
is positive and large. Then \eqn{COM} makes it clear that $X_g\ll x_p < 1$, thus confirming
that forward particle production explores the small-$X_g$ part of the nuclear wavefunction,
as anticipated in the Introduction.

\subsection{Dipole picture}
\label{sec:dip}

To LO in the CGC effective theory, the `quark multiplicity'  (i.e. the
distribution of the produced quarks in transverse momentum $\bk$ and COM rapidity $\eta$)
is computed as follows
\beq\label{LO}
 \frac{\rmd N^{pA\to qX}}{\rmd^2\bk\, \rmd \eta}\bigg|_{\lo}
=\, \frac{1}{(2\pi)^2}\,
 x_p q(x_p)\,
{\calS}(\bk,X_g)\,,
\eeq
where the kinematic variables $\bk$, $\eta$, $x_p$, and $X_g$ have already been introduced,
$x_p q(x_p)$ is the quark distribution in the proton for a collinear quark with
longitudinal momentum fraction $x_p$,
and $\calS(\bk,X_g)$ is the Fourier transform of the elastic $S$--matrix for the scattering
between a color dipole in the fundamental representation and the nucleus:
\beq\label{SF}
{\calS}(\bk,X_g)=\int \rmd^2\br\, \rme^{-\rmi \bk \cdot \br} {S}(\br,X_g)\,.
\eeq
The `color dipole' is a quark-antiquark pair in a color-singlet state. In the present context, 
this appears as merely a mathematical representation for the cross-section for the scattering 
between the produced quark and the nucleus: the `quark' component of the dipole
is the colliding quark viewed in the direct amplitude (DA) and the `antiquark' is the same physical 
quark, but viewed in the complex conjugate amplitude (CCA).

To the accuracy of interest, the dipole-nucleus scattering can be computed 
in the eikonal approximation, i.e. the transverse coordinates of the quark ($\bx$) and
the antiquark ($\by$) can be treated as fixed during the collision. Then the only effect of the 
collision are color rotations of the two fermions, as described by Wilson lines extending
along their trajectories:
 \beq\label{Sdip}
S(\bx,\by; X)\,\equiv\,\frac{1}{N_c}\left\langle \tr
 \big[U(\bx) U^\dagger(\by)\big] \right\rangle_{X}.
 \eeq
 Here, $U(\bx)$ and  $U^\dagger(\by)$ are Wilson lines in the fundamental representation,
e.g.,
\begin{align}\label{Udef}
 U(\bx) = \rmP \exp\left[\rmi g \int \dif x^+ A^-_a(x^+,\bx) t^a\right],
 \end{align}
and $A^-_a(x)$ is (the relevant component of) the color field representing the gluons from the target
with longitudinal momentum fraction $X\equiv q^-/P^-$. In general, this field is strong 
(corresponding to large gluon occupation numbers) and the path-ordered phase in \eqn{Udef} 
resums multiple scattering to all orders. In the Fourier transform in \eqn{SF}, the transverse momentum
$\bk$ of the produced quark is conjugated to the dipole size $\br\equiv \bx-\by$. Both the l.h.s. and
the r.h.s. of  \eqn{SF} depend upon the impact parameter $\bb\equiv (\bx+\by)/2$, but this
dependence is unessential for what follows and will be omitted: for our purposes, the target
can be treated as quasi-homogeneous in the transverse plane.

The brackets in the r.h.s. of \eqn{Sdip} denote the target
average over the color field $A^-_a(x)$, as computed with the CGC weight 
functional  \cite{Iancu:2002xk,Iancu:2003xm,Gelis:2010nm}.
By using the JIMWLK equation for the latter, or directly the Balitsky equations for the
color dipole operator, one finds an equation for the evolution of the dipole $S$-matrix 
with decreasing $X$. In general, this is just the first equation from an infinite hierarchy, but the
situation simplifies in the limit of a large number of colors $N_c\gg 1$, where the dipole $S$-matrix
obeys a closed, non-linear, equation, known as the Balitsky-Kovchegov (BK) equation \cite{Balitsky:1995ub,Kovchegov:1999yj}. This equation will be later needed, so let us display it here:
\begin{align}\label{BK}
 \frac{\del }{\del Y} \,S(\bx,\by; Y)=
 \frac{\abar}{2\pi}\, \int \rmd^2\bz\,
\frac{(\bx-\by)^2}{(\bx-\bz)^2(\bz-\by)^2}\,\Big[
 S(\bx,\bz; Y) S(\bz, \by; Y)
 -S(\bx,\by; Y)\Big]\,,
 \end{align}
where $\abar\equiv\alpha_s N_c/\pi$ and 
$Y\equiv \ln(1/X)$. The integration variable $\bz$ in \eqn{BK} represents 
the transverse coordinate of a soft gluon with longitudinal fraction $X= q^-/P^-\ll 1$, 
which is emitted by `fast' color sources from the target (valence quarks and gluons 
from the previous generations, with momentum fractions $X'\gg X$) and 
absorbed by the projectile dipole.  \eqn{BK} must be integrated from some
lower value $Y_0\equiv \ln(1/X_0)$, where one can use a low-energy model for 
the nuclear gluon distribution, 
up to $Y_g\equiv \ln(1/X_g)=\ln(\hat s/\kt^2)$, where we compute the quark production. 
Typically, $X_0\sim 1\gg X_g$. For instance, if one
uses a valence-quark model for the nucleus,  like the McLerran-Venugopalan (MV) model 
\cite{McLerran:1993ni,McLerran:1993ka}, 
then $X_0$ must satisfy $\abar  \ln(1/X_0)\ll 1$ and the whole gluon distribution
is built up via evolution. 

In the above discussion, we have privileged the viewpoint of {\em target evolution},
that is, we have described \eqn{BK} as the result of a change in the gluon distribution of the target.
The complementary point of view, that of {\em projectile} evolution, will be useful too for what
follows and will be introduced in the next subsection.

Also, we implicitly assumed that the relation \eqref{LO} between
the quark transverse momentum broadening and the dipole $S$-matrix remains valid in the
presence of the high-energy evolution; that is, both sides of \eqn{LO} evolve in exactly 
the same way with increasing energy (i.e. with increasing $\eta$,
or decreasing $X_g$). This is known to be true,
at least, up to NLO accuracy, as demonstrated  in Ref.~\cite{Kovchegov:2001sc} for the 
LO BK  evolution  and in Ref.~\cite{Mueller:2012bn} for the NLO one. However, \eqn{LO} 
is not complete beyond leading order: to this `dipole' piece, one must add the 
`corrections to the impact factor', that is, the contributions from
partonic configurations which do not reduce to either the dipole, or its
high-energy evolution. Such corrections will represent a main topic of the NLO discussion
in Sect.~\ref{sec:NLO}.

\subsection{Target versus projectile evolution}
\label{sec:TvsP}

Previously, we have insisted that the physical picture of the high-energy evolution  
depends upon the choice of a frame, but as a matter of facts the
BK equation \eqref{BK} holds exactly as written in {\em any} frame that is obtained from
the COM frame via a boost. This includes the infinite momentum frame of the nucleus that
we have considered so far, but also the corresponding frame for the projectile, where the
incoming proton carries most of the total energy and the high-energy evolution of interest
refers to the emission of soft gluons in the wavefunction of the colliding dipole (or quark).
By `soft gluons' in this case, one means gluons which carry small fractions 
$x\equiv p^+/q^+_0\ll 1$ of the longitudinal momentum $q^+_0$ of the parent quark.

This `boost invariance' of the LO BK equation is not an automatic consequence of
the underlying Lorentz symmetry of the problem  --- after all, the respective evolution variables are 
different: $Y=\ln(1/X)$ for the target evolution and $y\equiv \ln(1/x)$ for that of the projectile.
Rather, it reflects approximations specific to the LLA at hand, whose effect is to identify 
these two variables $Y$ and $y$ to the accuracy of interest (up to a change of sign). In other
terms, at LLA, the fact of decreasing $Y$ is indeed equivalent with increasing $y$, meaning that
the evolution can be progressively transferred from the projectile to the target, and back.
This point will play an important role in our
subsequent discussion of the NLO contribution to particle production.
In preparation for that, let us briefly remind here the kinematical assumptions
underlying the LLA and thus expose their limitations.

\begin{figure}[t] \centerline{
\includegraphics[width=.55\textwidth]{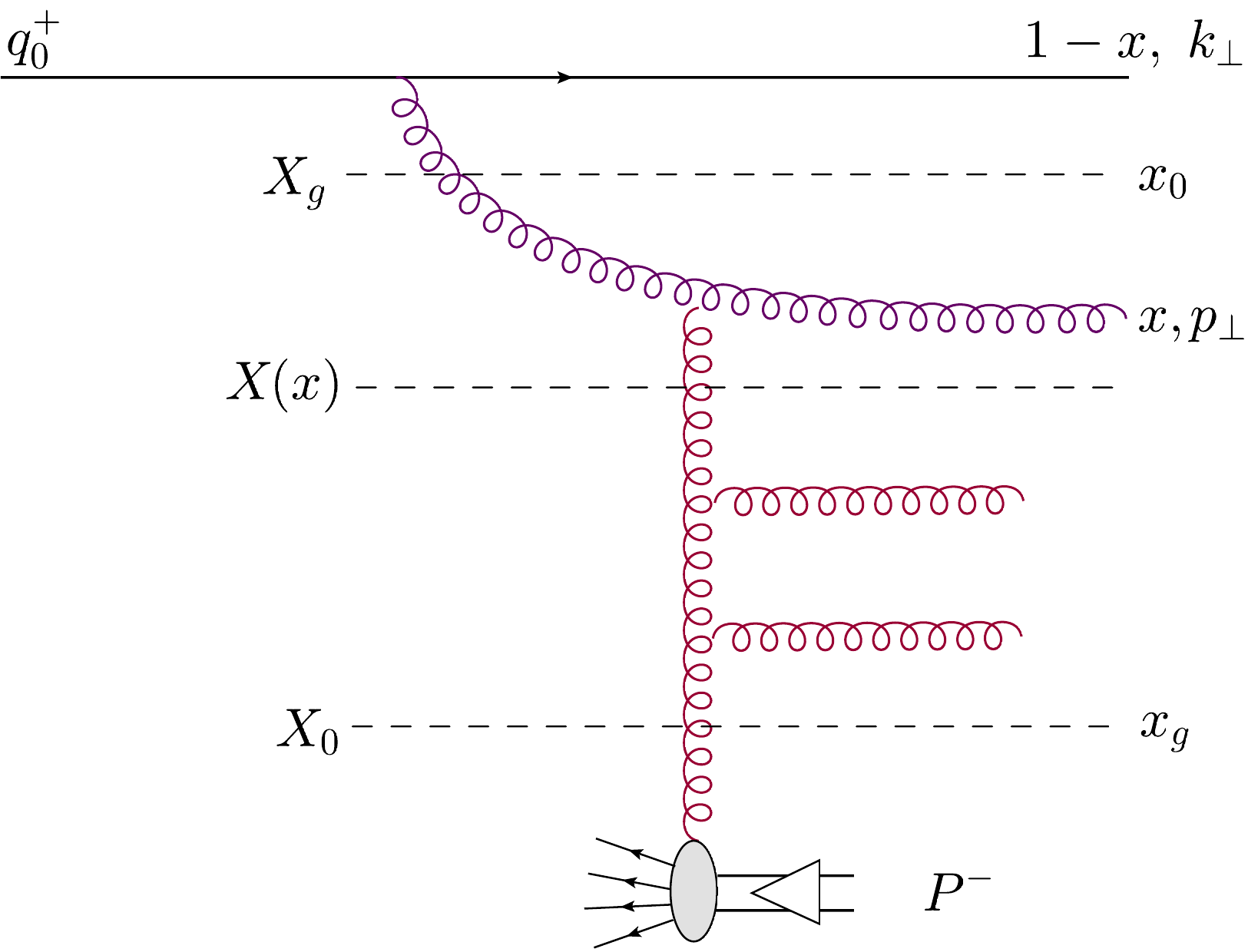}}
 \caption{\small An illustration of the LO high-energy evolution of the `cross-section for quark production' 
 $\calS(\bk,X_g)$ (the Fourier transform of the dipole $S$--matrix). The `primary gluon' (the first gluon 
 emission by the quark, which carries a longitudinal momentum fraction $x\ll 1$ is) viewed as a part of the wavefunction of the incoming quark (a right mover). The subsequent emissions are rather associated with 
 the evolution of the gluon distribution in the nuclear target (a left mover), 
 within the range $X(x) < X < X_0$, with $X(x)=X_g/x$ and $X_0\sim 1$. Alternatively,
 and equivalently at LLA, they can be associated with the evolution of the wavefunction of
 the primary quark-gluon pair (a right mover) within the `plus' momentum range $x_g < x' < x$. }
 \label{QtoQG}
\end{figure}

To that aim, we consider the situation where only one of the
soft gluons has been emitted by the quark, while all the other ones belong to 
the wavefunction of the target (see Fig.~\ref{QtoQG}). 
The gluon that has been singled out in this way is the one to be closest
in rapidity ($y$) to the incoming quark; we shall refer to it as the {\em
primary gluon} and write its longitudinal and transverse momenta
as $p^+=xq^+_0$ and respectively $\bp$.
Consider a `real' graph in which the primary gluon, albeit unmeasured, is
released in the final state\footnote{The associated `virtual' graphs are needed for the 
conservation of probability, hence
one can naturally assume that they must involve the same phase-space for gluon 
emission as the `real' ones. We shall later return to a more elaborate
discussion of this point, including an explicit computation of the `virtual' corrections
in App.~\ref{app:canc}.}. 
Then longitudinal momentum conservation implies
$k^+=(1\minus x)q^+_0=x_p Q^+$ and therefore $q_0^+/Q^+=x_p/(1\minus x)$.
[Recall that $x_p$ is defined as the boost-invariant ratio $x_p\equiv k^+/Q^+$,
which in the COM frame takes the form shown in \eqn{COM}.]

Both the quark and the primary gluon must be on mass-shell in the final state.
Then, light-cone energy conservation implies that the scattering 
off the nuclear target must transfer a total `minus' component $q^-=X(x,\pt)P^-$ with
(recall that $\hat s=x_p s$)
\beq\label{Xx}
X(x,\pt)\,=\,\frac{1}{P^-}\left(\frac{\kt^2}{2k^+}+\frac{\pt^2}{2p^+}\right)
=\frac{1}{\hat s}\left[\kt^2+\frac{1\minus x}{x}\, {\pt^2}\right]\,.\eeq
The LLA essentially relies on the two following kinematical assumptions: 

\texttt{(i)} The longitudinal fraction of the emitted gluon is small: $x\ll 1$. This allows one
to simplify the calculation, notably by computing
the quark-gluon vertex in Fig.~\ref{QtoQG} in the eikonal approximation.

\texttt{(ii)} The transverse momenta of the successive emissions are parametrically
of the same order: $\kt\sim \pt$ or, more precisely, $\ln(1/x)\gg |\ln(\kt^2/\pt^2)|$. This condition
is necessary to simplify the energy denominators (by neglecting $k^- \equiv 
{\kt^2}/{2k^+}$ compared to $p^-\equiv {\pt^2}/{2p^+}$) in the study of soft successive emissions
and thus obtain the LO BK (or BFKL) equation.

Under these assumptions, the second term in the r.h.s. of \eqn{Xx} 
(the light-cone energy of the primary gluon) dominates over
the first one and, moreover, one can ignore the difference between $\pt$ and $\kt$
when computing the evolution variables $Y=\ln(1/X)$ and $y=\ln(1/x)$.

The above argument can be immediately extended to 
an arbitrary separation of the LO high-energy evolution between the quark projectile and the 
nuclear target:
successive emissions in the projectile are strongly ordered in $x$ but have comparable
transverse momenta, hence both energy conservation and the energy 
denominators are controlled by the last emitted gluon --- the one with the smallest value of $x$
and a transverse momentum of order $\kt$.
Accordingly, instead of \eqn{Xx}, one can use the following, simpler, relation,
\beq\label{XxLO}
X(x)\,=\,\frac{\kt^2}{x \hat s}\,=\,\frac{X_g}{x}\,,\eeq
(or $Y=Y_g-y$) in order to connect the LO evolution of the projectile to that of the target.
The differences between \eqref{Xx} and \eqref{XxLO} become however important
starting with NLO, as we shall see.

The above discussion can be
summarized by the following integral representation of the solution to the BK equation,
illustrated in Fig.~\ref{QtoQG},
in which the total evolution is explicitly split between exactly one soft gluon ($x\ll 1$) in the quark
wavefunction and an arbitrary number of soft gluons ($X\ll 1$) in the wavefunction of the target:
\begin{align}\label{BKint}
 S\big(\bx,\by; X_g\big)=S_0(\bx,\by)+
 \frac{\abar}{2\pi}\, & \int_{x_g}^{1}\frac{\rmd x}{x}\int \rmd^2\bz\,
\frac{(\bx-\by)^2}{(\bx-\bz)^2(\bz-\by)^2}\nn
& \,\Big[
 S\big(\bx,\bz; X(x)\big) S\big(\bz, \by; X(x)\big)
 -S\big(\bx,\by; X(x)\big)\Big]\,.
 \end{align}
In this equation, $S_0$ is the initial condition at $X_0\simeq 1$ and the lower limit
$x_g\equiv \kt^2/\hat s$ for the integral over $x$ corresponds via \eqref{XxLO} to $X=1$, 
i.e. to the situation where the soft gluon from the projectile probes bare nucleons from the target.
\eqn{BKint} can also be viewed as purely target evolution provided one
changes the integration variable from $x$ to $X\equiv X(x)$.  Then it becomes obvious that
this is the same as \eqn{BK} integrated over $Y$, from $Y_0=0$ up to $Y_g$.

\comment{
Alternatively, \eqn{BKint} can be also viewed as purely projectile evolution; in that interpretation,
the rapidity argument $x_g/x$ of the dipole $S$-matrices is recognized as the ratio of two
`plus' momenta: $x_g/x=p^+_g/p^+$ 

 if one rewrites the rapidity argument of the dipole $S$-matrices 
in the r.h.s as $X(x)\to x_g/x$ (recall that $X_g=x_g$) and one observes that e.g.
$S(\bx,\bz; x_g/x)$ describes the evolution of the wavefunction of the right-moving dipole $(\bx,\,\bz)$ 
from a longitudinal momentum scale $xq^+_0$ (for the valence $q\bar q$ pair) down to 
$x_gq^+_0$ (for the softest gluons that matter for the scattering).
}

\begin{figure*}
\begin{center}
\begin{minipage}[b]{0.49\textwidth}
\begin{center}
\includegraphics[width=0.9\textwidth,angle=0]{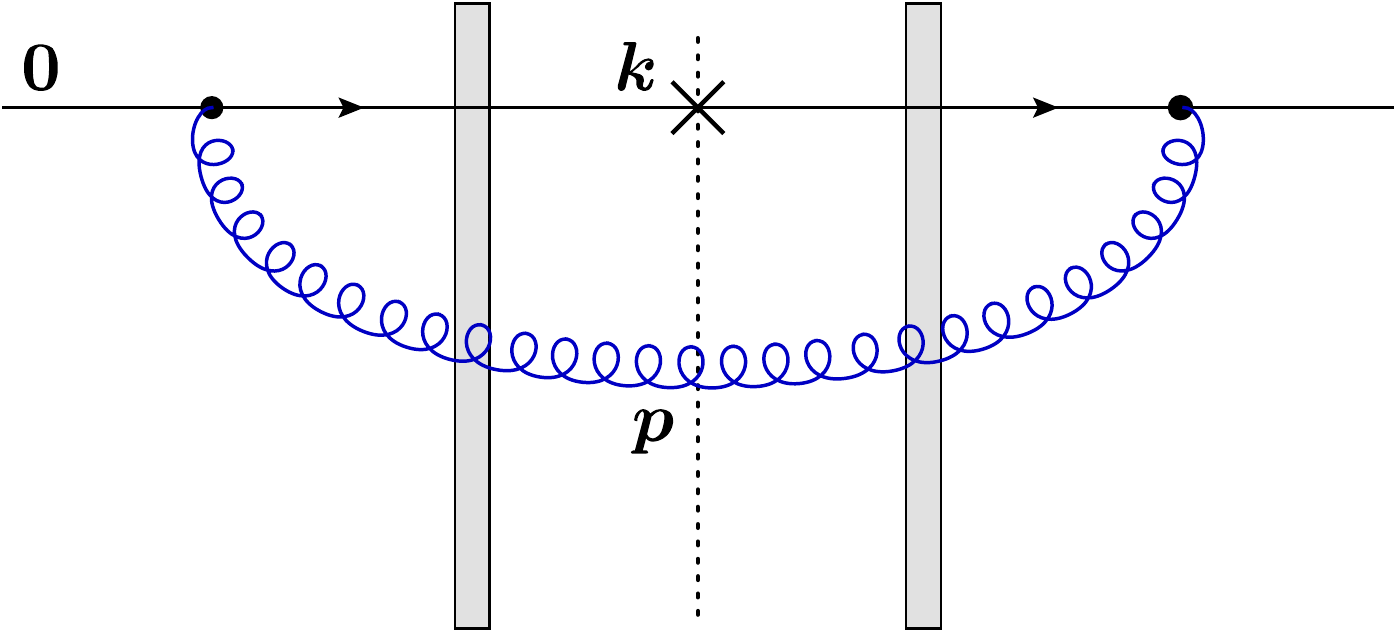}\\(a)\vspace{0.5cm}
\end{center}
\end{minipage}
\begin{minipage}[b]{0.49\textwidth}
\begin{center}
\includegraphics[width=0.9\textwidth,angle=0]{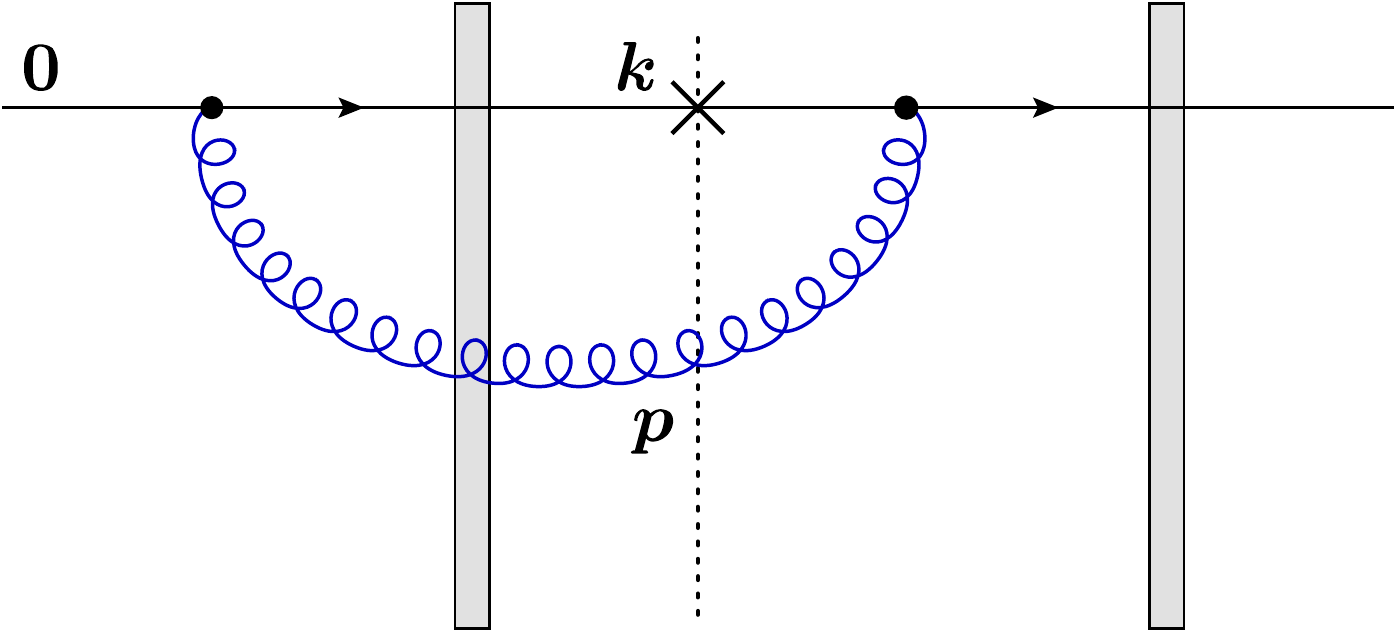}\\(b)\vspace{0.5cm}
\end{center}
\end{minipage}
\begin{minipage}[b]{0.49\textwidth}
\begin{center}
\includegraphics[width=0.92\textwidth,angle=0]{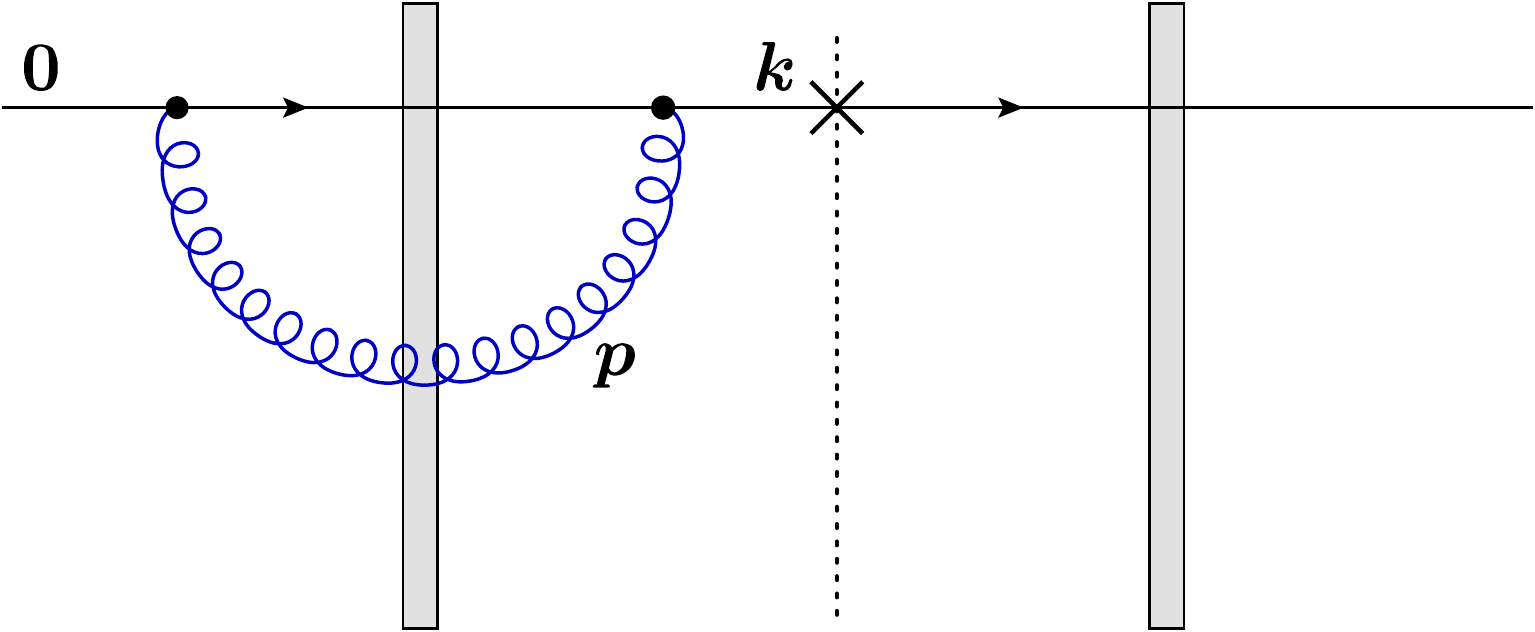}\\(c)\vspace{0cm}
\end{center}
\end{minipage}
\begin{minipage}[b]{0.49\textwidth}
\begin{center}
\includegraphics[width=0.92\textwidth,angle=0]{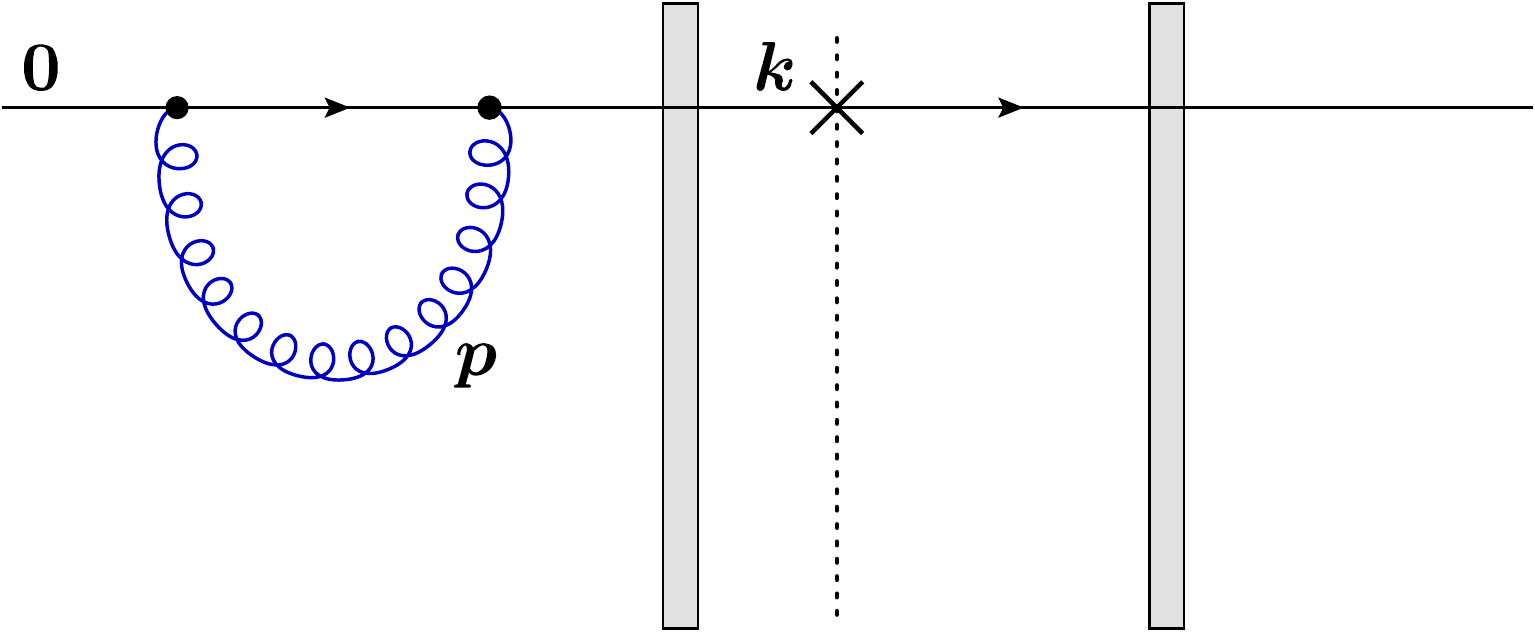}\\(d)\vspace{0cm}
\end{center}
\end{minipage}
\end{center}
\caption{\label{fig:shocks} \small
Typical diagrams contributing to the high-energy evolution of the quark production at leading order. (a) Real diagram, in which all possible interactions of the gluon with the target cancel one another. (b) Real diagram in which the gluon in the CCA is emitted before the collision with the shockwave, while in the DA is emitted after the collision. (c,d) Virtual diagrams in which the gluon interacts, or not, with the shockwave.  Only transverse momenta are shown in all the graphs.}
\end{figure*}

In Fig.~\ref{fig:shocks}, we present some Feynman graphs which contribute to the integral
term in \eqn{BKint}.
We do not use the dipole picture, rather we show graphs which enter the cross-section for quark production (so, in particular, we use the transverse momentum representation).
The nuclear target evolved up to $X(x)$ is represented as a shockwave (recall that we are still
in a frame where the target is ultrarelativistic). There are two types of graphs: `real', where 
the primary gluon appears in the final state --- it is emitted in the direct amplitude (DA) and
reabsorbed in the complex conjugate amplitude (CCA) --- and `virtual', where the gluon
is both emitted and reabsorbed on the same side of the cut (either in the DA, or in the CCA).
It is instructive to notice how such graphs are generated from the BK equation 
\eqref{BKint}: decomposing the dipole kernel there as
\beq\label{dipole}
{\mathcal M}_{\bx\by\bz}\equiv
\frac{(\bx-\by)^2}{(\bx-\bz)^2(\by-\bz)^2}=\frac{1}{(\bx-\bz)^2} + 
\frac{1}{(\by-\bz)^2} -  2\,\frac{x^i-z^i}{(\bx-\bz)^2}\frac{y^i-z^i}{(\by-\bz)^2}\,,
\eeq
one can check that the `virtual' terms are generated by the first 2 terms in the r.h.s.
of \eqn{dipole}, whereas the `real' terms come from the third one. Note that, for the particular
 `real' term where the gluon crosses the shockwave twice, cf. Fig.~\ref{fig:shocks}.a, the gluon
interactions cancel between the DA and the CCA, by unitarity, hence the respective
contribution to the r.h.s. of \eqn{BKint} involves only the scattering of the original
quark (as described by the dipole $S$-matrix $S\big(\bx,\by; X(x)\big)$).

Given the `boost-invariance' of the (LO) BK equation alluded to above, one may wonder
what can be the utility of dividing the evolution between target and projectile, as we
did above. As a matter of facts, there are several advantages for doing that. First, one
should keep in mind that the laboratory frame for `dilute-dense' (d+Au or p+Pb) collisions
at RHIC and the LHC coincides with the COM frame (at RHIC), or is close to it
(at the LHC). Hence the picture of the high-energy evolution which is directly visible
in the experiments is that of an evolution shared by the two incoming hadrons. Second,
we shall shortly argue that the first gluon emission by the incoming quark (the `primary
gluon') plays in fact a special role, at least for relatively large $\kt \gtrsim Q_s$. Because 
of that, it is preferable (and even compulsory, starting with NLO) to view this gluon
as a part of the quark evolution, like in \eqn{BKint}. Still beyond LO, it is 
conceptually simpler to associate the high-energy evolution with the target wavefunction.
As we shall see, the complete result for the quark multiplicity to NLO, to be
presented in Sect.~\ref{sec:NLO}, can be viewed as a natural generalization of \eqn{BKint}.

\subsection{Hard transverse momentum and di-jet events}
\label{sec:dijet}

In this subsection, we shall discuss the physical picture of forward quark production
in the IMF of the projectile, or, more generally, in any `mixed' frame, like
that illustrated in Fig.~\ref{QtoQG}, where the quark wavefunction contains at least
one soft gluon. We would like to show that, in any such a frame, the tail of the quark distribution 
at relatively high $\kt$ comes from the recoil in the emission of the {\em primary gluon}
(the first gluon emitted by the quark). That is, a forward quark with large transverse momentum
$\kt \gtrsim Q_s$ is produced in a {\em di-jet event} where the quark is accompanied by a recoil
gluon and the two particles propagate back-to-back in the transverse plane.
This point is important in that it will affect the NLO
calculation of the quark production at relatively high $\kt$, where the negativity problem in
the cross-section has been observed. 


At a first sight, the prominence of the di-jet configuration 
at large $\kt$ might look rather obvious, as an immediate
consequence of transverse momentum conservation at the emission vertex. But the situation
is a bit more subtle, since a large transverse momentum can also be transferred by the target,
via a sufficiently hard scattering. As a matter of facts, in the classical approximation at
low energy (i.e. in the absence of any evolution), a power-like tail $\propto 1/\kt^4$ 
in the quark distribution at high $\kt$ is generated via Coulomb scattering (see below). 
The same physical picture would also hold at high energy (within the limits of the LLA), 
but only in the target IMF, where there is no gluon emission by the quark. 
But in a frame where the quark itself is allowed to
radiate, the phase-space for high-energy evolution at high $\kt$ favors configurations 
where the momentum $\lt$ transferred from the target to the projectile (the quark together
with its small-$x$ radiation) is relatively low, $\lt\ll\kt$.
Because of that, the only way to produce a quark with very large $\kt$ is via a di-jet event,
as anticipated.

\begin{figure}[t] \centerline{
\includegraphics[width=.7\textwidth]{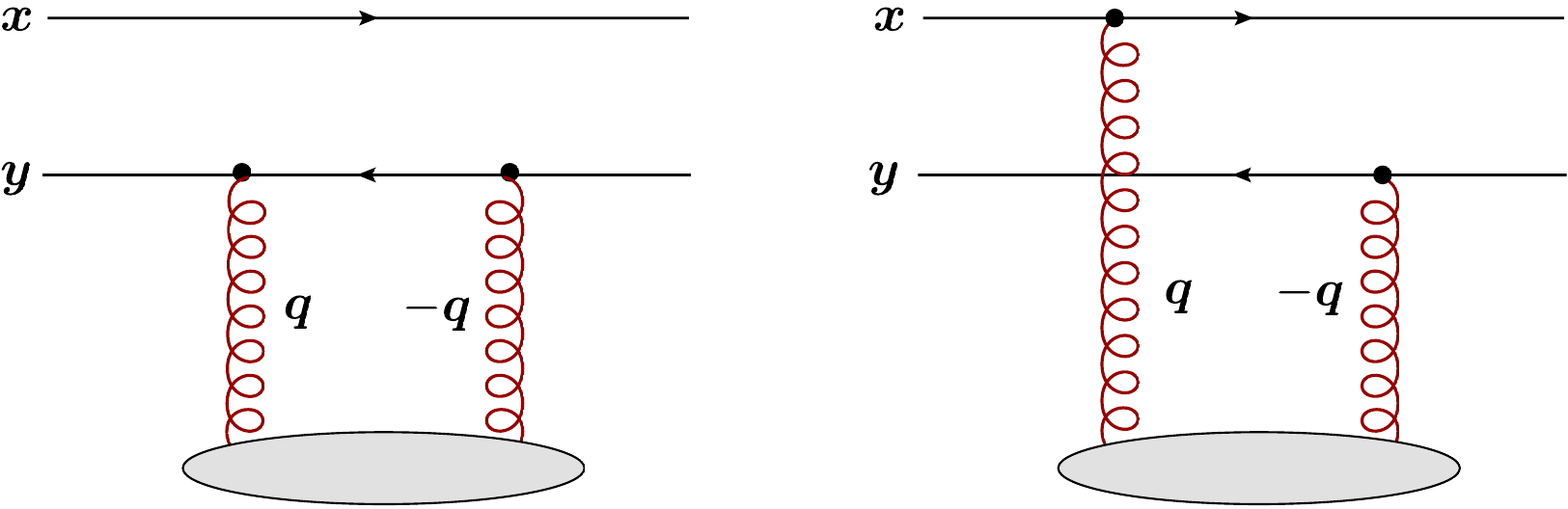}}
 \caption{\small Diagrams for the elastic scattering of the dipole in the single scattering
 approximation, or 2 gluon exchange. The blob at the bottom of the diagram refers to the
 average over the color fields in the target, which effectively generates the gluon
 distribution on the resolution scale $r$ of the dipole projectile.}
 \label{Dipole_2g}
\end{figure}

Consider first the semi-classical approximation (no evolution), that we shall treat within the
MV model. Since we are interested in a relatively hard quark with $\kt\gg Q_s$, we can limit
ourselves to the single-scattering approximation, as obtained by expanding the Wilson
lines in \eqn{Sdip} up to second order in the target color fields $A^-$ (see Fig.~\ref{Dipole_2g}). Writing 
$S=1-T$, one finds the dipole scattering amplitude in the 2-gluon exchange approximation as
(recall that $\br=\bx-\by$)
\begin{align}\label{T2g}
 \hspace*{-1.5cm}
 T_{0}(\br)&=
 \frac{g^2}{2N_c} \left\langle\big(A^-_a(\bx)-
A^-_a(\by)\big)^2\right\rangle\nonumber\\
&=g^2C_F\!
\int\frac{\rmd^2\bq}{(2\pi)^2}
  \frac{\mu^2}
  {q_\perp^4}\big[1-\rme^{i\bq\cdot\br}\big]
  \simeq\frac{\alpha_sC_F}{4}\,r^2\mu^2\ln\frac{1}{r^2\Lambda^2}\,,
\end{align}
where in the second line 
we have used the MV model expression for the 2-point correlator of the color fields
in a dense nucleus (with atomic number $A$ and transverse area $\pi R_A^2$), namely
\begin{align}\label{correlmed}
\int\rmd x^+\rmd y^+
\left\langle A^-_a(x^+,\bq)\,A^{-}_b(y^+,\bl)\right\rangle_{\mv}=
(2\pi)^2\delta^{(2)}(\bq+\bl)\delta_{ab}\,\frac{\mu^2}{\bq^4}\,,\qquad
 \mu^2= \frac{g^2C_F AN_c}{(N_c^2-1)\pi R_A^2}\,.
\end{align}
The quantity $\mu^2$ represents the color charge squared of the $AN_c$ valence quarks
(treated as uncorrelated color sources) per unit transverse area. 
The variable $\bq$ that is integrated over in \eqn{T2g} is the transverse 
momentum transferred from the target to the dipole and $\Lambda$ is an infrared cutoff
(say, the confinement scale). 
The unit term  within the square brackets corresponds to the case where the 
two exchanged gluons are attached to a same quark leg within the dipole, while the 
exponential $\rme^{i\bq\cdot\br}$ refers to attachments to both legs (see Fig.~\ref{Dipole_2g}).
For relatively small dipole sizes $r\ll 1/\Lambda$, the integral over $\qt$ develops a transverse
logarithm which can be isolated by expanding out the exponential to second order. This yields
the final result shown in \eqn{T2g}. When this result becomes of $\order{1}$, multiple scattering
becomes important and the above approximation breaks down. 
This condition defines the target saturation momentum  $Q_0$ at low energy:
$T_{0}(r)\sim 1$ for $r\simeq 1/Q_0$.

\begin{figure}[t] \centerline{
\includegraphics[width=.8\textwidth]{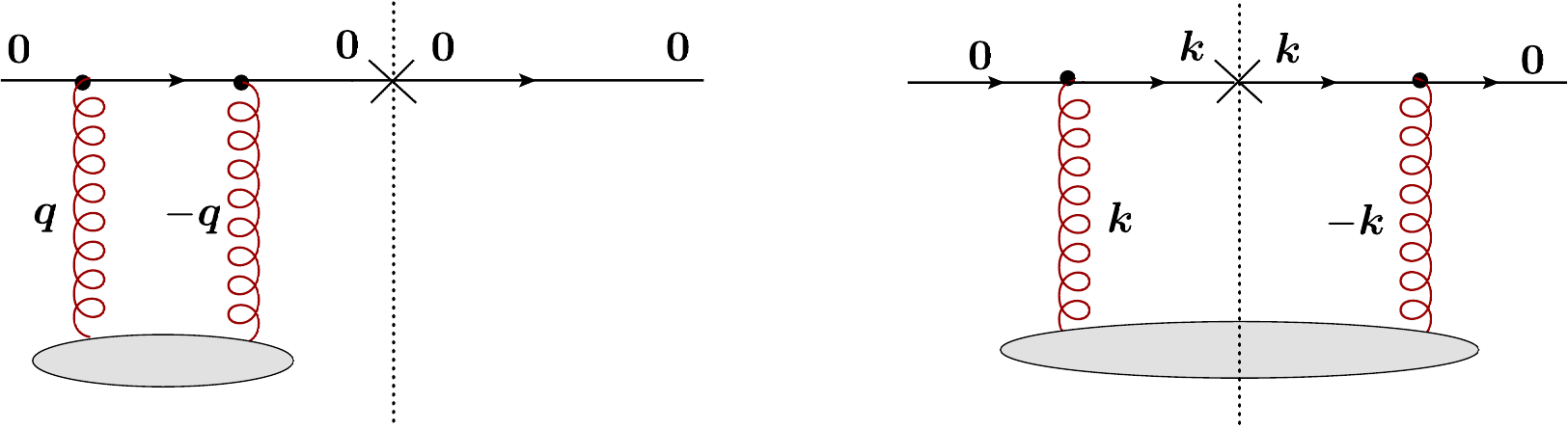}}
 \caption{\small Diagrams for quark production in the 2-gluon exchange approximation. 
 The diagram on the left describes an
 elastic scattering in the DA and no scattering in the CCA; hence it contributes
 to quark production only for $\bk=0$. The diagram on the right describes an inelastic
 scattering in the DA and another one in the CCA. The final momentum $\bk$ of the
 produced quark is transferred by the target.
Momenta are flowing from left to right both in the DA and in the CCA 
and from bottom to top in the exchange (red) gluons.}
 \label{Quark_2g}
\end{figure}

This simple calculation makes it clear that the scattering of a small dipole ($1/r\gg Q_0$) is
controlled by relatively soft gluon exchanges ($\Lambda \ll \qt\ll 1/r$) with the target. 
Let us similarly compute the
quark production, for a quark with large transverse momentum $\kt \gg Q_0$.
When taking the Fourier transform of $T_{0}(\br)$, the unit term within the square brackets in
 \eqn{T2g} does not matter (this would describe an elastic scattering without net momentum
transfer; see Fig.~\ref{Quark_2g}), whereas the exponential term there selects $\bq=\bk$. This is simply
the expression of momentum conservation and confirms that one needs a hard (inelastic)
scattering in order to produce a hight-$\kt$ quark. One thus finds\footnote{Notice that the
Fourier transform of the dipole scattering amplitude is defined with a minus sign, $T(\br) \to 
-\calT(\bk)$, in such a way that $\calT(\bk)$ has the same sign as $\calS(\bk)$ ; one therefore has
$\calS(\bk) = (2\pi)^2 \delta^{(2)}(\bk) + \calT(\bk)$.} 
$\calT_0(\kt)=g^2C_F\mu^2/\kt^4$,
which is recognized as the Rutherford cross-section for the Coulomb scattering between the quark
and the nucleus; therefore,
\beq\label{MV}
 \frac{\rmd N^{pA\to qX}}{\rmd^2\bk\, \rmd \eta}\bigg|_{\mv}
\simeq \,  x_p q(x_p)\,
\frac{\alpha_s C_F \mu^2}{\pi\kt^4}\,\qquad\mbox{for}\quad \kt\gg Q_0\,.
\eeq

We shall now study the high-energy evolution of the above results, in the double logarithmic
approximation (DLA) which is appropriate for sufficiently small dipole sizes, or large $\kt$. For the
present purposes, it is convenient to work in a frame where this is viewed as {\em projectile} evolution;
that is, the soft gluons belong to the wavefunction of the quark and they are all right movers.

\begin{figure}[t]
\begin{minipage}[b]{0.49\textwidth}
\begin{center}
\includegraphics[scale=0.51]{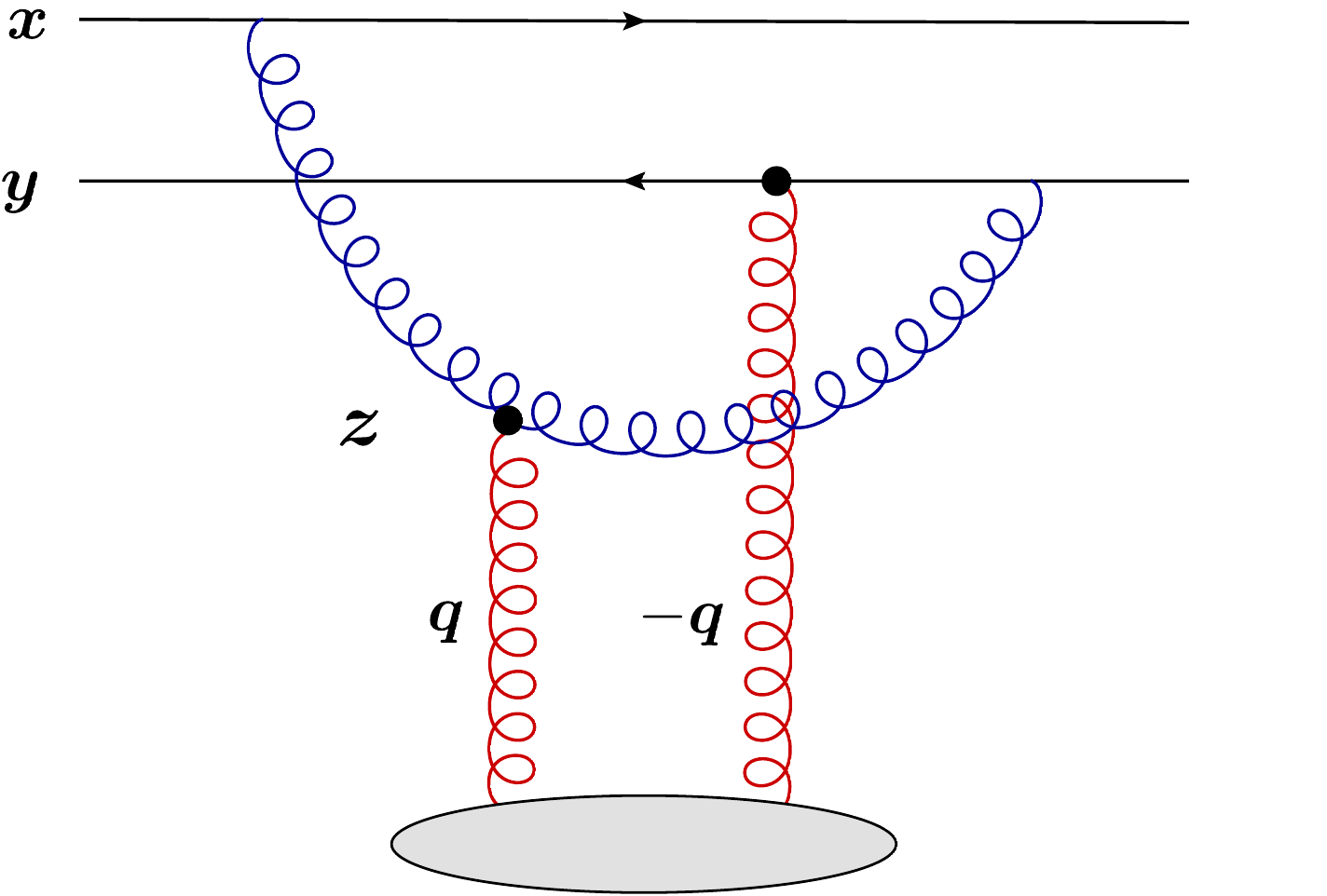} 
\end{center}
\end{minipage}
\begin{minipage}[b]{0.49\textwidth}
\begin{center}
\includegraphics[scale=0.53]{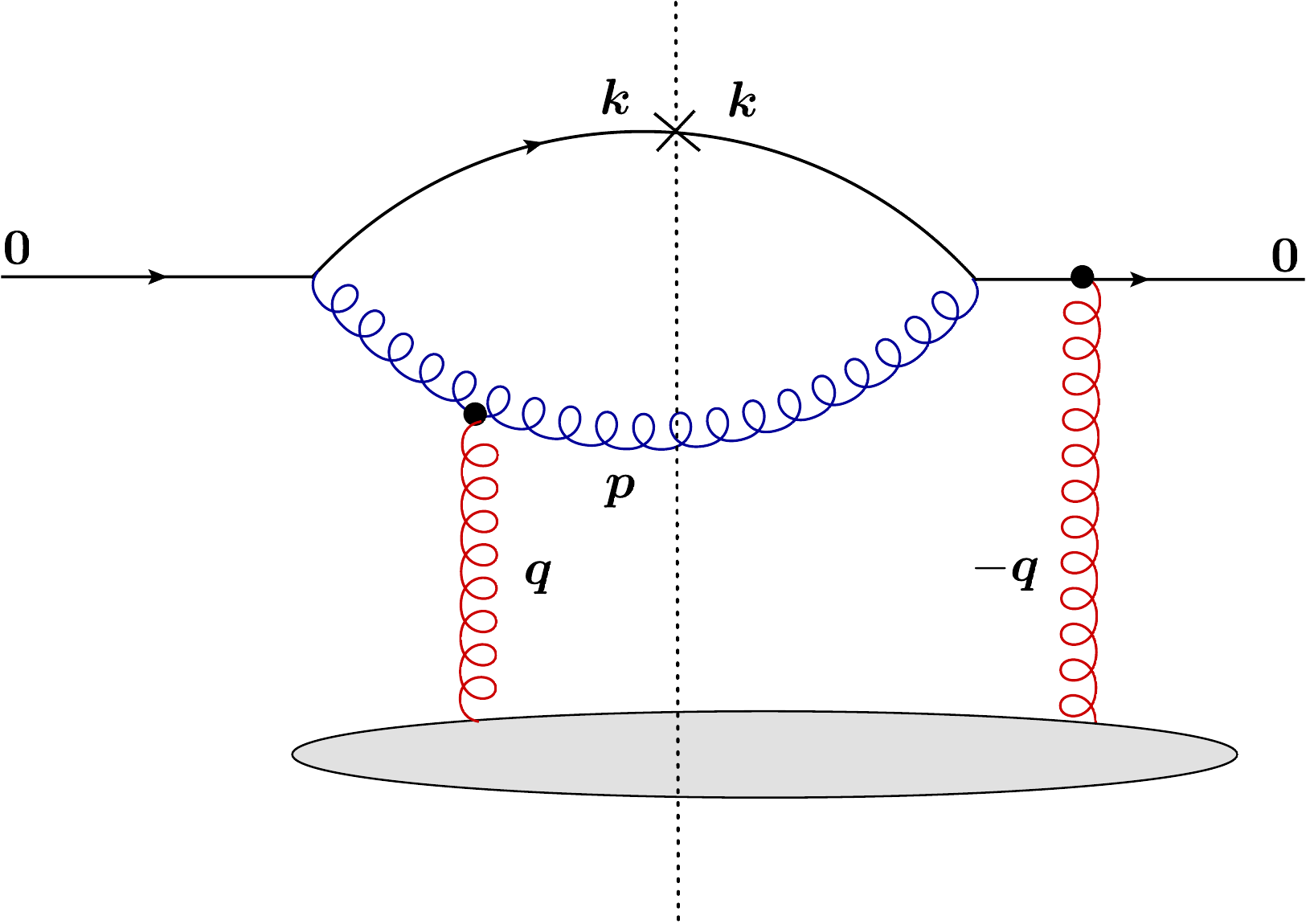} 
\end{center}
\end{minipage}
\caption{\small Left: one step in the DLA evolution of a small dipole, with size $r\ll 1/Q_s$. 
The daughter gluon is
typical soft and thus emitted at a large distance  $|\bz-\bx| \simeq |\bz-\by| \gg r$ from the
parent dipole. The gluon exchange $\bq$ with the nuclear target is even softer.
Right: one step
in the DLA evolution of the cross-section for quark production, at large transverse
momentum $\kt\gg Q_s$.
The primary gluon is as hard as the produced quark and they are both much harder
than the gluon exchanged with the target: $\kt\simeq\pt \gg \qt=|\bk+\bp|$.
The other diagrams contributing to this process at the level of the amplitude
are shown in Fig.~\ref{Recoil}.}
\label{fig:DLA}
\end{figure}

In transverse coordinate space, the DLA corresponds to the splitting of the original dipole 
$(\bx, \,\by)$ into two daughter dipoles, $(\bx, \,\bz)$ and $(\bz, \,\by)$,  whose transverse
sizes are much larger, but still small enough to undergo only single scattering: 
$r \ll \bar z \ll 1/Q_0$, with $\bar z\equiv |\bx-\bz| \simeq |\bz-\by|$.  The respective evolution 
equation is obtained from the general BK equation \eqref{BK} by \texttt{(i)} linearizing w.r.t. $T=1-S$
(by itself, this step yields the BFKL equation), then \texttt{(ii)} approximating the dipole kernel as
${\mathcal M}_{\bx\by\bz}\simeq r^2/\bar z^4$, and 
\texttt{(iii)} keeping only the scattering amplitudes $T(\bx,\bz)+T(\bz,\by)\simeq 2T(\bar z)$ 
for the two daughter dipoles, whose scattering is stronger (since $T(r)\propto r^2$
in this physical regime). One thus finds (as before, we use $y=\ln(1/x)$ for the evolution
`time' of the projectile)
\begin{align}\label{DLA}
 \frac{\del }{\del y} \,T(r, y)=
 {\abar}\,r^2 \int_{r^2}^{1/Q_0^2} \frac{\rmd\bar z^2}{\bar z^4}\,
T(\bar z, y)\,.
 \end{align}
Since $T(\bar z)\propto \bar z^2$, the integral in the r.h.s. is clearly logarithmic.
The first iteration of this equation, as obtained by evaluating its r.h.s. with the
amplitude $T_{0}$ from \eqn{T2g}, describes the first gluon emission by the parent dipole.
The physical picture of this emission follows from the previous discussion:
the original dipole with size $r$ emits a relatively soft gluon with transverse momentum $\pt\sim 1/\bar z$
within the range $Q_0\ll \pt\ll 1/r$, which then suffers an even softer scattering off the nuclear
target, with transferred momentum $\Lambda\lesssim \qt\ll \pt$ (see Fig.~\ref{fig:DLA} left). This picture extends
to the whole gluon cascade generated by iterating \eqn{DLA}: 
successive gluon emissions are strongly ordered not only in $x$ but also in
transverse momenta, and the final exchange with the target is even softer.

\begin{figure}[t] \centerline{
\includegraphics[width=.96\textwidth]{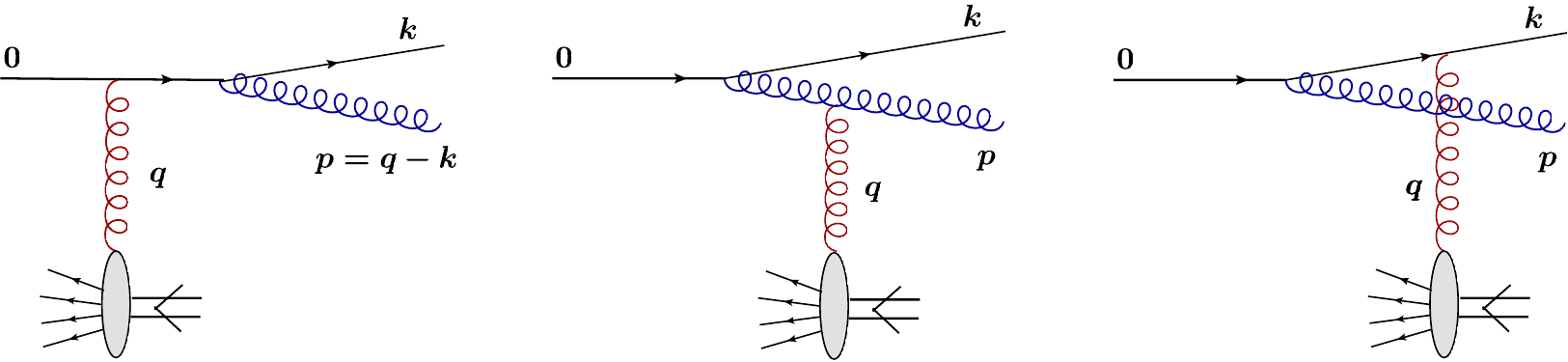}}
 \caption{\small The 3 diagrams which contribute to the production of a quark-gluon pair
 in the final state in the regime where both the quark and the gluon are relatively hard
 ($\kt\simeq\pt \gg \qt=|\bk+\bp|$).}
 \label{Recoil}
\end{figure}

We now turn to the corresponding picture in transverse momentum space, that is, 
to the problem of quark production (see Fig.~\ref{fig:DLA} right and also Fig.~\ref{Recoil}). 
The momentum-space DLA equation reads
\begin{align}\label{DLAk}
 \frac{\del }{\del y} \,\calT(\kt, y)=
 \frac{\abar}{\kt^4} \int_{Q_0^2}^{\kt^2} \rmd\qt^2\,\qt^2 \calT(\qt, y)\,,
 \end{align}
where the integral in the r.h.s. is indeed logarithmic, since $\calT(\qt, y)\propto 1/\qt^4$.
Within this integral, $\calT(\qt, y)$ should be interpreted as the cross-section for a single scattering, 
with transferred momentum $\qt$, between partons in the quark wavefunction
and the target. The factor ${\abar}/{\kt^4}$ in front of the integral does not represent
anymore a $t$-channel exchange with the target, as in \eqn{MV}, but rather it comes
from the propagator of the intermediate quark, or gluon, in the $s$-channel
(see Fig.~\ref{Recoil}). Hence, the physical picture of the first emission is now as follows: 
the original quark with zero transverse momentum emits a gluon
with  momentum $\bp$ and turns into a final quark with momentum $\bk$,
while at the same time receiving a momentum transfer $\bq$ from the target 
(via a scattering that can occur either before, or after the splitting). Transverse 
momentum conservation requires $\bq=\bk+\bp$. But the overall cross-section, as described
by \eqn{DLAk}, favors {\em soft} scattering, with transferred momenta $\qt\ll \kt$. Accordingly, 
the first emitted gluon must be hard, $\pt\simeq \kt$, to balance the momentum
of the produced quark.  

As for the subsequent gluon emissions, starting with the second one,
they follow the standard DLA ordering, in both $x$ and $\pt$, as in the respective calculation
in coordinate space, cf. \eqn{DLA}. This argument too shows that, when computing particle
production, it is quite natural to associate the primary gluon with the wavefunction of the produced
particle, whereas the other gluons are more conveniently included in the gluon distribution of
the target, as measured by the hard splitting process. While natural already at LLA, this viewpoint 
becomes almost unavoidable when moving to the next-to-leading order calculation, where 
the primary gluon is also allowed to have a large longitudinal momentum $p^+\sim q^+_0$.
The NLO calculation will be discussed in the next section.

\section{Next-to-leading order}
\label{sec:NLO}

In order to move on to next-to-leading order (NLO) accuracy,
one must relax some of the previous approximations and add new contributions
which start at NLO. By inspection of the LO result \eqref{LO}, it is clear that one 
ingredient required in that sense is the NLO version of the B-JIMWLK (or BK) equations
\cite{Balitsky:2008zza,Balitsky:2013fea,Kovner:2013ona}, together with their all-order
`collinear' resummations  \cite{Beuf:2014uia,Iancu:2015vea,Iancu:2015joa,Hatta:2016ujq}.
This in particular means that some gluon emissions must
be computed beyond the eikonal approximation: besides the effect of order $\alpha_s Y$, 
which dominates at high energy, one must also keep, for each such an emission, 
the `pure-$\alpha_s$' corrections which are not enhanced by the rapidity logarithm
$Y$ (but may be accompanied by transverse logarithms). So long as these
NLO corrections refer to generic gluons inside the cascade, they can be absorbed
into a renormalization of the kernel of the evolution equation. 
The same is true for the quark-antiquark loop which at NLO can be inserted within any 
of the gluon lines. But the NLO corrections associated with the  `primary gluon' 
(the very first emission by the leading quark) must rather be used to renormalize 
the `impact factor', i.e. the value of the cross-section in the absence of high-energy evolution. 

At LO, the impact factor is the cross-section for the inelastic scattering between
the leading quark and the low energy nucleus (say, as described by the MV model).
Equivalently (to the accuracy of interest), it can be written as the $S$--matrix for the 
elastic scattering of a $q\bar q$ dipole. At NLO, one must add the impact factor encoding 
the inelastic scattering of the quark-gluon pair made with 
the leading quark and the primary gluon. Unlike the emission of the primary gluon,
which must be computed exactly, the scattering between the quark-gluon pair and the target
can still be computed in the eikonal approximation and thus related to
elastic scattering amplitudes for color multipoles
\cite{Marquet:2007vb,Dominguez:2011wm,Chirilli:2012jd}. 

So, it may look like, in order to compute quark production at NLO, one must dress 
the two contributions to the impact factor aforementioned with the high-energy
evolutions of the respective scattering amplitudes 
(themselves computed at NLO) and then add the results. But a moment of
thinking reveals that the two pieces of the impact factor mix with each other under the
high-energy evolution: a part of the primary gluon emission that we have explicitly 
included in the NLO impact factor is also included (within the limits of the eikonal 
approximation) as the first small-$x$ gluon in the evolution of the dipole $S$--matrix 
from the LO cross-section \eqref{LO}. This is the problem of over-counting. 
Previous papers in the literature \cite{Chirilli:2011km,Chirilli:2012jd} proposed a solution 
to this problem, in the form of a `plus' prescription which subtracts the LO
evolution from the NLO impact factor. This prescription however appears
to be responsible for the problem with the negativity of the cross-section 
discussed in the Introduction.

In what follows, we shall propose a different way to organize the calculation,
which avoids the over-counting without performing any subtraction. Our strategy
will naturally exploit the structure of perturbation theory at high energy.
As we shall see, the contribution to the cross-section which includes the NLO correction 
to the impact factor does also encode, completely and faithfully, the LO evolution
of the dipole $S$-matrix. Hence, by computing this contribution as it stands,
one can simultaneously include both effects without any ambiguity, or over-counting.
On top of that, there is a NLO correction to the evolution of the color dipole; this
will be clearly identified and related to recent results concerning the NLO
version of the BK equation \cite{Balitsky:2008zza}  and its collinear resummations
\cite{Iancu:2015vea,Iancu:2015joa}.

\subsection{Revisiting the NLO calculation by Chirilli, Xiao, and Yuan}
\label{sec:CXY}

In this subsection, we shall exhibit, discuss, and adapt to our present purposes the result of the NLO
calculation of the impact factor by Chirilli, Xiao, and Yuan \cite{Chirilli:2011km,Chirilli:2012jd}.
First, we shall display their `bare', or `unsubtracted', result, where the soft  
divergence\footnote{We recall that $x=p^+/q^+_0$ is the longitudinal
momentum fraction of the primary gluon relative to the incoming quark. In 
Refs.~\cite{Chirilli:2011km,Chirilli:2012jd}, one has rather used the variable $\xi\equiv 1-x$,
hence our `soft divergence' at $x=0$ appears there as the `rapidity divergence' at $\xi=1$. 
To facilitate the comparison, in this section we shall use both notations, $x$ and $\xi$.} at $x\to 0$ is explicit. Then we shall briefly mention the `plus' prescription advocated in 
Refs.~\cite{Chirilli:2011km,Chirilli:2012jd} in order to subtract the rapidity divergence.
(We shall return to this point in Sect.~\ref{sec:sub}.) Finally, we shall explain our strategy
to deal with this problem, which is to use kinematical constraints like energy 
conservation in order to cut off the soft divergence and at the same time fix the rapidity variables
for the evolution of the dipole $S$-matrices. The only subtle point here is the treatment of the
virtual corrections, where the phase-space for the emission of the primary gluon is not directly 
constrained by the kinematics. Yet, as we shall demonstrate via explicit calculations
(in Appendix \ref{app:canc}), the same 
lower limit on $x$ applies in that case too, albeit its emergence is now {\em dynamical}. 
 
 \begin{figure}[t] \centerline{
\includegraphics[width=.55\textwidth]{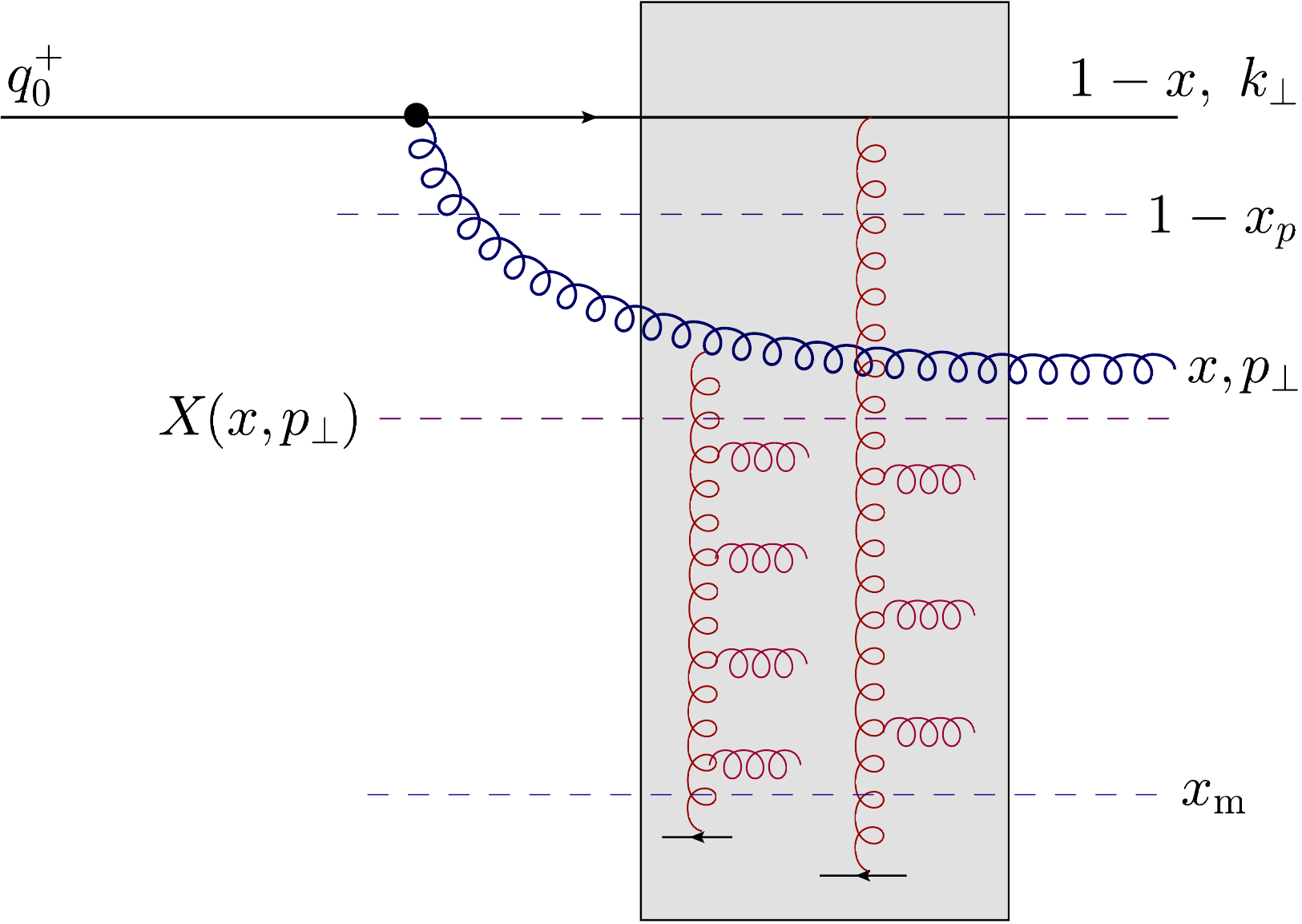}}
 \caption{\small Pictorial representation of a typical amplitude contributing to the NLO piece of the impact factor. 
 This is a  `real' amplitude,
 in the sense that the primary gluon is released in the final state. }
 \label{fig:NLO}
\end{figure}

\begin{figure}[t]
\begin{minipage}[b]{0.49\textwidth}
\begin{center}
\includegraphics[scale=0.53]{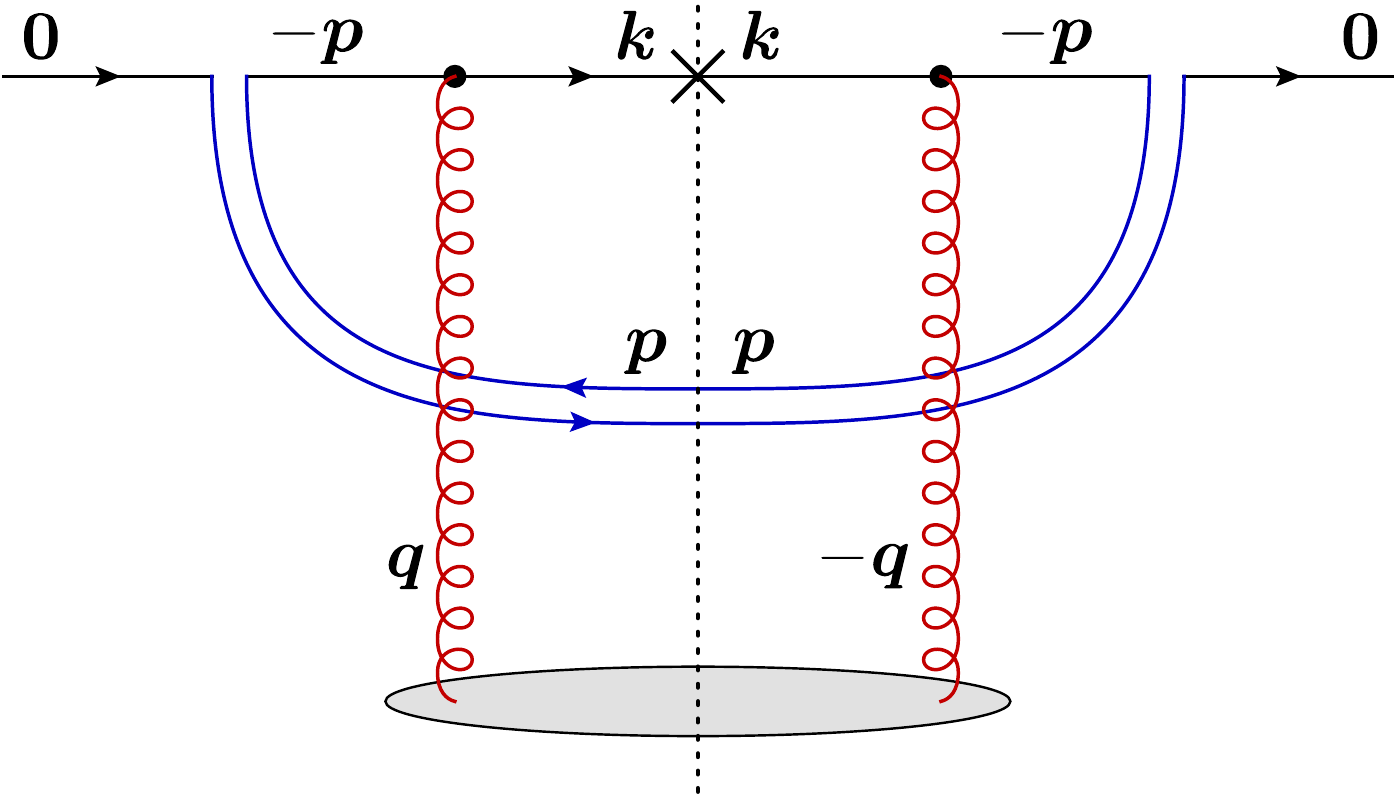}\\{\small (a)}
\end{center}
\end{minipage}
\begin{minipage}[b]{0.49\textwidth}
\begin{center}
\includegraphics[scale=0.53]{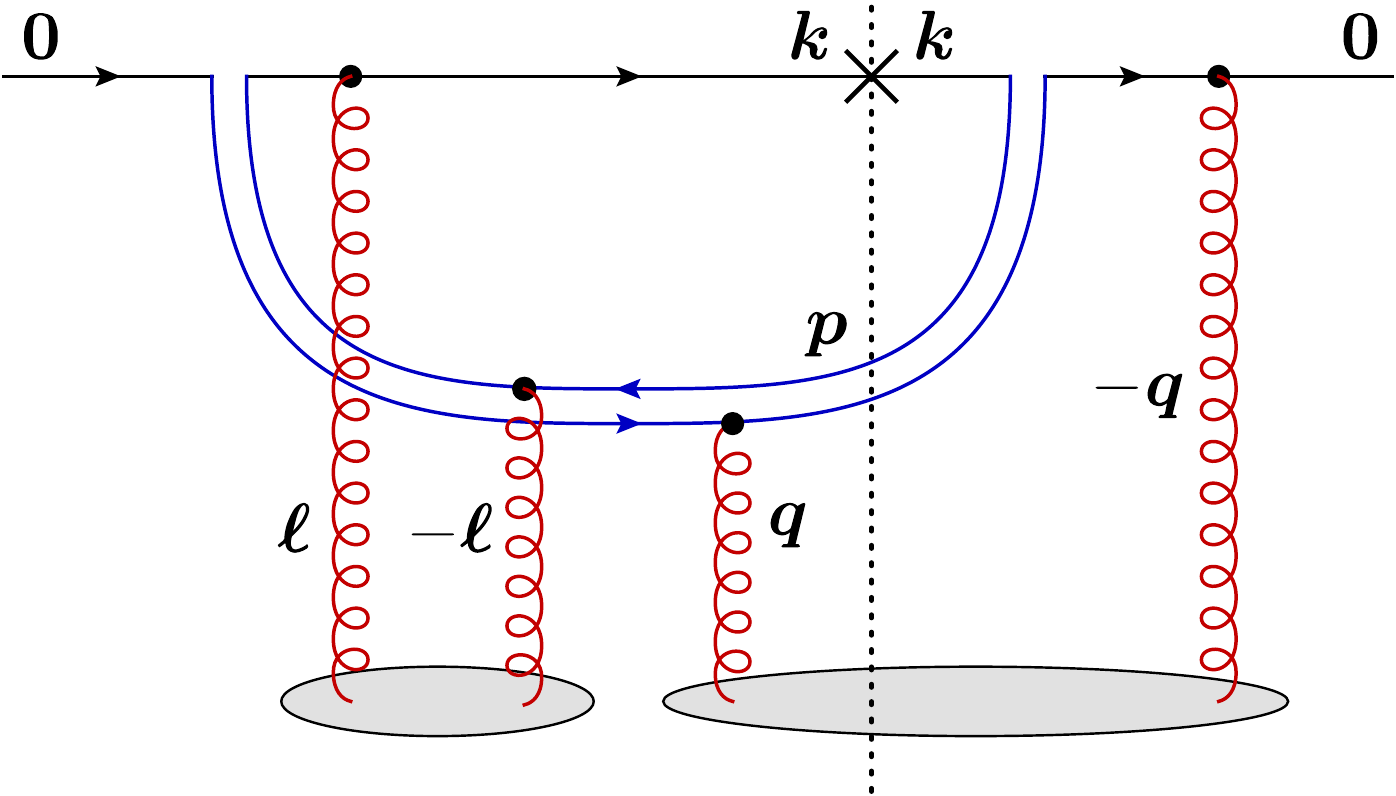}\\{\small (b)}
\end{center}
\end{minipage}
\caption{\small Production of a quark with a transverse momentum $\bm{k}$. Typical real diagrams, i.e.~diagrams in which the gluon is crossing the cut, but at the same time is integrated (cf. \eqn{Jbare}). 
The primary gluon is represented as a
$q\bar q$ pair, as appropriate at large $N_c$.
(a) Left: Diagram contributing to the real term proportional to $\calS(\bm{q})$ and which originates from $S(\bm{x},\bm{y})$ in coordinate space. All possible interactions of the gluon with the target cancel each other. (b) Right: Diagram contributing to the real term proportional to $\calS(\bm{q}) \calS(\bm{\ell})$ and which originates from $S(\bm{x},\bm{z}) S(\bm{z},\bm{y})$ in coordinate space. In both diagrams the target transfers momentum $\bm{q} = \bm{p} + \bm{k}$ to the final state. Momenta are flowing from left to right both in the DA and in the CCA and from bottom to top in the exchange (red) gluons.}
\label{fig:real}
\end{figure}

\begin{figure}[t]
\begin{minipage}[b]{0.49\textwidth}
\begin{center}
\includegraphics[scale=0.53]{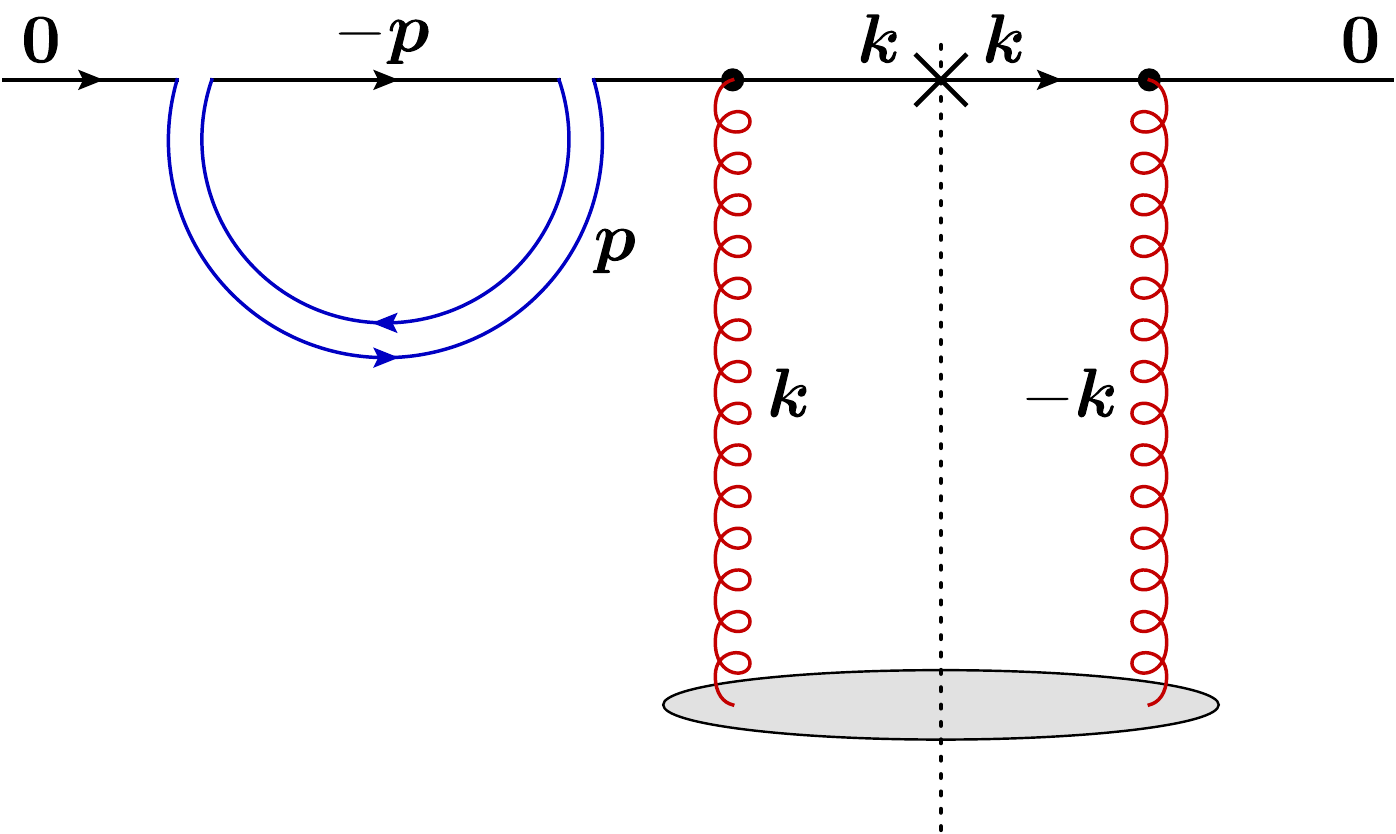}\\{\small (a)}
\end{center}
\end{minipage}
\begin{minipage}[b]{0.49\textwidth}
\begin{center}
\includegraphics[scale=0.53]{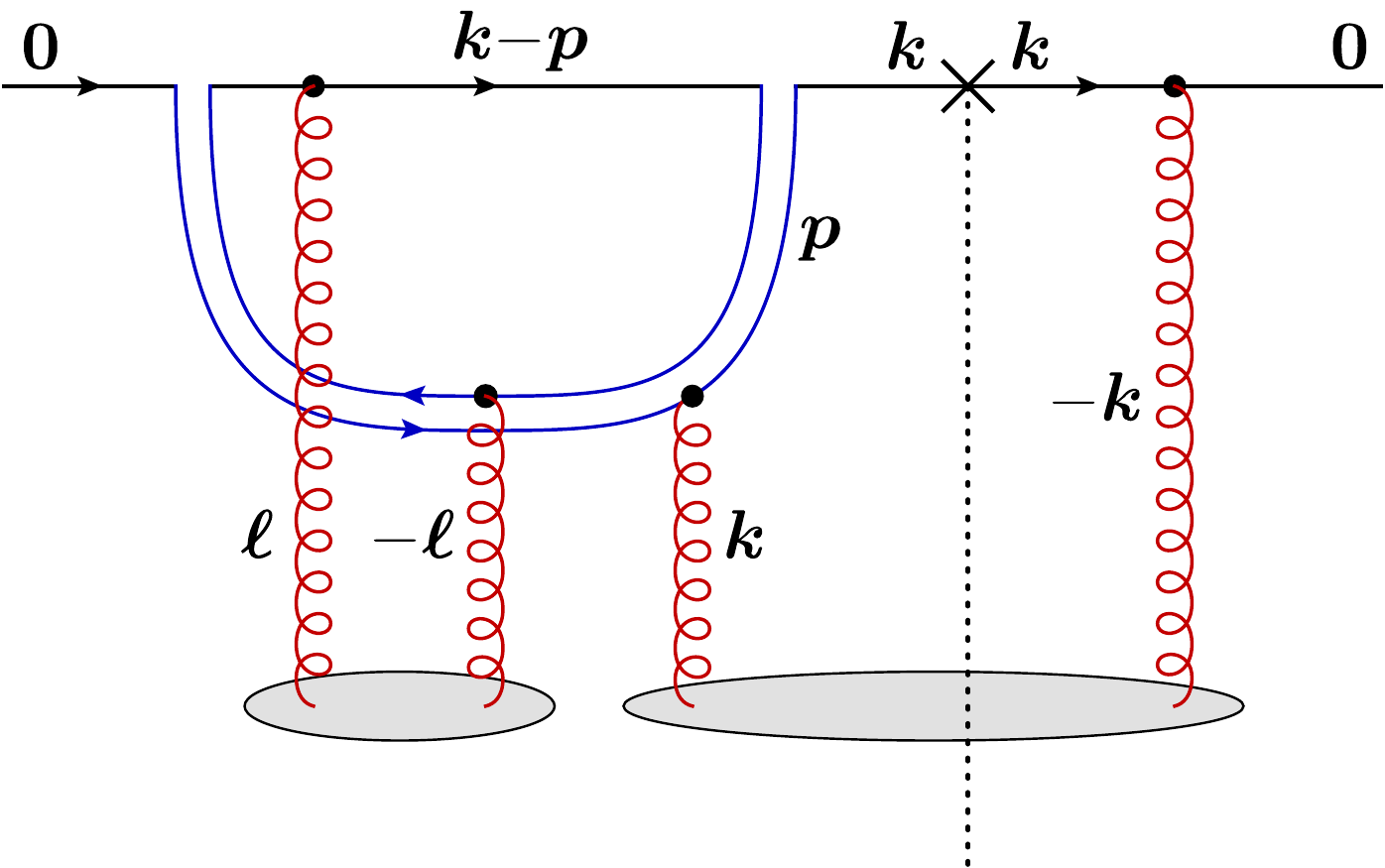}\\{\small (b)}
\end{center}
\end{minipage}
\caption{\small
Production of a quark with a transverse momentum $\bm{k}$. Typical virtual diagrams, i.e.~diagrams in which the gluon is not crossing the cut (cf. \eqn{Jvbare}). (a) Left: Diagram contributing to the virtual term proportional to $\calS(\bm{k})$, which originates from $S(\bm{x},\bm{y})$ in coordinate space (b) Right: Diagram contributing to the real term proportional to $\calS(\bm{k}) \calS(\bm{\ell})$ and which originates from $S(\bm{x},\bm{z}) S(\bm{z},\bm{y})$ in coordinate space. Momenta are flowing from left to right both in the DA and in the CCA and from bottom to top in the exchange (red) gluons.}
\label{fig:virtual}
\end{figure}

The NLO result in Refs.~\cite{Chirilli:2011km,Chirilli:2012jd} has been obtained by evaluating
Feynman graphs like that illustrated in Fig.~\ref{fig:NLO} in which the emission
of the primary gluon is treated exactly. There is a similar graph where the
gluon emission occurs after the scattering between the quark and the target. And there
are of course virtual graphs, whose evaluation is somewhat subtle as just mentioned 
and that we shall deal with in some detail.  (See Figs.~\ref{fig:real} and \ref{fig:virtual}
below for more examples of Feynman graphs.)
After the scattering, both the quark and the gluon will fragment into hadrons and
thus contribute to single-inclusive hadron production.  There is also
another channel where the original collinear parton is a gluon, which splits into
a pair of gluons, or into a quark-antiquark pair, 
in the process of scattering. As before, we shall omit the discussion of the 
fragmentation process and concentrate on quark production alone (see
Refs.~\cite{Chirilli:2011km,Chirilli:2012jd} for a complete discussion and also
\cite{Altinoluk:2014eka} for an alternative calculation, whose precise relation to the
original results in \cite{Chirilli:2011km,Chirilli:2012jd} is still unclear). That is, the primary
gluon is not measured, so one needs to integrate out its kinematics --- the longitudinal
momentum fraction $x=p^+/q_0^+$ and the transverse momentum $\bp$.

The NLO result in Refs.~\cite{Chirilli:2011km,Chirilli:2012jd} can be conveniently 
written as the sum of 2 pieces\footnote{\label{footFierz} 
These 2 different color structures are generated when 
using Fierz identities to rewrite the adjoint Wilson lines which describe the eikonal scattering
of the primary gluon in terms of fundamental Wilson lines. Accordingly, all the scattering operators
which appear in the final result are built with fundamental Wilson lines alone. At large $N_c$,
they are either linear, or bi-linear, in the dipole $S$-matrix 
(see Eqs.~\eqref{Jbare} and \eqref{Jvbare} below).}:

\texttt{(A)} A piece proportional to the quark Casimir $\CF$ which develops no logarithm
at small $x$ (the respective integrand vanishes as $x\to 0$), but has collinear divergences
in the transverse momentum integrations. In \cite{Chirilli:2011km,Chirilli:2012jd}, these divergences 
have been isolated with the help of dimensional regularization and reabsorbed into the leading-order 
DGLAP evolution of the quark distribution function $q(x_p)$ (if the primary emission occurs prior
to scattering) and of the quark-to-hadrons fragmentation function (if the emission occurs after
the scattering). This prescription leaves a finite remainder of NLO order whose explicit
evaluation poses no special problem.

\texttt{(B)} A piece proportional to the gluon Casimir $\Nc$ which is free of collinear problems but
develops a logarithm at small $x$ (the respective integral over $x$ exhibits a logarithmic divergence
at $x\to 0$ in the absence of any physical regulator). The proper way to deal with this
`rapidity divergence' at small $x$ represents our main concern in this paper. 
To better focus on this problem while avoiding cumbersome notations,
we shall omit the piece proportional to $\CF$ in what follows.
(This piece can be easily added to our main result shown in \eqn{NLO} below.)
As for the second piece, proportional to $\Nc$, we start by displaying the original result,
as presented in Refs.~\cite{Chirilli:2011km,Chirilli:2012jd} :
\beq
\label{eq:nlosigma}
 \frac{\rmd N^{pA\to qX}}{\rmd^2\bk\, \rmd \eta}\bigg|_{\nlo}^{\rm unsub}
=  \frac{\alpha_s \Nc}{(2\pi)^2}
\int_{0}^1 \rmd \xi \,\frac{1+\xi^2}{1-\xi}\left\{
\frac{x_p}{\xi} q\left(\frac{x_p}{\xi}\right)\jcal(\bk ,\xi)-
x_p q\left(x_p \right) \jcal_v(\bk ,\xi) \right\},
\eeq
where $\xi\equiv 1-x$ 
and the two functions $\jcal(\bk ,\xi)$
and $\jcal_v(\bk ,\xi) $ correspond to real and virtual contributions 
to the process illustrated in Fig.~\ref{fig:NLO}. They read
(our present notations are slightly different from the original ones
Refs.~\cite{Chirilli:2011km,Chirilli:2012jd}, but follow closely the recent paper
\cite{Ducloue:2016shw})
\beq\label{Jbare}
\jcal(\bk ,\xi) =
\!\int \frac{\rmd^2 \bq}{(2\pi)^2} 
\frac{2(\bk-\xi\bq)\cdot(\bk-\bq)}{(\bk-\xi\bq)^2(\bk-\bq)^2}
\calS(\bq)
-\int\! \frac{\rmd^2 \bq}{(2\pi)^2} \frac{ \rmd^2\bl}{(2\pi)^2}
\frac{2(\bk-\xi\bq)\cdot(\bk-\bl)}{(\bk-\xi\bq)^2(\bk-\bl)^2}
\calS(\bq)\calS(\bl),
\eeq
and respectively
\beq\label{Jvbare}
\jcal_v(\bk ,\xi) =
\calS(\bk)
\left[
\int\! \frac{\rmd^2 \bq}{(2\pi)^2} 
\frac{2(\xi\bk-\bq)\cdot(\bk-\bq)}{(\xi\bk-\bq)^2(\bk-\bq)^2}
-\int \!\frac{\rmd^2 \bq}{(2\pi)^2} \frac{  \rmd^2\bl}{(2\pi)^2}
\frac{2(\xi\bk-\bq)\cdot(\bl-\bq)}{(\xi\bk-\bq)^2(\bl-\bq)^2}
\calS(\bl)
\right].
\eeq
As before, the dipole $S$-matrices like $\calS(\bk)$ or $\calS(\bq)$ refer to dipoles in the fundamental
representation (cf. footnote \ref{footFierz}).
To simplify writing, we have considered the large $\Nc$ limit, in which the 
scattering of a system of two dipoles factorizes as the product of
two individual dipole $S$-matrices, but this limit is not essential for what follows.

The variables $\bq$ and $\bl$ which appear in the above integrations
represent transverse momenta exchanged between the target and the quark-gluon pair.
For what follows, it is important to understand their precise meaning and notably
their relation with the transverse momentum $\bp$ taken by the primary gluon.
By following the derivation of these results in Refs.~\cite{Chirilli:2011km,Chirilli:2012jd}, 
one can check
that $\bq=\bp+\bk$ whereas $\bl$ is independent of $\bp$.
For more clarity, let us briefly discuss the physical interpretation 
of the various terms in Eqs.~\eqref{Jbare} and \eqref{Jvbare}.

The `real'  terms in \eqn{Jbare} represent processes where the 
primary gluon, albeit not measured, is released in the final state (see Fig.~\ref{fig:real}).
For such processes, longitudinal momentum conservation implies $\xi= k^+/q^+_0$.
The first term in \eqn{Jbare}, which is linear in $\calS(\bq)$, represents situations where the 
hard splitting occurs  either after the collision, or prior to it, in {\em both} the DA and the CCA.
In these cases, the gluon either does not interact with the target at all (emissions after the
collision), or the effects of its interaction cancel out from the final result, by unitarity, because
the gluon is not measured (emissions before the collision). Accordingly, there
is only one dipole $S$-matrix, $\calS(\bq)$, which physically describes the {\em inelastic} scattering
of the quark. This scattering transfers a non-zero transverse momentum $\bq$ to the quark;
then momentum conservation implies $\bq=\bp+\bk$, as aforementioned.

The second term in \eqn{Jbare}, bilinear in the dipole $S$-matrix, corresponds to interference 
processes, where the primary gluon is emitted prior to scattering in the 
direct amplitude (DA) and after the scattering in the complex conjugate
amplitude (CCA), or vice-versa. In such processes, both the quark and the gluon can
participate in the collision. At large $N_c$, this yields 2 dipole $S$-matrices:
one made with the quark in the DA and the antiquark piece of the gluon, the other
one with the quark piece of the gluon and the antiquark in CCA. One of these $S$-matrices,
denoted as $\calS(\bl)$ in \eqref{Jbare}, describes the {\em elastic} scattering
of a {\em physical} dipole --- i.e. a dipole whose both fermion legs exist on the same side
of the cut (either in the DA, or in the CCA). For this elastic scattering, there is no net transfer
of transverse momentum; e.g., if $\calS(\bl)$ is computed in the 2-gluon exchange approximation,
then the momentum $\bl$ transferred by the first exchanged gluon towards the dipole is 
subsequently taken back by the second exchanged gluon. The other
dipole $S$-matrix, $\calS(\bq)$, describes an inelastic scattering with net momentum transfer 
$\bq=\bp+\bk$.

Consider similarly the `virtual' contributions encoded in  \eqref{Jvbare} (see Fig.~\ref{fig:virtual}). 
In that case,
the primary gluon is both emitted and reabsorbed on the same side of the cut, hence
the momentum $\bk$ of the produced quark fully comes via inelastic scattering (and
$k^+=q_0^+$).
In the first term in \eqref{Jvbare}, the gluon fluctuation has no overlap with the target,
hence the (inelastic) scattering refers to the quark alone. In the second term, the gluon
can scatter too. Accordingly, this term involve 2 dipole $S$-matrices, one describing
an elastic scattering ($\calS(\bl)$), the other one an inelastic one ($\calS(\bk)$).

The following observations will be useful for the subsequent arguments:

\texttt{(i)} In \eqn{eq:nlosigma} one recognizes the full LO DGLAP quark-to-quark
splitting function $P_{qq}(\xi)$, in line with the fact that the gluon emission has been
treated exactly, and not in the eikonal approximation.

\texttt{(ii)} In Eqs.~\eqref{Jbare} and \eqref{Jvbare}, the splitting fraction $\xi$ is visible only 
in the various kernels describing the transverse momentum structure of the hard splitting,
which in turn have been generated by combining the light-cone energy denominator with
factors coming from the splitting vertex.

\texttt{(iii)} The various dipole $S$-matrices in Eqs.~\eqref{Jbare} and \eqref{Jvbare} are
supposed to describe scattering off the nuclear gluon distribution evolved up to the right
`rapidity' ($Y=\ln(1/X)$) scale, but this scale is left unspecified in the above equations. 
For the `real' contributions at least, we know by now what is the typical longitudinal momentum 
fraction $X$ of the gluons from the target which are probed by this scattering: this is the value 
$X(x,\pt)$ given by \eqn{Xx}. Hence, the dipole $S$-matrices in 
Eq.~\eqref{Jbare} must be evaluated at $X\simeq X(x,\pt)$,
where it is understood that $\bp=\bq-\bk$. We shall later demonstrate that $X(x,\pt)$
with $\bp=\bq-\bk$ is also the appropriate choice
for the rapidity argument of 
$S$-matrices which enter the `virtual' terms in \eqn{Jvbare}. This means
that, strictly speaking, one cannot factorize the $S$-matrix $\calS(\bk)$ in front of
the integrals in \eqn{Jvbare}, in contrast to the results in \cite{Chirilli:2011km,Chirilli:2012jd}.

\texttt{(iv)} The integral over $\xi$ in \eqn{eq:nlosigma}
seems to develop a logarithmic singularity at $\xi=1$, meaning an infrared
divergence associated with the emission of very soft ($x\to 0$) gluons. (This is the meaning 
of the upper label `unsub' in the l.h.s. of \eqn{eq:nlosigma}.) As already mentioned,
Refs.~\cite{Chirilli:2011km,Chirilli:2012jd} proposed to eliminate this divergence via
the `plus' prescription, defined as (for a generic function $F(\xi)$)
\beq\label{plus}
\int_{0}^1 \rmd \xi \,\frac{F(\xi)}{1-\xi}\,\longrightarrow\,
\int_{0}^1 \rmd \xi \,\frac{F(\xi)}{(1-\xi)_+}\,\equiv\,
\int_{0}^1 \rmd \xi \,\frac{F(\xi)-F(1)}{1-\xi}\,.
\eeq
After this subtraction, the result in Eqs.~\eqref{eq:nlosigma}-- \eqref{Jvbare} is supposed 
to represent a purely NLO correction, to be added to the respective LO result in \eqn{LO}.
We shall further discuss this particular prescription in Sect.~\ref{sec:sub}, 
but already at this level it should be clear that, as a matter of facts, 
there is no physical singularity in \eqn{eq:nlosigma}: for the `real' terms at least,
the integral over $\xi$ is cut off near $\xi=1$ by energy conservation, cf. \eqn{Xx}.
Specifically,  by using \eqn{Xx} together with the kinematical limit $X(x,\pt)\le 1$, 
one finds the following lower limit on $x=1-\xi$:
\beq\label{xmM}
x\,\gtrsim\, \xm(\pt) \equiv\, \frac{\pt^2}{\hat s}
 \,,\eeq
where we also used $\xm\ll 1$. Still for the `real' terms, there is
also an upper limit $x \le 1-x_p$, coming from the condition 
$x_p/(1-x) \le 1$ on the longitudinal fraction of the incoming quark.

This lower limit $x\gtrsim \xm$, can be recognized as the condition that the lifetime 
$\Delta x^+\sim 2p^+/\pt^2$ of the softest primary
gluon emission be at least as large as the longitudinal width $1/P^-$ of the target (a
necessary condition for having significant scattering). This condition has been previously 
emphasized in \cite{Altinoluk:2014eka} (the `Ioffe time') and numerically implemented in
\cite{Watanabe:2015tja} (where however the dependence of the various $S$-matrices upon
the target rapidity $X(x,\pt)$  has not been taken into account).

The existence of a physical lower limit on $x$ is indeed crucial for our subsequent 
construction, which will not involve the `plus' prescription, or any other infrared regularization
of the integral over $x$. It is therefore important to demonstrate 
that such a limit exists also for the `virtual' terms in  \eqn{Jvbare}, 
for which the previous argument on energy conservation does not apply.
We shall do that in Appendix \ref{app:canc}, where we demonstrate that the same lower limit 
$x\gtrsim \xm$ holds also for the `virtual' terms, as a consequence 
of fine cancellations among the virtual gluon graphs 
which occur in the complementary region at $x < \xm$.
Whereas mathematically subtle, the occurrence of such cancellations has a clear
physical interpretation: gluon fluctuations with $x<\xm$ cannot interact with the target,
since their lifetime is too short. Accordingly the 
emissions of such short-lived gluons cannot modify the $S$-matrix of the 
projectile. Since `real' emissions with $x < \xm$  are anyway forbidden by energy
conservation, it follows that the respective `virtual' graphs must cancel among themselves.
The precise way how such cancellations occur is demonstrated in Appendix \ref{app:canc}
(for the somewhat simpler, but general enough, situation where the target 
itself is a quark).


\texttt{(v)} If one takes the limit $\xi\to 1$ (i.e. $x\to 0$) in the {\em transverse kernels}
in Eqs.~\eqref{Jbare} and \eqref{Jvbare} (but not necessarily also in their implicit dependence 
upon $x$ via the rapidity cutoff $X(x,\pt)$), then the combination of these two terms which
enters \eqn{eq:nlosigma} with $\xi\to 1$ reduces
to the Fourier transform of the r.h.s. of the BK equation \eqn{BK}.  Specifically,
\beq\label{xi1}
2\pi\abar\big[\jcal(\bk ,1)- \jcal_v(\bk ,1)\big]=
\int \rmd^2\br\, \rme^{-\rmi \bk \cdot \br}\, 
 \frac{\del }{\del Y} \,S(\br; Y)
 \eeq
where the notation $\del S(\br; Y)/\del Y$ is merely a shortcut for
the r.h.s. of \eqref{BK} with $\br=\bx-\by$. The appearance of the BK equation
was in fact to be expected: when $\xi\to 1$,  Eqs.~\eqref{eq:nlosigma}--\eqref{Jvbare} 
describe the emission of a soft primary gluon by the incoming quark, 
in the eikonal approximation. By definition, the effect of this soft emission on the quark
multiplicity is equivalent to the first step in the BK evolution of the respective 
LO result in \eqn{LO}. Note however that in general the $S$-matrices
implicit in the r.h.s. of  \eqn{xi1} are meant to be computed {\em beyond} the LLA
and their rapidity argument $X(x,\pt)$  is a complicated function of the kinematics
of the emitted gluon.


\subsection{CGC factorization at NLO}
\label{sec:main}

\begin{figure}[t]
\begin{minipage}[b]{0.46\textwidth}
\begin{center}
\includegraphics[scale=0.45]{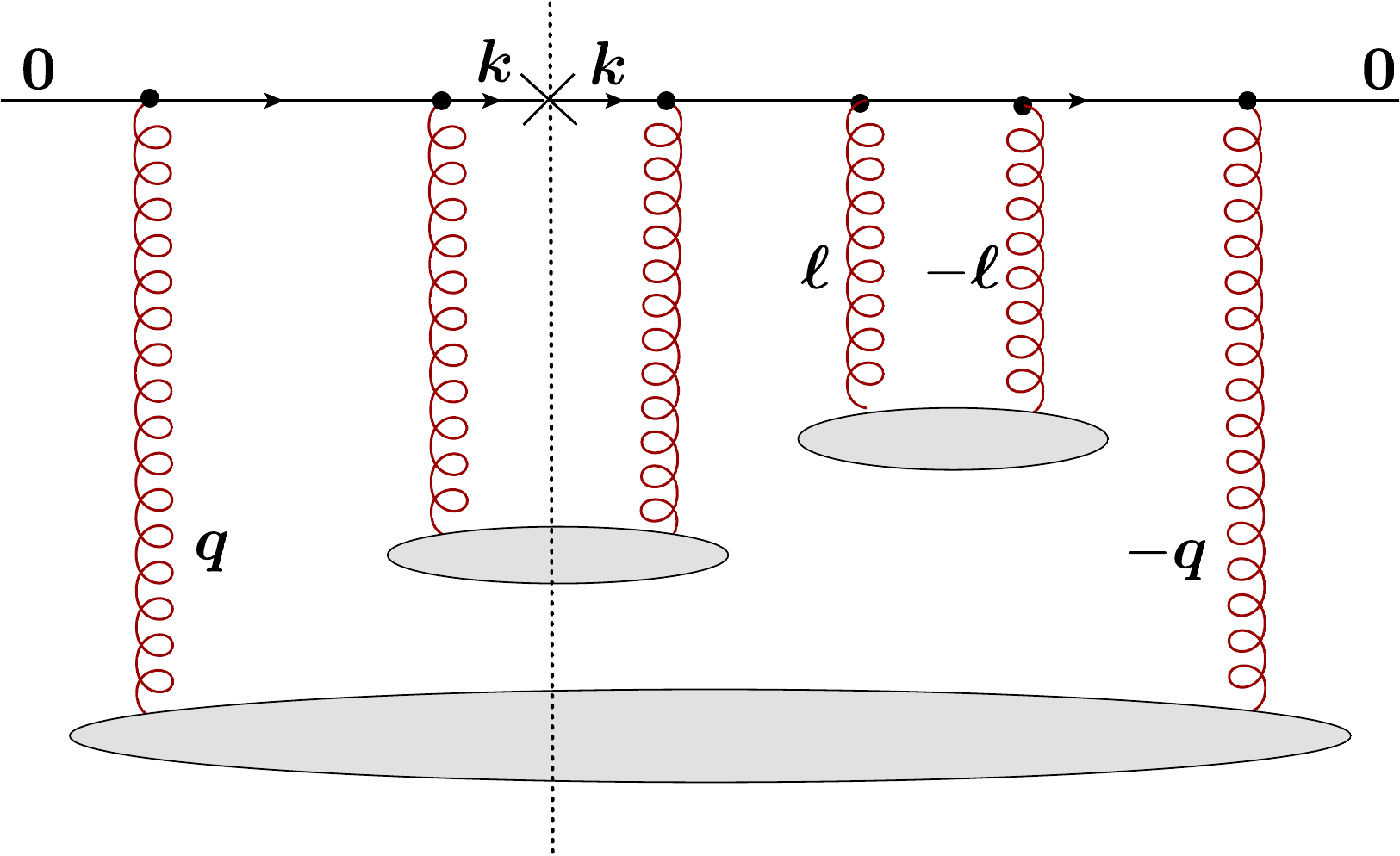}
\end{center}
\end{minipage}\ \
\begin{minipage}[b]{0.46\textwidth}
\begin{center}
\includegraphics[scale=0.56]{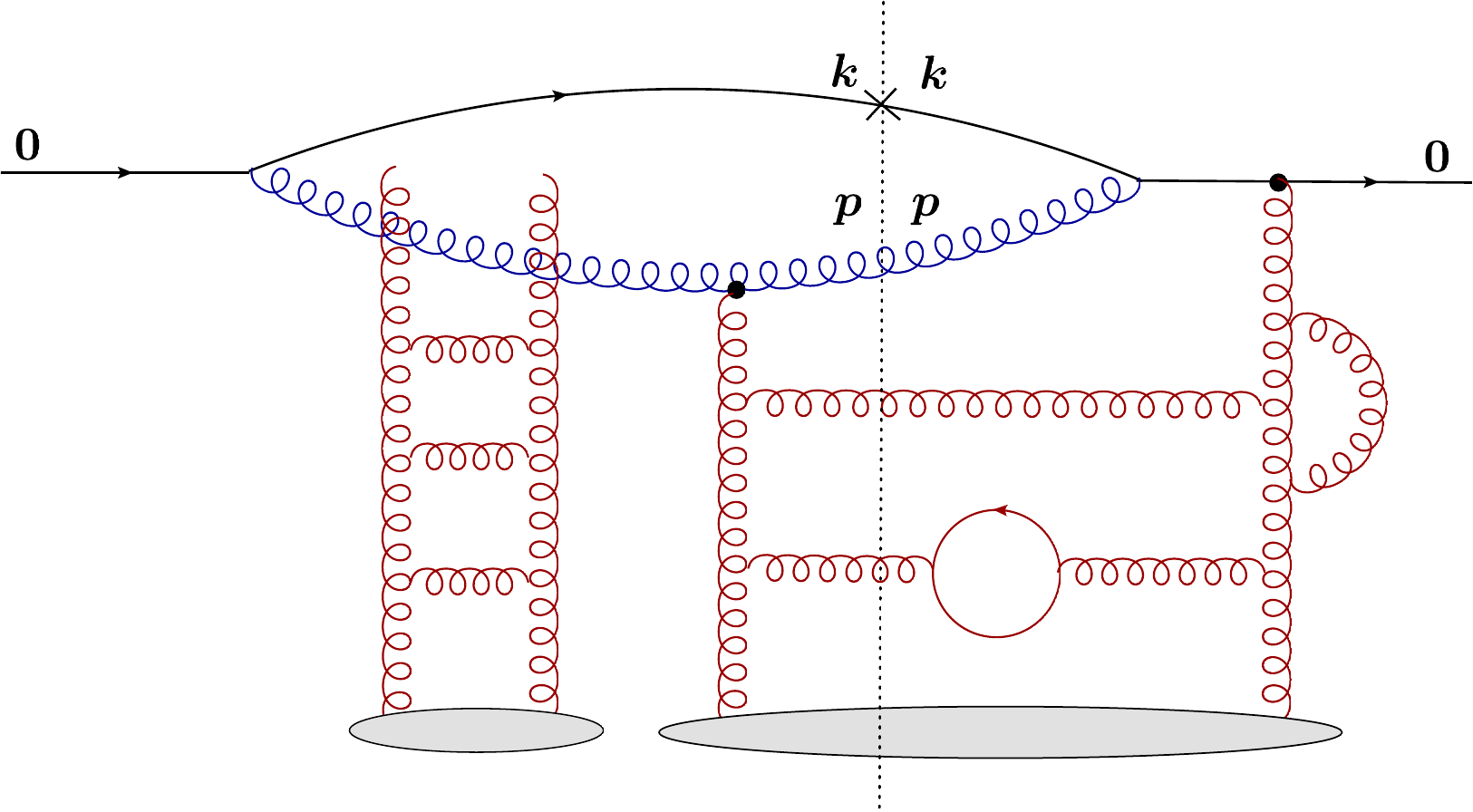}
\end{center}
\end{minipage}
\caption{\small Graphical illustration of the factorization in \eqn{NLO}.
Left: a diagram describing multiple scattering in the MV model; 
this is representative for the tree-level term $\calS_0$ in \eqn{NLO}.
Right: a diagram which exhibits the primary quark-gluon pair (the NLO impact
factor) and its multiple scattering off the gluon distribution of the target, itself
evolved to NLO; this is representative for the second term in \eqn{NLO}, which encodes all the
radiative corrections to the quark multiplicity through NLO.}
\label{fig:fact}
\end{figure}

After this preparation, we are now in a position to present our master formula for
the single-inclusive quark multiplicity valid through next-to-leading order
(i.e. which includes both the LO and the NLO contributions). The relation between
this formula and the factorization scheme proposed in Refs.~\cite{Chirilli:2011km,Chirilli:2012jd} 
will be discussed later, in Sect.~\ref{sec:sub}. As already
mentioned, we systematically omit the NLO corrections proportional to the quark Casimir $\CF$,
which play no special role for the high-energy evolution. These corrections
can be taken over from Refs.~\cite{Chirilli:2011km,Chirilli:2012jd} and simply added to
our master formula, which reads
\begin{align}
\label{NLO}\hspace*{-0.7cm}
 \frac{\rmd N^{pA\to qX}}{\rmd^2\bk \,\rmd \eta}\bigg|_{\lo+\nlo}\!
& =  \frac{ x_p q(x_p)}{(2\pi)^2}\Big[{\calS}_0(\bk)+\Delta{\calS}(\bk,X_g)\Big]
+ \int\! \frac{\rmd^2 \bp}{(2\pi)^2} 
\int_{\xm(\pt)}^{1} \!\rmd x \ \frac{\abar}{2\pi} \,
\frac{1+(1-x)^2}{2x}\nonumber\\*[0.2cm]
 &\times\left\{
\frac{x_p}{1-x} q\left(\frac{x_p}{1-x}\right)\tilde \jcal\big(\bk ,x; \bp, X(x,\pt)\big)-
x_p q\left(x_p \right)\tilde \jcal_v\big(\bk ,x; \bp, X(x,\pt)\big) \right\}.
\end{align}
This formula is illustrated in Fig.~\ref{fig:fact}.
As before, ${\calS}_0(\bk)$ denotes the tree-level contribution to the dipole $S$-matrix,
say as given by the MV model (see e.g. 
\cite{Iancu:2002xk,Iancu:2003xm,Gelis:2010nm,Kovchegov:2012mbw} for an explicit expression).
The quantity $\Delta{\calS}(\bk,X_g)$ within the square brackets
denotes a NLO correction to the dipole $S$-matrix, to be specified in 
Sect.~\ref{sec:evol}. The last term in \eqn{NLO}, which 
involves a double integration --- over the longitudinal fraction $x=1-\xi$
and the transverse momentum $\bp$ of the primary gluon --- is the main term for our present purposes. It encodes the impact factor to NLO accuracy, the LO evolution of the dipole $S$-matrix
and also a part of the respective NLO contribution (namely, the part which is not included
in $\Delta{\calS}(\bk,X_g)$; see Sect.~\ref{sec:evol} for details).

The new functions $\tilde\jcal\big(\bk ,x; \bp, X(x,\pt)\big)$
and $\tilde \jcal_v\big(\bk ,x; \bp, X(x,\pt)\big)$ are essentially the integrands in
Eqs.~\eqref{Jbare} and respectively \eqref{Jvbare}, in which we replaced $\xi\to 1-x$
and $\bq \to \bp+\bk$. (As compared to \eqref{Jbare} and \eqref{Jvbare},
we now use $\bp$ and $\bl$ as integration variables; the integral over $\bp$ is
explicit in \eqn{NLO}, while that over $\bl$ is included in the definitions of 
$\tilde\jcal$ and $\tilde\jcal_v$.) The notation emphasizes that the various dipole 
$S$-matrices implicit in these functions should be evaluated at a target 
rapidity $Y=\ln(1/X)$ with $X=X(x,\pt)$, cf. \eqn{Xx}. 
The lower limit $\xm(\pt)$ is shown in \eqn{xmM}. In the real term,
it is understood that the support of the quark distribution limits the integration
to $x < 1-x_p$. 

For more clarity,
let us exhibit here the function $\tilde\jcal\big(\bk ,x; \bp, X(x,\pt)\big)$
which enters the `real contribution (the corresponding expression for  $\tilde \jcal_v$ can 
be similarly written):
\begin{align}\hspace*{-.5cm}
\label{Jreal}
\tilde\jcal\big(\bk ,x; \bp, X(x,\pt)\big)&= 
\frac{2\bp \cdot \big[(1-x)\bp-x\bk\big]}{\bp^2 \big[(1-x)\bp-x\bk\big]^2}\,
\calS\big(\bp+\bk, X(x,\pt)\big)\nonumber\\*[0.2cm]
&+\int\! \frac{ \rmd^2\bl}{(2\pi)^2}
\frac{2(\bk-\bl)\cdot  \big[(1-x)\bp-x\bk\big]}{(\bk-\bl)^2
 \big[(1-x)\bp-x\bk\big]^2}\,
\calS\big(\bp+\bk, X(x,\pt)\big)\,\calS\big(\bl, X(x,\pt)\big).
\end{align}
It is perhaps interesting to notice that the linear combination ${\bm P}\equiv
(1-x)\bp-x\bk$ which appears in the above integrand is the momentum
conjugated to the transverse separation $\bx-\bz$ between the quark and 
the primary gluon. Similarly, the total momentum
${\bm q}\equiv\bk+\bp$ is conjugated to the center-of-mass $(1-x)\bx + x\bz$ of the quark-gluon pair.
(As in \eqn{BK}, $\bx$ and $\bz$ denote the transverse positions of the quark and the
primary gluon, respectively.)

To convincingly demonstrate the validity of \eqn{NLO} to the NLO accuracy of interest,
one still needs to describe the correction $\Delta{\calS}(\bk,X_g)$ to the dipole $S$-matrix,
which we shall do in Sect.~\ref{sec:evol}. In the remaining part of this
subsection, we shall merely check that
\eqn{LO} properly encodes the LO result, cf. \eqn{LO}, together with 
the NLO correction to the impact factor discussed
in Sect.~\ref{sec:CXY}, without any over-counting. 

The LLA limit of \eqn{NLO} is obtained by making approximations appropriate
at small $x$, that is, by treating the emission of the primary gluon in the eikonal
approximation and by replacing
$\pt\sim \kt$ in the kinematical limits and the various rapidity variables;
that is, one approximates $\xm(\pt)\simeq x_g= \kt^2/\hat s$ and $X(x,\pt)\simeq
X(x)=X_g/x$,  with $X_g=x_g$ (cf. \eqn{XxLO}). Also, all dipoles $S$-matrices are now understood to
obey the LO BK evolution, from $X_0\simeq 1$ down to $X(x)$. Under these assumptions, one can
first commute the integrations over $\bp$ and $x$ in \eqn{NLO} and then use the identity \eqref{xi1} to
rewrite the simplified version of this equation in the following, suggestive, form
\begin{align}
\label{NLOtoLO}
 \frac{\rmd N^{pA\to qX}}{\rmd^2\bk \,\rmd \eta}\bigg|_{\lo+\nlo}^{(0)}
 & =   x_p q(x_p)\,\frac{\calS_0(\bk)}{(2\pi)^2}\,+\,x_p q(x_p)\,
 \frac{1}{(2\pi)^2}\int_{x_g}^{1}
 \frac{\rmd x}{x}\, \frac{\del }{\del Y} \,\calS\big(\bk; Y=\ln1/X(x)\big)\,.
\end{align}
The above integral runs over the longitudinal momentum fraction $x=p^+/k^+$
of the right-moving gluon, whereas the evolution of the dipole
$S$-matrix has been rather performed w.r.t. the longitudinal fraction $X$ of the gluons
in the target. However,
to LLA, $x$ and $X$ are related via $X(x)=X_g/x$, so one can 
change the integration variable from $x$ to $Y\equiv \ln(1/X)$ and thus identify a total derivative
\beq
\int_{x_g}^{1}
 \frac{\rmd x}{x}\, \frac{\del }{\del Y} \,\calS\big(\bk; Y=\ln1/X(x)\big)=
 \int_0^{Y_g}\rmd Y\, \frac{\del }{\del Y} \,\calS(\bk; Y)\,=\,\calS(\bk; Y_g)-{\calS}_0(\bk)\,.
\eeq
After also adding the initial condition in \eqn{NLOtoLO}, one recognizes the LO result \eqref{LO},
as anticipated.

The r.h.s. of \eqn{NLOtoLO} is recognized as the `integral' version of the LO BK equation 
introduced in \eqn{BKint}.
Hence, the full result \eqref{NLO} can be viewed as the generalization of 
that integral representation to NLO and to the exact kinematics for the primary
gluon emission. 
(This will be confirmed by the discussion in Sect.~\ref{sec:approx}.)
The explicit separation of the first gluon emission
from the remaining evolution, as operated by this representation, 
has allowed us to promote the calculation of the impact factor 
to NLO accuracy, while at the same time avoiding over-counting.

At this point, it is important to more precisely specify the perturbative content of
\eqn{NLO}. As just explained, the integral term there fully encodes the LO evolution of
the dipole $S$-matrix, that is, it resums corrections of the type ${(\abar Y_g)^n}$ to all orders.
It obviously encodes NLO corrections due
to the fact that the emission of the primary gluon is treated exactly; that is, the integral
over $x$ also generates corrections of $\order{\abar}$ besides the dominant contribution
of $\order{\abar Y_g}$, which counts for the LO evolution. To ensure NLO accuracy, the
evolution of the various dipole $S$-matrices must be computed to NLO as well. Indeed,
the NLO BK kernel includes corrections of $\order{\abar}$; hence, the solution to the NLO
BK equation involves corrections proportional to ${\abar(\abar Y_g)^n}$, which count to NLO.

By a similar argument, one must include the (one-loop) running coupling corrections 
within the QCD coupling $\abar$ associated with the primary gluon vertex. 
This is why, in writing \eqn{NLO}, we have
inserted  the factor $\alpha_s$  {\em inside} the double integral over $\bp$ and $x$:
after including the running coupling corrections, this factor will depend upon the 
transverse momenta $\bk$ and $\bp$ which enter the emission vertex
and possibly also upon $x$ (via the gluon kinematics). Specifying this dependence
requires a prescription, which is most conveniently formulated in the transverse coordinate
representation (since this is the representation in which the BK equation is generally 
solved in practice). Such prescriptions will be discussed in Sect.~\ref{sec:evol}
(see e.g. \eqn{azero} there), together with the other NLO corrections to the dipole
evolution. Notice however that, in order to evaluate \eqn{NLO}, one also needs
a prescription for the running coupling which is directly formulated
in momentum space. In general such a prescription
will be different from the one in coordinate space. This mismatch could have consequences
for the fine-tinning issue to be discussed in Sect.~\ref{sec:sub}.

The above arguments show that, strictly speaking, the integral term in \eqn{NLO} also includes
terms of NNLO, as generated by the product between the NLO correction to the impact
factor and the NLO effects in the high-energy evolution, or in the running coupling.
As we shall explain in Sect.~\ref{sec:sub}, these various types of NLO effects can be disentangled
from each other via a reorganization of the perturbation theory which involves a 
`rapidity subtraction', as in Refs.~\cite{Chirilli:2011km,Chirilli:2012jd}. Yet, this procedure has
some inconveniences, as we shall see (notably, it introduces the `fine-tuning' issue anticipated in
the Introduction). So it is important to stress here that, although going beyond a strict NLO 
approximation, the result \eqref{NLO} is in fact the natural outcome of perturbative QCD --- that is,
the direct result of evaluating Feynman graphs at the loop-order of interest, before performing
additional manipulations like the  `rapidity subtraction'.

We conclude this subsection with a discussion of potential difficulties
with using \eqn{NLO} in practice. All these issues
will be addressed in more detail in the next two subsections, where we shall provide 
solutions to them, at least at the expense of further approximations.

First,  \eqn{NLO} looks very cumbersome, notably due the intricacy of the 
multiple  integrations, over both transverse ($\bp$, $\bl$) and longitudinal ($x$) momenta, 
which are entangled with each other. It is not clear to us whether these integrations
can be computed as such, not even numerically. The calculations 
might be further complicated by the need to compute the Fourier transform
of the dipole $S$-matrix, as numerically obtained by solving the BK equation 
in coordinate space.

Second, \eqn{NLO} is somewhat formal, in that the dipole $S$-matrices implicit
there are supposed to encode the evolution of the target gluon distribution at NLO. 
However, the high-energy evolution of a dense nucleus has not been explicitly computed
beyond LO. (All NLO calculations to date refer to the evolution  of a dilute projectile, like a dipole 
\cite{Balitsky:2008zza,Balitsky:2013fea,Kovner:2013ona}.) Besides,
a purely NLO approximation to the high-energy evolution
is likely to become unstable (and thus require resummations)
in the `collinear' regime where the transverse momentum $\kt$ is relatively large ($\kt\gg Q_s$).

Finally, the NLO calculation based on \eqn{NLO} is formally sensitive
to the physics of the nuclear wavefunction at large values of $X$ 
(recall that $x\sim x_g$ corresponds to $X\sim 1$),
which is not really under control within the present, high-energy approximations.
This cannot be a serious difficulty in the physical context at hand: the NLO corrections
to the impact factor are controlled by relatively hard primary gluons
with $x\sim \order{1}$ and hence $X(x)\ll 1$; such a hard emission by the projectile
should be well separated from the valence structure of the target at $X\sim 1$. At the
end of the next subsection we shall describe an explicit procedure which implements this
separation.

\subsection{Simplifying the kinematics}
\label{sec:approx}

In this subsection, we shall propose strategies to deal with some of the problems alluded to
at the end of the previous subsection. First, we shall argue
that one can approximate $\pt\sim \kt$ within the rapidity variables for the high-energy evolution
and thus greatly simplify the structure of the transverse and longitudinal integrations in \eqn{NLO}.
Second, we shall discuss the prescription for the running of the coupling in the emission
of the primary gluon. Third, we shall reformulate the initial condition at low energy in such a 
way to reduce the sensitivity to the large-$X$ region in the target wavefunction.

Concerning the first point --- the dependence of the kinematical constraints on the
high-energy evolution upon the transverse momenta
of the primary quark-gluon pair ---, we note that there are two interesting physical regimes:

\texttt{(i)} {\it Hard quark production and di-jet configurations:} $\kt\gg Q_s(X)$. 
This is the regime which is primarily concerned by the negativity problem discussed in 
Refs.~\cite{Stasto:2013cha,Stasto:2014sea,Kang:2014lha,Xiao:2014uba,Altinoluk:2014eka,Watanabe:2015tja,Ducloue:2016shw}. In this case, we have already argued in Sect.~\ref{sec:dijet} 
that the transverse momentum of the primary gluon and that of the produced quark
must balance each other : $\pt\simeq\kt$. 


\texttt{(ii)} {\it Semi-hard quark production:} $\kt\gtrsim Q_s(X)$. This regime includes the  
`geometric scaling' window \cite{Stasto:2000er}, where the scattering is weak, but the $S$-matrix 
is still influenced by non-linear effects, via the `saturation boundary' at $Q_s(X)$  
\cite{Iancu:2002tr,Mueller:2002zm,Munier:2003vc}. In this regime, all the relevant transverse
momenta --- the $\kt$ of the produced quark, the $\pt$ of the primary gluon, and the $\qt$
transferred by the nuclear target --- take typical values of $\order{Q_s}$, since
this is the value naturally acquired via rescattering off the saturated gluon distribution of the
target.

We see that in both cases the quantities $\pt$ and $\kt$ cannot be very
different from each other, so one can approximate $\pt\simeq \kt$ when evaluating the rapidity
variables (cf. Eqs.~\eqref{Xx} and \eqref{xmM}) :
\beq\label{xmk}
\xm(\pt)\,\to\,\xm(\kt)=\frac{\kt^2}{\hat s}\equiv x_g\quad\mbox{and}\quad
X(x,\pt)\,\to\,X(x,\kt)=\frac{\kt^2}{x\hat s}\equiv X(x)\,.
\eeq
Remarkably, thanks to this approximation, we have returned to the same
expressions for the rapidity variables $x_g$ and  $X(x)$ as at LO, cf. \eqn{Xx}.
This greatly simplifies \eqn{NLO} since the transverse momentum
integrations can now be performed prior to the integral over $x$. Then \eqn{NLO} takes 
a form closer to that in \eqn{eq:nlosigma}, namely,
\begin{align}
\label{NLOkt}
 \frac{\rmd N^{pA\to qX}}{\rmd^2\bk \,\rmd \eta}
 & =    \frac{ x_p q(x_p)}{(2\pi)^2}\Big[{\calS}_0(\bk)+\Delta{\calS}(\bk,X_g)\Big]
+\frac{\abar(\kt^2)}{2\pi}  \int_{x_g}^{1} \rmd x \ \frac{1+(1-x)^2}{2x}\nonumber\\*[0.2cm]
 &\qquad\qquad\times\left\{
\frac{x_p}{1-x} q\left(\frac{x_p}{1-x}\right) \jcal\big(\bk ,x; X(x)\big)-
x_p q\left(x_p \right) \jcal_v\big(\bk ,x; X(x)\big) \right\},
\end{align}
where the functions $\jcal\big(\bk ,x; X(x)\big)$ and $\jcal_v\big(\bk ,x; X(x)\big) $
have the same formal expressions as in Eqs.~\eqref{Jbare}--\eqref{Jvbare} [with $\xi=1\minus x$,
of course], except for the fact that the rapidity variable for the evolution of 
the dipole $S$-matrices is now clearly identified, namely $Y=\ln \big(1/X(x)\big)$. 
Once again, it is understood that the integral over $x$ in the real term
is restricted to $x < 1-x_p$.

In writing \eqn{NLOkt}, we have also identified the argument of the running coupling for the primary
vertex as $\kt^2$. This is unambiguous under the present assumptions, since $\kt\sim\pt$ is the
only hard scale involved in that splitting\footnote{Recall that  the momentum
conjugated to the transverse separation between the quark and 
the primary gluon is ${\bm P}=(1-x)\bp-x\bk$ ; when $\pt\sim\kt$, we have $P_\perp\sim \kt$
for any $x$, hence the transverse separation is of order $1/\kt$.}. 
\eqn{NLOkt} also shows that the natural `LO approximation' for the high-energy evolution in
the problem at hand is the LO BK equation with running coupling (rcBK). Indeed, when evaluating
the second term in \eqn{NLOkt} within the eikonal approximation, as appropriate for $x\ll 1$,
the r.h.s. of this equation becomes proportional to the integral version of rcBK; that is,
this is tantamount to evaluating the LO formula \eqref{LO} with the solution $\calS_\rcBK(\bk, X_g)$
to rcBK.

The considerably simpler structure of \eqn{NLOkt} also allows us to reformulate the
initial condition at low-energy, in such a way to avoid the `dangerous' region at $X\sim 1$.
To that aim, let us introduce a `better' separation scale $X_0$ for the rapidity 
evolution of the target, which is such that the high-energy approximations are indeed justified
for any $X\le X_0$.  (For instance the value $X_0=0.01$
is often used in the fits to the HERA data for deep inelastic scattering; see e.g.
 \cite{Albacete:2010sy,Iancu:2015joa}.) In particular, this scale should obey $x_g \ll X_0 \ll 1$.
We then separate the integral over $x$ into two regions, $x_g< x < x_{g}/X_0$
and $x_g/X_0 < x <1$, which in terms of $X=X_g/x$ correspond to $1 > X > X_0$ and respectively 
$X_0 > X > X_g$. (We recall that $X_g=\kt^2/\hat s= x_g$.) Within the first region,
one has $x\ll 1$, hence one can replace 
$x\to 0$ within the transverse kernels which enter the functions $\jcal$ 
and $\jcal_v$  [e.g., $\jcal\big(\bk ,x; X(x)\big)\to \jcal\big(\bk ,0; X(x)\big)$]
and also within the quark distribution. 

Under these assumptions, the sum between the `initial condition' $\calS_0$ 
in \eqn{NLOkt} and the part of the integral there which runs over the
small-$x$ interval at $x_g< x < x_{g}/X_0$ is formally the same as the
r.h.s. of the LO BK equation with running coupling (rcBK) integrated from $X=1$ 
down to $X_0$ (recall \eqn{xi1}).  It might be tempting to interpret this sum
as the solution to rcBK evaluated at $X=X_0$, but we shall not adopt this point of
view: after all,
this `rcBK evolution' refers to the large-$X$ interval at $1 > X > X_0$, where 
the high-energy approximations are not valid. 
We shall rather {\em replace} the result of this fictitious  `rcBK evolution'
with a {\em new} initial condition, denoted as $\calS(\bk, X_0)$, 
which is formulated directly at $X_0$. That is, we replace
\eqn{NLOkt} by\footnote{Strictly speaking, the integral over $x$ implicit within
the NLO quantity $\Delta{\calS}(\bk,X_g)$, as visible e.g. in \eqn{toyNLO2} below,
should be cutoff at $x_g/X_0$ as well when $\Delta{\calS}(\bk,X_g)$ is inserted into \eqn{NLOX0}.}
\begin{align}
\label{NLOX0}
 \frac{\rmd N^{pA\to qX}}{\rmd^2\bk \,\rmd \eta}
 & =   \frac{ x_p q(x_p)}{(2\pi)^2}\Big[{\calS}(\bk,X_0)+\Delta{\calS}(\bk,X_g)\Big]
+\frac{\abar(\kt^2)}{2\pi}  \int_{x_g/X_0}^{1} \rmd x \ \frac{1+(1-x)^2}{2x}\nonumber\\*[0.2cm]
 &\qquad\qquad \times\left\{
\frac{x_p}{1-x} q\left(\frac{x_p}{1-x}\right) \jcal\big(\bk ,x; X(x)\big)-
x_p q\left(x_p \right) \jcal_v\big(\bk ,x; X(x)\big) \right\},
\end{align}
where it is now understood that all the $S$-matrices implicit in the second term
in the r.h.s. are obtained by solving appropriate evolution equations with the initial
condition formulated at $X=X_0$. The evolution equations to be used in this
context will be discussed in the next subsection.


\subsection{The evolution of the color dipole beyond leading order}
\label{sec:evol}

In this subsection, we shall describe the NLO evolution of the dipole
$S$-matrices to be used in conjunction with the factorization scheme in \eqn{NLO} or
\eqref{NLOkt}. In particular, we shall present an explicit expression for
the NLO correction $\Delta{\calS}$ which enters this scheme but so far has not 
been specified.  To simplify the arguments and the notations, we shall work
at the level of the kinematic approximations introduced in Sect.~\ref{sec:approx},
that is, we shall built on top of \eqn{NLOkt}.

There are two aspects which are rather subtle and that we shall try to clarify
in what follows. The first refers one to the proper inclusion of NLO corrections within
$\Delta{\calS}$ without any over-counting: this quantity must contain only those
corrections to the high-energy evolution that are not already included in
the integral term in \eqn{NLOkt}. The second aspect refers to the relation between
the viewpoint of {\em target} evolution, that was explicitly used in our previous
arguments (for instance, when specifying the rapidity arguments of the various $S$-matrices),
and that of the evolution of the {\em projectile} (a color dipole), 
for which the evolution equation is currently known at
NLO accuracy \cite{Balitsky:2008zza}, including the collinear improvement
\cite{Beuf:2014uia,Iancu:2015vea,Iancu:2015joa,Hatta:2016ujq}.
Indeed, the two evolutions refer to different variables (`evolution times')
--- $x=q^+/q_0^+$ for the right-moving projectile and respectively $X=q^-/P^-$ 
for the left-moving target ---, so the corresponding equations cannot be identical
beyond leading order, when the differences between various transverse momenta 
and also the off-shell effects start to play a role.

As mentioned after \eqn{NLO}, the unknown quantity $\Delta{\calS}(\bk,X_g)$
represents a part of the NLO corrections to the evolution of the dipole $S$-matrix. 
Since these corrections are fully known by now \cite{Balitsky:2008zza}, the simplest
way to obtain $\Delta{\calS}(\bk,X_g)$ is by clarifying its relation to the NLO calculation in
Ref.~\cite{Balitsky:2008zza}. We first observe that, by definition, the full NLO
result for the quark multiplicity must involve two types of NLO corrections: 
those associated with the impact factor and those related
to the high-energy evolution of the color dipoles. Hence, if one `switches off' the first type
of corrections (which one can do by computing the emission of the 
primary gluon in the eikonal approximation), then the r.h.s. of Eq.~\eqref{NLOkt} 
should be proportional to
${\calS}(\bk, X_g)$ --- the dipole $S$-matrix computed at NLO accuracy and for the
kinematics of interest. This argument implies 
\beq\label{Sfull}
{\calS}(\bk, X_g)={\calS}_0(\bk)+\Delta{\calS}(\bk,X_g)
+2\pi\abar(\kt^2)  \int_{x_g}^{1}\ \frac{\rmd x}{x}
\left\{\jcal\big(\bk ,0; X(x)\big)-
\jcal_v\big(\bk ,0; X(x)\big) \right\},
\eeq
where the function $\jcal\big(\bk ,0; X(x)\big)$ is obtained from  \eqn{Jbare} by letting 
$x\equiv 1-\xi\to 0$ and by evaluating the $S$-matrices there to
NLO accuracy and for a rapidity argument $Y=\ln(1/X(x))$ 
[and similarly for the function $\jcal_v\big(\bk ,0; X(x)\big)$]. Using
this condition together with the NLO result for the dipole $S$-matrix which emerges
from Ref.~\cite{Balitsky:2008zza}, it is possible to identify the quantity $\Delta{\calS}(\bk,X_g)$.
By the `NLO result for ${\calS}$', we mean the integral representation for
the NLO $S$-matrix, as obtained by formally solving the respective evolution equation
--- that is, the generalization of \eqn{BKint} to NLO. 

\comment{This whole procedure may look a bit formal, for the reasons aforementioned --- the
NLO version of the BK equation is written for the projectile evolution in $x$, whereas 
the $S$-matrix ${\calS}(\bk, X_g)$ depends upon the rapidity argument $X=X_g$
for the target. Yet, it should be possible to express at least the first step in this evolution
in terms of gluon emissions by the projectile. We already did so for the LO evolution
(cf. Sect.~\ref{sec:TvsP}). The only new ingredient which appears
when moving to NLO accuracy
is that the first evolution step can also involve the emission of a set of 2 gluons.
These 2 gluons have similar rapidities ($x_1\sim x_2$), but they are both soft as 
compared to the parent quark ($x\equiv x_1+x_2\ll 1$).
}

Before we proceed, it is useful to, first, clarify what we precisely mean by the
`LO evolution' and, second, introduce some simplified notations, which focus on
the essential and help making the subsequent arguments more transparent.

As already mentioned after \eqn{NLOkt}, our LO approximation to the dipole 
$S$-matrix is the solution $\calS_\rcBK(\bk, X)$ to the LO BK equation with running coupling 
(rcBK) \cite{Kovchegov:2006wf,Kovchegov:2006vj,Balitsky:2006wa}. This choice deserves
some comment, in that it already includes a subset of the NLO (and
higher-order) corrections, via the running of the coupling. But precisely because of that, 
rcBK offers a better starting point for a perturbative expansion than the {\em strict} 
leading-order approximation --- the LO BK equation with fixed coupling. This is related
to the poor convergence of the perturbative expansion at high energy: the
LO BK equation with fixed coupling is well known to predict an unrealistically fast evolution 
with increasing energy, meaning a too large value for the saturation exponent.
Hence, for sufficiently high energies, the strict LO estimate \eqref{LO} for the multiplicity 
becomes {\em exponentially larger} (in the sense of an exponential in $Y=\ln(1/X)$) 
than the actual result at NLO. This problem is considerably alleviated if one instead 
uses rcBK as the `leading order' evolution: this approximation 
predicts a significantly slower evolution \cite{Mueller:2002zm,Albacete:2007yr} 
and offers a reasonably good description
for the phenomenology \cite{Albacete:2010bs,Albacete:2010sy,Albacete:2014fwa}.

The all-order resummation of running coupling corrections requires a prescription.
Here, we shall mention two popular such prescriptions, both built with 
the one-loop approximation for the running coupling and which are roughly equivalent to
each other (see Ref.~\cite{Iancu:2015joa} for a recent discussion).  
Consider the splitting of the parent dipole $(\bx,\by)$ into two daughter dipoles
$(\bx,\bz)$ and $(\bz,\by)$, as described by the LO BK equation \eqref{BK}.
The {\em smallest dipole prescription} consists in replacing\footnote{Clearly,
after such a replacement, the coupling $\abar(r_{\rm min})$ must be moved {\em inside}
the integral over $\bz$ in  \eqn{BK}.} 
$\abar \to \abar(r_{\rm min})$,  where
$r_{\rm min} \equiv \min\big\{|\bx \minus\by|,|\bx \minus\bz|,|\by \minus\bz|\big\}$ and
\begin{equation}\label{1Lrun}
 \abar(r) =
 \frac{1}{\bar b\ln\big[4/r\Lambda_{\rm QCD}^2\big]}\,,\qquad
 \bar b=(11\Nc- 2 \Nf)/12\Nc\,.
 \end{equation}
The other prescription, known as {\it fastest apparent convergence (fac)},
amounts to $\abar \to \bar{\alpha}_{\rm fac}$, with
 \begin{equation}
 \label{azero}
 	\bar{\alpha}_{\rm fac}
 \equiv \left[\frac{1}{\abar(|\bx \minus \by|)} 
 + \frac{(\bx \minus \bz)^2 - 
 (\by \minus \bz)^2}{(\bx \minus \by)^2} \,
 \frac{\abar(|\bx \minus \bz|) - \abar(|\by \minus \bz|)}{\abar(|\bx \minus \bz|)\abar(|\by \minus \bz|)}\right]^{-1}.
 \end{equation}      
This last prescription is particularly useful for what follows, in that it simplifies the
expression that we shall obtain for $\Delta{\calS}$.

After a Fourier transform to momentum space, the solution to rcBK can be given
the following integral representation:
\beq
\label{rcBKint}
{\calS}_\rcBK(\bk,X_g)={\calS}_0(\bk)+2\pi\abar(\kt^2) \int_{x_g}^{1}\frac{\rmd x}{x}\,
\big[\jcal_\rcBK\big(\bk,0;X(x)\big)- \jcal_{v,\rcBK}\big(\bk,0;X(x)\big)\big]\,,\eeq
where the factorization of the running coupling $\abar(\kt^2)$ was possible
because of our assumption that $\kt$ is sufficiently hard (recall the discussion 
after \eqn{NLOkt}). The functions $\jcal_\rcBK\big(\bk,0;X(x)\big)$ and 
$\jcal_{v,\rcBK}\big(\bk,0;X(x)\big)$ are obtained from the respective functions
in \eqn{Sfull} after replacing $\calS\to {\calS}_\rcBK$.

 \begin{figure}[t] \centerline{
\includegraphics[width=.7\textwidth]{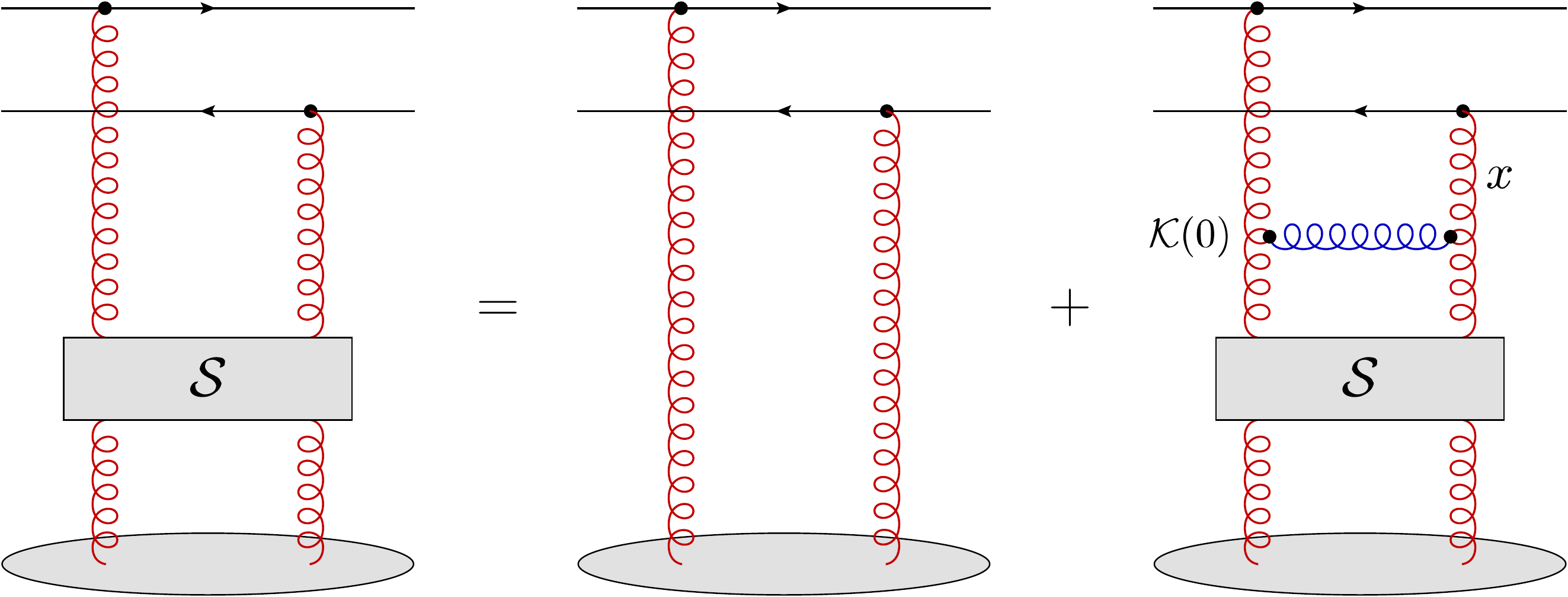}}
 \caption{\small A graphical illustration of the integral version of the LO BK equation \eqref{toyLO}.
 Running-coupling corrections and non-linear effects in $\calS$ describing multiple scattering
 are implicitly assumed, but not explicitly depicted.}
 \label{fig:toyBK}
\end{figure}

We now introduce more schematic notations, as anticipated.
Specifically, let us ignore (just in our notations) the transverse momentum convolutions, the inessential numerical factors, and the non-linear structure of functions
like $\jcal$ w.r.t. the dipole $S$-matrix. Also, in writing the cross-section, we shall omit
the quark distribution function. That is, we shall rewrite \eqn{NLOkt} simply as
\beq\label{toyNLO}
\calN_{\lo+\nlo}=\calS_0 +\Delta{\calS}(X_g)+ \abar(\kt^2) \int_{x_g}^{1} \frac{\rmd x}{x} 
\,\calK(x) \,\calS\big(X(x)\big)
\,,\eeq
where the kernel $\calK(x)$ encodes all the momentum space variables and convolutions
(but no dipole $S$-matrix) and $\calS\big(X(x)\big)$ succinctly denotes all the factors involving
the dipole $S$-matrix, which could be either linear, or bi-linear, in $\calS$. Both `real' and `virtual'
contributions are implicitly added in the above integral.
With these new notations, Eqs.~\eqref{Sfull} and \eqref{rcBKint} become
\beq\label{toySfull}
{\calS}(X_g)={\calS}_0+\Delta{\calS}(X_g)
+ \abar(\kt^2) \int_{x_g}^{1} \frac{\rmd x}{x} \,\calK(0) \,
\calS\big(X(x)\big)\,,\eeq
and respectively 
\beq\label{toyLO}
{\calS}_\rcBK(X_g)=\calS_0 + \abar(\kt^2) \int_{x_g}^{1} \frac{\rmd x}{x} \,\calK(0) \,
\calS_\rcBK\big(X(x)\big)\,,\eeq
where the kernel is now evaluated at $x=0$ (or $\xi=1$), that is, in the eikonal
approximation. Clearly, the compact 
notation $\calK(0)$ stays for the LO BK (or dipole) kernel. In particular, the rcBK equation
\eqref{toyLO} is graphically illustrated in Fig.~\ref{fig:toyBK}.

 \begin{figure}[t] \centerline{
\includegraphics[width=.9\textwidth]{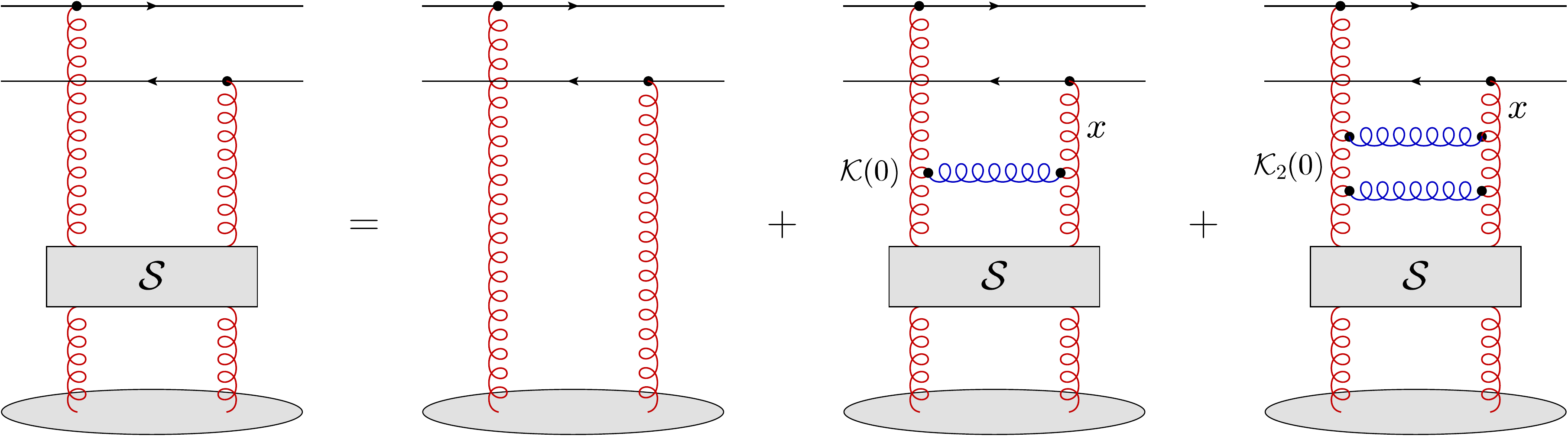}}
 \caption{\small A graphical illustration of the integral version of the NLO BK equation 
 \eqref{toyNLOBK}. The middle term with kernel $\calK(0)$ describes a soft gluon emission
 and corresponds to the LO BK kernel, but with a running coupling. The last term
 with kernel $\calK_2(0)$ represents the NLO piece of the BK kernel, with the
 running-coupling corrections excluded.  The two $s$-channel gluons included within
 $\calK_2(0)$ can be close in rapidity.
 The non-linear effects in $\calS$ describing multiple scattering
 are kept implicit.}
 \label{fig:toyNLOBK}
\end{figure}

These simpler notations hopefully make clear that the quantity denoted as $\Delta{\calS}$
must encode all NLO corrections to the BK kernel except for those expressing
the running of the coupling. These corrections should be 
computed in the large-$N_c$ limit, for consistency with our previous approximations.
They can be inferred by inspection of the NLO 
version of the BK equation shown in Eq. (5) of Ref.~\cite{Balitsky:2008zza}.
For convenience, we display the large-$N_c$ version of this equation
in Appendix \ref{sec:NLOBK}, where we also discuss its collinear improvement,
along the lines of Refs.~\cite{Iancu:2015vea,Iancu:2015joa}.
For simplicity, we shall stick here to our schematic notations and refer to
Appendix \ref{sec:NLOBK} for more explicit formulae.

The integral version of the NLO BK equation reads (in momentum space and adapted 
to the kinematics at hand)
\beq\label{toyNLOBK}
{\calS}(X_g)={\calS}_0+ \abar(\kt^2) \int_{x_g}^{1} \frac{\rmd x}{x} \,\calK(0) \,
\calS\big(X(x)\big)+
 \abar^2(\kt^2) \int_{x_g}^{1} \frac{\rmd x}{x} \,\calK_2(0) \,
\calS\big(X(x)\big)
\,,\eeq
where $\calK_2(0)$ is a compact notation for the NLO piece of the BK kernel
alluded to above: its action on $\calS$ generates 
 all the NLO terms visible in the r.h.s. of
\eqn{nlobkint} {\em except for the running coupling corrections},
which are explicitly included in the middle term of \eqn{toyNLOBK}.
More precisely, if one uses the prescription \eqref{azero} for the running coupling,
then in constructing $\calK_2(0)$ one should omit the NLO terms in
\eqn{nlobkint} which are proportional to $\bar b$
(but keep all the other ones). Besides, the large NLO corrections enhanced by double or
single collinear logarithms require all-order resummations  (`collinear improvement'),
to be shortly described (see also the discussion in Appendix \ref{sec:NLOBK}).

By comparing  Eqs.~\eqref{toySfull} and \eqref{toyNLOBK}, it is now obvious that
$\Delta{\calS}(X_g)$ must be identified with the third term in the r.h.s. of
\eqn{toyNLOBK}, that we now describe in some detail. As a general rule,
the NLO corrections to the BK kernel are obtained by evaluating all the 2-loop
graphs which yield a contribution of order $\alpha_s^2 Y$ and hence count for
a single step in the high-energy evolution (see Ref.~\cite{Balitsky:2008zza}
for an exhaustive list of diagrams and explicit calculations). A typical such a graph
involves a sequence of two gluon emissions,
whose longitudinal fractions $x_1$ and $x_2$ obey
$x_1\sim x_2\ll 1$; that is, the two gluons have similar rapidities, but they
are both soft compared to the projectile. 
The integration variable $x$ in the last term in \eqn{toyNLOBK} can be
interpreted as $x\equiv x_1+x_2$. The other independent rapidity integration,
say over the variable $u\equiv x_2/(x_1+x_2)$ with $0 < u \le 1$, is implicitly included
in the structure of the NLO kernel $\calK_2(0)$. This is possible since the scattering
between the 2-gluon system and the target is independent of $u$ to the accuracy of 
interest\footnote{The double integral over $x$ and $u$ also generates contributions
of order $(\alpha_s Y)^2$, which count for 2 steps in the LO evolution. Such contributions
are exlicitly subtracted in Ref.~\cite{Balitsky:2008zza}, 
via the `plus' prescription, since already generated via 2 iterations
of the middle term in \eqn{toyNLOBK}.}. A rather schematic, but  intuitive, graphical
illustration of \eqn{toyNLOBK} is shown in Fig.~\ref{fig:toyNLOBK}.

At large $N_c$, two successive gluon emissions
from the original $q\bar q$ dipole can generate up to three dipoles
in the fundamental representation. Accordingly, the third term in \eqn{toyNLOBK}
involves contributions which are cubic  in the dipole
$S$-matrix, together with quadratic and linear terms.
All these contributions are visible in \eqn{nlobkint}.
To the accuracy of interest, the kinematics of the primary gluon
(with energy fraction $x=x_1+x_2$) can be treated in the LLA. This explains
why we were able to use the same lower limit $x_g$ in the integral over $x$
and also the same rapidity variable $X(x)$ for the relevant $S$-matrices as
in the middle term in \eqn{toyNLOBK}, which encodes the LO evolution.

 \begin{figure}[t] \centerline{
\includegraphics[width=.85\textwidth]{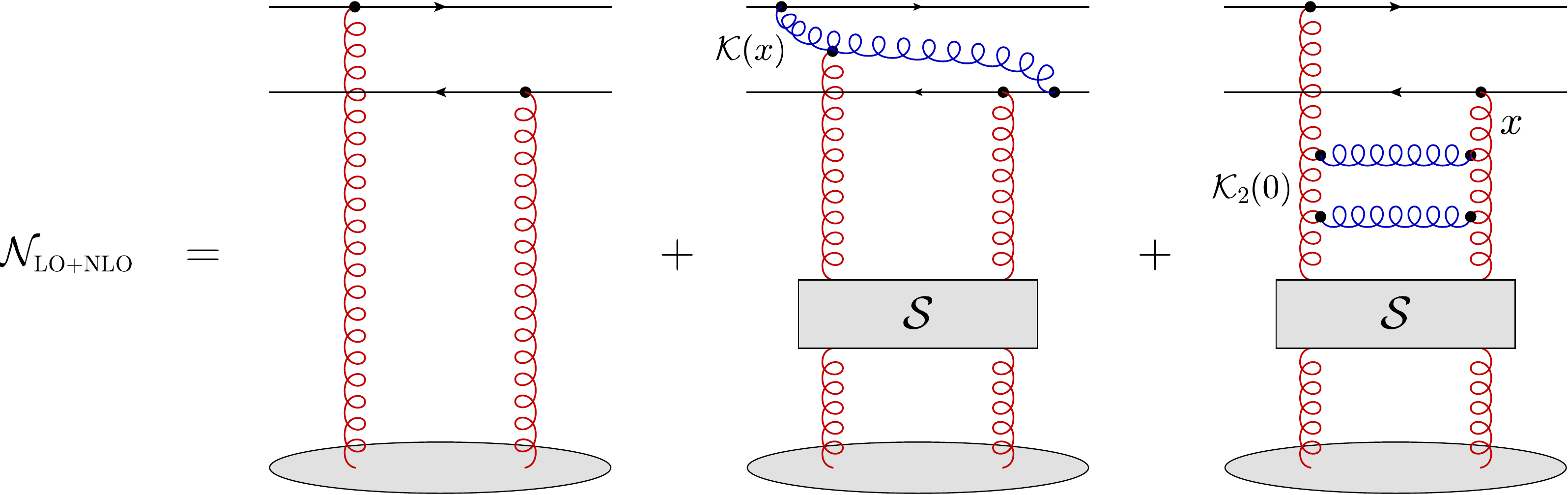}}
 \caption{\small A graphical illustration of the factorization of quark production 
 at NLO,  as schematically encoded in \eqn{toyNLO2}. The primary gluon emission, 
 with kernel $\calK(x)$, is included with exact kinematics. The evolution step depicted
 as two-gluon emission with kernel $\calK_2(0)$ succinctly represents the NLO
 piece of the BK kernel (but without the running-coupling corrections, which were
 already included in the middle term). Non-linear effects in $\calS$,
 corresponding to multiple scattering, are implicitly included but not explicitly shown.}
 \label{fig:toyNLO}
\end{figure}

To conclude, the result for the quark multiplicity which is complete to NLO accuracy
can be compactly, but schematically, written as
\beq\label{toyNLO2}
\calN_{\lo+\nlo}=\calS_0 + \abar(\kt^2) \int_{x_g}^{1} \frac{\rmd x}{x} 
\,\calK(x) \,\calS\big(X(x)\big)+
 \abar^2(\kt^2) \int_{x_g}^{1} \frac{\rmd x}{x} \,\calK_2(0) \,
\calS\big(X(x)\big)
\,,\eeq
and is illustrated in Fig.~\ref{fig:toyNLO}. To deduce a more explicit expression,
one should use \eqn{NLOkt} together with the formula for $\Delta\calS$ shown in
 Appendix \ref{sec:NLOBK}.
 
Although the above factorization scheme is formally correct to NLO, it might be still
too hard, if not impossible, to achieve a full NLO accuracy in a practical calculation.
Indeed, as already mentioned, the target evolution is not known to NLO and
the relation between the function $\calS\big(X(x)\big)$ which enters the above
factorization and the solution $\calS(x)$ to the NLO BK equation for 
the projectile is not known at the accuracy of interest. Besides, numerical calculations
might be hindered by the complexity of the NLO BK equation and of the transverse 
convolutions implicit in the two integral terms in \eqn{toyNLO2}.
In view of that, we would like to propose two approximation schemes which we
believe have more chances to be transposed in practice. In both schemes,
the approximations refer to the dipole evolution, whereas the NLO impact factor,
as represented by the kernel $\calK(x)$ in the middle term in \eqn{toyNLO2},
should be treated exactly.

The simplest approximation which is still physically meaningful is 
the LO approximation (in the sense of rcBK) to the dipole evolution.
This is obtained by neglecting the third term in \eqn{toyNLO2} and replacing
$\calS\simeq\calS_\rcBK$ within the integrand of the middle term.
A similar strategy has been used \cite{Stasto:2013cha,Stasto:2014sea,Watanabe:2015tja}
in relation with the `plus' prescription \cite{Chirilli:2011km,Chirilli:2012jd} ; the respective
numerical calculations, albeit very complex, turned out to be tractable
\cite{Zaslavsky:2014asa}.

The second approximation, which is more ambitious, refers to the use of the collinearly-improved
version of the BK equation, as proposed in Refs.~\cite{Iancu:2015vea,Iancu:2015joa}.
We recall that, besides the running coupling corrections, this equation also resums
double-collinear logarithms together with a subset of the single collinear logarithms
(which includes the respective contribution at NLO). The corresponding approximation
to \eqn{toyNLO2} involves two aspects. 
On the one hand, one must relate the function $\calS\big(X(x)\big)$, which encodes the
evolution of the target, to the solution to the collinearly-improved BK equation (which
refers to the evolution of the projectile). This aspect is particularly important
for the middle term in \eqn{toyNLO2} and will be discussed in  Appendix \ref{sec:proj}. 
On the other hand, one must use a simplified
version of the NLO BK kernel $\calK_2(0)$ which keeps only those corrections
to the LO kernel which refer to the collinear improvement. 
This will be described in what follows.

To that aim, it is convenient to use the coordinate representation: the
last term in \eqn{toyNLO2}, which we recall corresponds to the piece
$\Delta{\calS}(\bk,X_g)$ in \eqn{NLOkt}, will be written as
\beq\label{DeltaSF}
\Delta{\calS}(\bk,X_g)=\int \rmd^2\br\, \rme^{-\rmi \bk \cdot \br} {\Delta S}(\br,x_\sT)\,,
\eeq
where $\br=\bx-\by$ and
\begin{align}\hspace*{-0.5cm}
\label{Scoll}
{\Delta S}(\bx,\by; x_\sT)\equiv  \int_{x_\sT}^{1}\frac{\rmd x}{x} \int  \frac{\rmd^2\bz}{2\pi}\,
\bar{\alpha}_{\rm fac}\,{\mathcal M}_{\bx\by\bz}\big(\mathcal{K}_{\sdla}
\mathcal{K}_{\ssl}-1\big)
\Big[
 S\Big(\bx,\bz; \frac{x_\sT}{x}\Big) S\Big(\bz, \by;  \frac{x_\sT}{x}\Big)
 -S\Big(\bx,\by;  \frac{x_\sT}{x}\Big)\Big].
 \end{align}
In the above equation,
we recognize the dipole kernel ${\mathcal M}_{\bx\by\bz}$ and the running coupling
$\bar{\alpha}_{\rm fac}$ that were introduced before, together with two multiplicative 
corrections to the
kernel, $\mathcal{K}_{\sdla}$ and $\mathcal{K}_{\ssl}$, which encode the resummations
of double and respectively single collinear logarithms, as alluded to above. Physicswise, 
$\mathcal{K}_{\sdla}$ implements the condition of time-ordering for the successive soft
gluon emissions by the projectile, whereas $\mathcal{K}_{\ssl}$ resums a subset of the DGLAP
logarithms (see Refs.~\cite{Iancu:2015vea,Iancu:2015joa} for details).
The new rapidity argument $x_\sT$ which appears too in  \eqn{Scoll} will be later explained.

Specifically, $\mathcal{K}_{\sdla}$ is defined as the function
\begin{equation}\label{kdla}
 	\mathcal{K}_{\sdla}(\rho) = \frac{\rmJ_1
 	\big(2\sqrt{\abar \rho^2}\big)}{\sqrt{\abar \rho^2}} = 
 	1- \frac{\abar \rho^2}{2} + \frac{(\abar\rho^2)^2}{12} + \cdots,
 \end{equation}
evaluated at $\rho^2={L_{\bx\bz r}L_{\by \bz r}}$, with $L_{\bx\bz r} \equiv \ln[(\bx-\bz)^2/r^2]$.
If the double logarithm $L_{\bx\bz r}L_{\by \bz r}$ is negative, then one uses its absolute value
and the Bessel function $\rmJ_1$ gets replaced by the modified Bessel function $\rmI_1$.
Note however that if, e.g., $(\bx-\bz)^2 \ll r^2$, so that $L_{\bx\bz r} <0$, then
$(\by-\bz)^2 \simeq r^2$ and hence $L_{\by \bz r}\simeq 0$. Accordingly, the relatively small 
daughter dipoles bring no significant contributions to the difference
$\mathcal{K}_{\sdla}-1$. Furthermore,
\beq\label{ksl}
\mathcal{K}_{\ssl}=\exp\left\{-\abar A_1\left|\ln
\frac{(\bx \minus \by)^2}{\min\{(\bx \minus \bz)^2,(\by \minus \bz)^2\}}\right|\right\},\qquad
A_1 =\frac{11}{12} + \frac{\Nf}{6\Nc^3}\,,
\eeq
where $A_1$ is the `gluonic anomalous dimension' of the DGLAP evolution
(see e.g. \cite{Ciafaloni:1999yw} for details). Note that the difference 
$\mathcal{K}_{\sdla}\mathcal{K}_{\ssl}-1$ starts at $\order{\abar}$, as expected.
However, keeping only that lowest-order term in the expansion would artificially enhance
the importance of the `collinear' regions in phase-space --- the regions where
 the successive gluon emissions
(or dipole splittings) are strongly ordered in transverse sizes, or momenta. This is the origin
of the instability of the strict NLO approximation to the high-energy evolution,
as previously mentioned. 

Vice-versa, the all-order resummation of such corrections within the 
factor $\mathcal{K}_{\sdla}\mathcal{K}_{\ssl}$ suppresses the contributions from
the `collinear' regions and thus restores the convergence of perturbation theory.
This is rather obvious for the second factor $\mathcal{K}_{\ssl}$, which exponentially
cuts off the configurations where the daughter dipoles are either much smaller, or much
larger, than the parent dipole. But this is also true for the other factor  $\mathcal{K}_{\sdla}$,
which, as already mentioned, becomes important only when the daughter dipoles are
sufficiently large, such that $\abar \rho^2\gg 1$. 
In that case, the Bessel function $\rmJ_1\big(2\sqrt{\abar \rho^2}\big)$ is rapidly oscillating
when varying the position $\bz$ of the emitted gluon, hence the integral over 
the regions in space where `$\bz$ is large' (in the sense
that $|\bz-\bx|\sim |\bz-\by|\gg r$) averages to zero. 

To conclude, for gluon emissions which are strongly ordered in transverse momenta,
we have $\mathcal{K}_{\sdla}\mathcal{K}_{\ssl}\simeq 0$ and then the overall kernel 
in \eqn{Scoll} reduces to {\em minus} the LO kernel $-\calK(0)$. In turn, the latter
subtracts the soft and collinear contributions to the middle term in \eqn{toyNLO2}, 
that is, it implements the collinear improvement for the LO evolution, as it should.

There is one more aspect of \eqn{Scoll} which requires a few words of explanation: the rapidity 
arguments of the various $S$-matrices and, related to them, the
lower limit $x_\sT$ for the integral over $x$. If the evolution of these $S$-matrices is
computed to LO, i.e. according to rcBK, then one can identify 
$x_\sT\simeq x_g=\kt^2/\hat s$ and there is no distinction between projectile and target
evolutions. Albeit such an approximation would be formally justified 
to NLO accuracy, it is still preferable to use the collinearly-improved
evolution, i.e. the BK equation with the resummed
kernel ${\mathcal M}_{\bx\by\bz}\mathcal{K}_{\sdla} \mathcal{K}_{\ssl}$. 
Indeed, this would ensure a smooth matching 
with the middle term in \eqn{toyNLO2}, where the collinearly-improved version of the
evolution becomes compulsory. Since the latter has been formulated for projectile
evolution alone \cite{Iancu:2015vea,Iancu:2015joa}, one must understand what is
the longitudinal phase-space for the evolution in $x=q^+/q_0^+$ which corresponds 
to the physical range for the evolution in $X=q^-/P^-$. In coordinate space
and to the accuracy of interest, one can write (see e.g. \cite{Iancu:2015vea})
\beq\label{SPST}
S_\sT\big(\br,  X\big) = S_\sP\big(\br, x_\sT\big)\,\qquad
\mbox{with}\qquad X=\frac{Q^2}{\hat s}\, \gg\, x_\sT=\frac{Q^2_0}{\hat s}\,,
\eeq
where we have temporarily introduced the subscripts T (`target') and P (`projectile'),
to make the discussion more transparent and we recall that $\hat s=2q_0^+P^-$.
 In this equation, $Q=1/r$ is the dipole
resolution scale in the transverse plane and can be also identified with the 
transverse momentum $\kt$ of the produced quark, via the Fourier transform 
 \eqref{DeltaSF}; hence, $X\simeq X_g=x_g$. Furthermore,
$Q_0$ is the saturation scale in the nuclear target at low energy (say, within the MV model). 
The interesting situation is such that $\kt \gtrsim Q_s(X_g)\gg Q_0$
and therefore $x_\sT \ll x_g$, as indicated too in \eqn{SPST}. 

The relation \eqref{SPST} can be understood as follows. The scattering between 
the small dipole and the nuclear target probes the evolution of the latter
down to values of $X$ such that the longitudinal extent $\Delta x^+=1/(XP^-)$ of the
softest target fluctuations is still smaller than the lifetime $2q_0^+/Q^2$ of the dipole projectile.
This argument, which selects $X={Q^2}/{\hat s}$ as anticipated, is merely a variation
of our earlier derivation of \eqn{Xg} in Sect.~\ref{sec:kin}. If on the other hand the evolution
is encoded in the wavefunction of the projectile, then one should allow for the small-$x$
fluctuations with transverse momenta $\bq$ within the range $Q^2 > \qt^2 > Q_0^2$
and with large enough lifetimes $2xq_0^+/\qt^2\gtrsim 1/P^-$ (indeed, $1/P^-$ is the
longitudinal extent of the un-evolved target). These conditions imply $x\gtrsim \qt^2/\hat s
\ge Q_0^2/\hat s$, 
or $x \ge x_\sT$, which explains the lower limit in the integral over $x$ in \eqn{Scoll}.

One may find surprising that the rapidity interval $\Delta y=\ln(1/x_\sT)$
available for the evolution of the projectile is much larger than that, $\Delta Y=\ln(1/X_g)$,
allowed for the evolution of the target. But one should keep in mind that the projectile
evolution with decreasing $x$ is strongly constrained by the condition of time-ordering,
which limits the corresponding transverse phase-space (via the multiplicative
correction $\mathcal{K}_{\sdla}$ to the kernel) and thus reduces the evolution speed.
By contrast, there is no similar constraint for the evolution of the target with
decreasing $X$, since in that case the physical condition of time-ordering is
automatically satisfied.

\subsection{Subtracting the leading order evolution: why is this subtle}
\label{sec:sub}

The factorization scheme that we have constructed in the previous sections,
cf. Eqs.~\eqref{NLOkt} or \eqref{toyNLO2}, does not involve any subtraction:
there is no over-counting of the relevant perturbative contributions and hence
no need for a subtraction. We therefore expect that calculations based 
on this factorization should yield a positive result for the quark multiplicity.
More explicit arguments in that sense will be presented later in this section.
This represents a significant improvement over the previous proposal in
Refs.~\cite{Chirilli:2011km,Chirilli:2012jd}, so it is interesting to better understand
the relation between these two schemes.

From the discussion in Sect.~\ref{sec:CXY}, we recall that in the approach
\cite{Chirilli:2011km,Chirilli:2012jd} on explicitly subtracts the LO evolution of
the dipole $S$-matrix from the NLO correction to the impact factor. Such a 
subtraction can also be performed within our present approach. To synthetically 
describe that, we shall again use the schematic notations introduced in the previous
subsection. Using  \eqn{toyNLOBK}, we can express the last term in the NLO cross-section
\eqref{toyNLO2}  (a NLO correction to the dipole evolution)
as the difference between the dipole $S$-matrix at NLO and its LO evolution:
\beq\label{toyK2}
 \abar^2(\kt^2) \int_{x_g}^{1} \frac{\rmd x}{x} \,\calK_2(0) 
\calS\big(X(x)\big)={\calS}(X_g) - \left[{\calS}_0+ \abar(\kt^2) \int_{x_g}^{1} \frac{\rmd x}{x} \,\calK(0) \,
\calS\big(X(x)\big)\right]
\,.\eeq
The r.h.s. of this equation is not exactly the same as the difference $\calS-\calS_\rcBK$, 
but it is very close to it.
(The expression between the square brackets would reduce to $\calS_\rcBK$
if the $S$-matrix under the integral would be itself approximated by rcBK; recall
\eqn{toyrcBK}.) So, clearly, \eqn{toyK2} expresses a NLO correction which is {\em a priori}
small as the difference between two quantities which are individually large: each
of them includes the LO contribution $\calS_\rcBK$. 
Inserting \eqn{toyK2} into \eqn{toyNLO2}, one deduces an alternative expression for 
the NLO cross-section,
\beq\label{toyNLO1}
\calN_{\lo+\nlo}=\calS(X_g)
\,+ \abar(\kt^2) \int_{x_g}^{1} \frac{\rmd x}{x} \,\big[\calK(x)-\calK(0)\big]\,\calS\big(X(x)\big),
\eeq
in which the LO evolution (as represented by the kernel $\calK(0)$) is explicitly subtracted
from the impact factor. Since the difference $\calK(x)-\calK(0)$ 
vanishes as $x\to 0$, the above integral is controlled by large values $x\sim 1$
and it does not generate a small-$x$ logarithm anymore. 
That is, the second term in the r.h.s. of \eqn{toyNLO1} is a pure $\abar$ correction,
that can be viewed as the NLO contribution to the impact factor. The
NLO correction to the dipole evolution is now fully encoded in the first term
$\calS(X_g)$. Indeed,  to the NLO accuracy of interest, the $S$-matrix within the integral 
can be evaluated by using the LO approximation, $\calS\simeq\calS_\rcBK$.

\eqn{toyNLO1} is very similar to the proposal  in 
Refs.~\cite{Chirilli:2011km,Chirilli:2012jd}, which in our schematic notations reads
\begin{align}\label{toyCXY}
\calN_{\CXY}=
\calS(X_g)+ \abar(\kt^2) \int_{0}^{1} \frac{\rmd x}{x} \,\big[\calK(x)-\calK(0)\big]\,\calS(X_g)\,.
\end{align}
The main difference w.r.t. \eqn{toyNLO1} is that  
the $S$-matrix within the above integral over $x$
is not evaluated at the $x$-dependent rapidity argument $X(x)$,
but rather at its endpoint value $X_g=X(1)$. In other terms, \eqn{toyCXY}
is  {\em local} in the target rapidity $X_g$, as standard for the $k_\perp$-factorization.
In spite of this difference, the results in Eqs.~\eqref{toyCXY}  and  \eqref{toyNLO1}
are perturbatively equivalent to NLO accuracy. 
Indeed, since the integral over $x$ in  \eqref{toyNLO1} is controlled by $x\sim 1$,
there is no loss of NLO accuracy if one replaces $\calS\big(X(x)\big)\to \calS(X_g)$.
This can be easily checked by using the dominant energy-behavior of the BK
solution in the  weak scattering regime, namely  $\calS(X)\propto 1/X^\lambda$
with $\lambda=\order{\abar}$.

This being said, one should keep in mind that the limiting value $\calS(X_g)$ 
is strictly larger than $\calS\big(X(x)\big)$ for any $x<1$, because the function
$\calS(X)$ increases when decreasing $X$.
Hence, albeit formally allowed to the accuracy of interest,
the replacement $\calS\big(X(x)\big)\to \calS(X_g)$ could still be troublesome, 
in that it might result in an {\em over-subtraction}. Indeed,
as discussed  in \cite{Ducloue:2016shw}, the difference 
$\calK(x)-\calK(0)$ is strictly negative for sufficiently large $\kt\gtrsim Q_s$.
Hence, the negative correction to the impact factor at large $\kt$ is over included in
\eqref{toyCXY} as compared to \eqref{toyNLO1}, a feature which might 
contribute to the `negativity' problem under consideration.

In order to discuss this problem --- the fact that the cross-section computed according
to \eqn{toyCXY} becomes negative at sufficiently large (but still semi-hard) $\kt$
---, it is useful to keep in mind that previous numerical evaluations of \eqn{toyCXY}
used approximate versions of the dipole $S$-matrix, 
like $\calS_\rcBK$ \cite{Stasto:2013cha,Stasto:2014sea,Watanabe:2015tja}.
Hence, it would be interesting to understand why \eqn{toyCXY} with 
$\calS\to\calS_\rcBK$ can lead to a negative cross-section at large transverse momenta.

To that aim, let us consider the rcBK approximation to the `subtracted' version of our
factorization scheme, cf.  \eqref{toyNLO1}. This reads
\beq\label{toyrcBK}
\calN_{\rcBK}=  \calS_{\rcBK}+ \abar(\kt^2)  \int_{x_g}^{1} \frac{\rmd x}{x} \,
\big[\calK(x)-\calK(0)\big]\,\calS_\rcBK\big(X(x)\big)\,.
\eeq
Via the rcBK equation \eqref{toyLO}, this is equivalent to
\beq\label{toyrcBK2}
\calN_{\rcBK} = \calS_{0}+ \abar(\kt^2)  \int_{x_g}^{1} \frac{\rmd x}{x} \,
\calK(x)\,\calS_\rcBK\big(X(x)\big),
\eeq
which is of course the same as our `un-subtracted' result in \eqn{toyNLO} 
with $\Delta\calS=0$ and $\calS\to\calS_\rcBK$. 

We believe that the r.h.s. of \eqn{toyrcBK2} should be positive.
Indeed, when the primary gluon emission is treated in the eikonal approximation
($\calK(x)\to \calK(0)$), \eqn{toyrcBK2} is the same as the integral representation
of $\calS_\rcBK$, which is well known to be positive definite (at least in coordinate space).
With the actual kernel $\calK(x)$, the value of the integral term should be somewhat
reduced (since the correct phase-space for the primary gluon emission is smaller
than its eikonal estimate), but this should remain positive, as it describes the
growth of the dipole $S$-matrix via gluon emissions.

Hence, the `subtracted' result in \eqn{toyrcBK} should be positive as well. 
In spite of that, we believe that explicit numerical calculations based on \eqn{toyrcBK} 
may run into difficulties (in particular, yield a negative result)
because of the high degree of `fine-tuning' inherent in the subtraction method:
in going from \eqn{toyrcBK2} to \eqn{toyrcBK}, we have added and subtracted 
the same quantity ($\calS_\rcBK$), but we have done that in a very peculiar
way: we have added the l.h.s. of  \eqn{toyLO}, that is, $\calS_\rcBK$ itself, but we
have subtracted the r.h.s. of  \eqn{toyLO}, that is, the integral representation of $\calS_\rcBK$.
This procedure leaves the result unchanged if and only if the function
$\calS_\rcBK$ is an exact solution to the integral equation \eqref{toyLO}.
But any numerical approximation in solving rcBK, or in the Fourier transform
$S(\br,X)\to\calS(\bk,X)$, may lead to an imbalance between
the terms that have been added and respectively subtracted.  

For instance, such an imbalance
could be introduced by the treatment of the running coupling corrections,
which in general is not fully coherent between the coordinate-space 
and the momentum-space representations.
In practice, the momentum-space version of the rcBK equation, as shown
in \eqref{toyLO}, is not the {\em exact} Fourier transform of the respective equation in
coordinate space: the Fourier transform is not also applied to the
running of the coupling. So, strictly speaking, there is some
mismatch between Eqs.~ \eqref{toyrcBK} and \eqref{toyrcBK2} even when 
using the exact solution to the rcBK equation, as obtained in coordinate space.
When computing the cross-section according to \eqn{toyCXY}, 
this mismatch can be further enhanced by the fact that the function $\calS_\rcBK\big(X(x)\big)$
gets replaced by its maximal value $\calS_\rcBK(X_g)$ (cf. the discussion after 
\eqn{toyCXY}).

From the above discussion, we see that the rcBK equation plays the role of a `self-consistency
condition' for rewriting the cross-section in a `subtracted' form.
To illustrate the importance of having `good' solutions to this equation, let us consider a
somewhat extreme example, where this condition is strongly violated.
(A similar discussion can be found in \cite{Ducloue:2016shw}.)  
Namely, we consider the popular approximation in which the dipole $S$-matrix is taken from 
the GBW model \cite{GolecBiernat:1998js}, which is a Gaussian:
\beq\label{GBW}
{\calS}_\gbw(\bk, X)\,=\,\frac{4\pi}{Q_s^2}\,\rme^{-\frac{\kt^2}{Q_s^2}}\,,\eeq
where $Q_s^2(X)=Q_0^2(X_0/X)^\lambda$ with $\lambda\simeq 0.3$. Clearly, this is a very poor
approximation at high $\kt\gtrsim Q_s$, where it decays exponentially, in sharp contrast with
the power law tail $1/\kt^4$ predicted by pQCD.
Using this particular model within \eqn{toyNLO1}, one finds 
\beq\label{toyrcGBW}
\calN_{\gbw}= \calS_{\gbw}(X_g)+ \abar \int_{x_g}^{1} \frac{\rmd x}{x} \,
\big[\calK(x)-\calK(0)\big]\,\calS_\gbw\big(X(x)\big)\,.
\eeq
The `LO' piece $\calS_{\gbw}$ exponentially vanishes at large $\kt\gg Q_s$, whereas
the `subtracted' piece shows a tail $\propto 1/\kt^4$, since obtained by iterating once the LO BK equation.
So, the `subtracted' piece is not only
larger than the `LO' one, but it is the only one to survive at large $\kt$. Hence,
at $\kt\gtrsim Q_s$ the overall result reduces to the second, `NLO', piece, which is
negative in that particular region of phase-space \cite{Ducloue:2016shw}, as already mentioned.
This is in agreement with the numerical findings in \cite{Watanabe:2015tja,Ducloue:2016shw}. 

To summarize, albeit Eqs.~\eqref{toyrcBK} and \eqref{toyrcBK2} are in principle equivalent with each
other, the second equation is probably safer to use in practice. A similar discussion applies to the full NLO cross-section: the `unsubtracted' factorization \eqref{toyNLO2} should provide a meaningful result which is positive when evaluated with either the NLO $S$-matrix $\calS$, or with its collinearly-improved version $\calS_{\collBK}$. On the other hand, the `subtracted' version \eqref{toyNLO1}, which may look appealing since structurally simpler, is likely to be more tricky to use in practice.

\section{Summary and conclusions}
\label{sec:conc}

In this paper, we have established a factorization scheme allowing the computation of single-inclusive
particle production at forward rapidities in proton-nucleus collisions at next-to-leading order
in pQCD, in the presence of the non-linear effects associated with the high gluon
density in the nuclear target. The main difference with respect to the previous proposal in 
\cite{Chirilli:2011km,Chirilli:2012jd} is that our result involves no rapidity subtraction, 
meaning that it is free of the fine-tuning problem inherent 
in any such a subtraction scheme. The `fine-tuning' refers to the fact that
the numerical solution to the evolution equation for the dipole $S$-matrix
must be precisely known to ensure that
the quantity which is included as  `LO evolution' is properly subtracted from the
`NLO correction to the impact factor'. Further approximations
within the subtraction scheme, which are formally
allowed to NLO accuracy, or even small numerical errors in the
associated calculations, may spoil such a fine cancellation and lead to unphysical results. 
In our opinion, this is the reason why 
the cross-section computed within this scheme
appears to turn negative at sufficiently large transverse momenta 
\cite{Stasto:2013cha,Stasto:2014sea,Watanabe:2015tja,Ducloue:2016shw}.
By contrast, our scheme is more robust and should converge to physical,
positive-definite, results with considerably less numerical efforts.

Our factorization scheme relies on the skeleton structure of perturbative QCD
up to two-loop order, which is the relevant loop-order for the NLO calculation.
The most general version of our result is exhibited in \eqn{NLO}.
This version however is probably too complicated to be used in practice.
Fortunately, important simplifications become possible in the interesting situation
where the transverse momentum $\kt$ of the
produced quark is relatively large, $\kt\gtrsim Q_s$. In that case, the primary gluon is hard as
well, $\pt\sim\kt$, and the general formula \eqref{NLO} can be then replaced by
\eqn{NLOkt}  [or \eqref{NLOX0}], which is considerably simpler. The latter shows the same degree of
complexity, in terms of transverse integrations, like the formulae already used
in practice in relation with the subtraction method. So, we are confident that \eqn{NLOkt}
can indeed by explicitly evaluated, via the numerical tools developed in 
\cite{Zaslavsky:2014asa,Stasto:2013cha,Stasto:2014sea,Watanabe:2015tja,Ducloue:2016shw}.
As discussed in Sect.~\ref{sec:dijet}, the fact that we can approximate $\pt\sim\kt$ is specific 
to the problem of particle production. This would not apply, say, to deep inelastic scattering,
where the dipole evolution must be computed in transverse coordinate space.
In that case, if the parent dipole is sufficiently small ($r\ll 1/Q_s$) ---  as appropriate for DIS at 
relatively high virtuality $Q^2\gg Q_s^2$ ---, then the first gluon emission in the high-energy evolution
is considerably {\em softer} (it typically carries a transverse momentum $\pt \ll 1/r$) and the 
dependence upon its transverse kinematics cannot be simplified at NLO. This is visible in the NLO 
calculations of the DIS impact factor \cite{Balitsky:2010ze,Beuf:2011xd,Balitsky:2012bs,Beuf:2016wdz}.

To explicitly evaluate the quark multiplicity according to \eqn{NLOkt}, one also needs a suitable
approximation for the high-energy evolution of the dipole $S$-matrix. 
The simplest such an approximation which is still meaningful for
 phenomenology is the LO BK equation with running coupling
(rcBK) \cite{Kovchegov:2006wf,Kovchegov:2006vj,Balitsky:2006wa}. A more accurate 
treatment of the evolution can be obtained by using the collinearly-improved BK equation
\cite{Beuf:2014uia,Iancu:2015vea,Iancu:2015joa,Albacete:2015xza,Lappi:2016fmu,Hatta:2016ujq},
which resums the double collinear logarithms inherent in the `hard-to-soft' evolution
of the projectile together with a subset of single collinear logs. The inclusion of the
collinear improvement in the problem at hand is quite subtle, due to the need to properly
identify the rapidity phase-space for the evolution of the dilute projectile 
(the incoming quark, or the pair made with this quark and the
hard primary gluon). This is clarified in Sect.~\ref{sec:evol} and Appendix \ref{sec:proj}.

An important lesson emerging from our analysis is that one should not always insist in
writing the result of a NLO calculation in pQCD at high-energy in a `$\kt$-factorized form',
which is local in rapidity. This can be best appreciated by comparing our 
results in Eqs.~\eqref{toyNLO2}, \eqref{toyNLO1}, and \eqref{toyCXY}. \eqn{toyNLO2} is  obtained
via a direct evaluation of the relevant, one-loop and two-loop, diagrams. As such, it 
does not involve any rapidity subtraction at one-loop level: the would-be NLO
correction to the impact factor and the first step in the high-energy evolution of the color dipole 
are both encoded in the middle term in \eqn{toyNLO2}.  In \eqn{toyNLO1},  the two 
contributions that we just mentioned are explicitly separated from each other,
via a rapidity subtraction, but the result is still not `factorized': the 
$\order{\alpha_s}$-correction to the impact factor and the dipole $S$-matrix are still entangled
by the rapidity integral over $x$ (besides the transverse momentum convolutions which are
implicit in our schematic notations). Finally, \eqn{toyCXY} expresses a genuine 
$\kt$-factorization: the r.h.s. is local in the target rapidity $X_g$ 
and the NLO impact factor is explicitly factorized (in so far as the rapidity
dependence is concerned) from  the dipole $S$-matrices,
which encode the high-energy evolution\footnote{Notice that in this high-density context where
gluon saturation and multiple scattering are important, the Fourier transform
of the dipole $S$-matrix plays the role of a generalized
unintegrated gluon distribution for the nuclear target; see e.g. \cite{Dominguez:2011wm}.}.
These three representations for the NLO cross-section, \eqref{toyNLO2}, \eqref{toyNLO1}, 
and \eqref{toyCXY}, are all consistent with each other to NLO accuracy. Yet, the two 
representations involving a rapidity subtraction,  Eqs.~\eqref{toyNLO1} and \eqref{toyCXY},
are potentially affected by the issue of fine-tuning that we have identified in this paper.
By contrast, the original result in \eqn{toyNLO2} is free
from this problem and therefore it should provide a positive-definite estimate for the cross-section.

We would like to conclude this section with some prospects for the extension of this
factorization program beyond NLO. Our factorization scheme suggests an interesting
pattern which is likely to survive beyond the present approximations. The distinguished
feature of this pattern is the fact that the higher-order corrections to the impact factor are
not explicitly separated from the relevant corrections to the high-energy evolution.
To illustrate this point, let us consider quark production at NNLO. Without any claim
to completeness and leaving aside the possibility of new contributions which go beyond
the dipole picture, let us here indicate the generic structure that we expect in view of
our previous analysis in this paper. Namely, we expect the NNLO cross-section to include
the following four terms (in schematic notations, 
similar to those introduced in Sect.~\ref{sec:evol})
\begin{align}\label{toyNNLO}
\calN_{\lo+\nlo+\nnlo}= \calS_0 & + \abar(\kt^2) \int_{x_g}^{1} \frac{\rmd x}{x} 
\,\calK(x) \,\calS\big(X(x)\big) \nn & +
 \abar^2(\kt^2) \int_{x_g}^{1} \frac{\rmd x_1}{x_1}  \int_{x_g}^{x_1} \frac{\rmd x_2}{x_2} \,
 \big[\calK_2(x_1,x_2/x_1)-\calK_2(x_1,0)\big] \,
\calS\big(X(x_1,x_2)\big)\nn & +
 \abar^3(\kt^2) \int_{x_g}^{1} \frac{\rmd x}{x} \,\calK_3(0) \,
\calS\big(X(x)\big)
\,,\end{align}
where it is now understood that the dipole $S$-matrices evolve according to the
(presently unknown) NNLO version of the BK equation.  The argument of the running 
couplings is taken as $\kt^2$, for illustration, but this should be the right choice
only in the limit where the transverse momentum of the produced quark is larger than
any other transverse scale in the problem.  Also, the lower limit $x_g$ on the
integrals over $x$ is purely illustrative: the actual limit should depend upon the
exact kinematics. The first two terms in the r.h.s.
of \eqn{toyNNLO} have the same structure as the respective terms in \eqn{toyNLO2};
in particular, the second term encodes the NLO impact factor and the LO piece of the
BK kernel. The third term is a generalization of the last term in \eqn{toyNLO2}: it encodes
the NLO piece of the BK kernel together with the NNLO correction to the impact factor.
This term is generated by 2-loop graphs where the 2 emitted partons (say, gluons) can have
arbitrarily energy fractions $x_1$ and $x_2$ (with $x_2\le x_1$ for definiteness),
so their emissions must be computed with exact kinematics. The sum of all such graphs
is schematically represented by the kernel  $\calK_2(x_1,x_2/x_1)$. The subtraction of 
$\calK_2(x_1,0)$ is needed to avoid the over inclusion of the first two steps in the
high energy evolution, that were already included in the second term, with kernel 
$\calK(x)$. Finally, the fourth term in \eqn{toyNNLO} represents the NNLO correction
to the BK kernel, as generated by 3-loop graphs
in which the 3 gluons are all soft, but close in rapidity to each other:
$x_1\sim x_2\sim x_3\ll 1$. Clearly, in the limit where the very first emission by the projectile
is soft and computed in the eikonal approximation, the r.h.s. of \eqn{toyNNLO} becomes
proportional to the integral representation of the NNLO BK equation.

\section*{Acknowledgments}
\vspace*{-0.3cm}
We would like to thank Tuomas Lappi for inspiring discussions which triggered
our interest on this problem. We are grateful to Bertrand Duclou\'e, Dmitry Ivanov,
Tuomas Lappi, and Yan Zhu for useful comments on the manuscript.
A.H.M. would like to thank Bowen Xiao for useful and informative discussions.
D.N.T. would like to thank Guillaume Beuf for insightful remarks on the interplay
between evolution and factorization at next-to-leading.
A.H.M. and D.N.T. would like to acknowledge l'Institut de Physique Th\'eorique de Saclay for 
hospitality during the early stages of this work.
The work of E.I. is supported in part by the European Research Council 
under the Advanced Investigator Grant ERC-AD-267258.  The work of A.H.M.
is supported in part by the U.S. Department of Energy Grant \# DE-FG02-92ER40699.

\appendix
\section{Cancellation of shortly lived virtual fluctuations}
\label{app:canc}

As we have seen, the minus longitudinal momentum $p^-= p_{\perp}^2/2p^+$ of the primary gluon is constrained
by light-cone energy conservation, that is, by \eqn{Xx}. This constraint however exists only for the gluons which are crossing the cut (albeit eventually non measured), but not also for those which are emitted and reabsorbed on the same side of the cut. In principle, such `virtual' gluons are allowed to have very short lifetimes
$2 p^+/p_{\perp}^2 < 1/P^-$, however we shall show here that their respective contributions cancel exactly. 
To that aim, we shall separately consider the two possible regimes in terms of the ratio $p_{\perp}/k_{\perp}$.
We recall that $k_{\perp}$ is the transverse momentum of the measured quark and for the virtual diagrams 
it coincides with the momentum transferred by the target via scattering.

\begin{figure}[t]
\begin{minipage}[b]{0.33\textwidth}
\begin{center}
\includegraphics[scale=0.45]{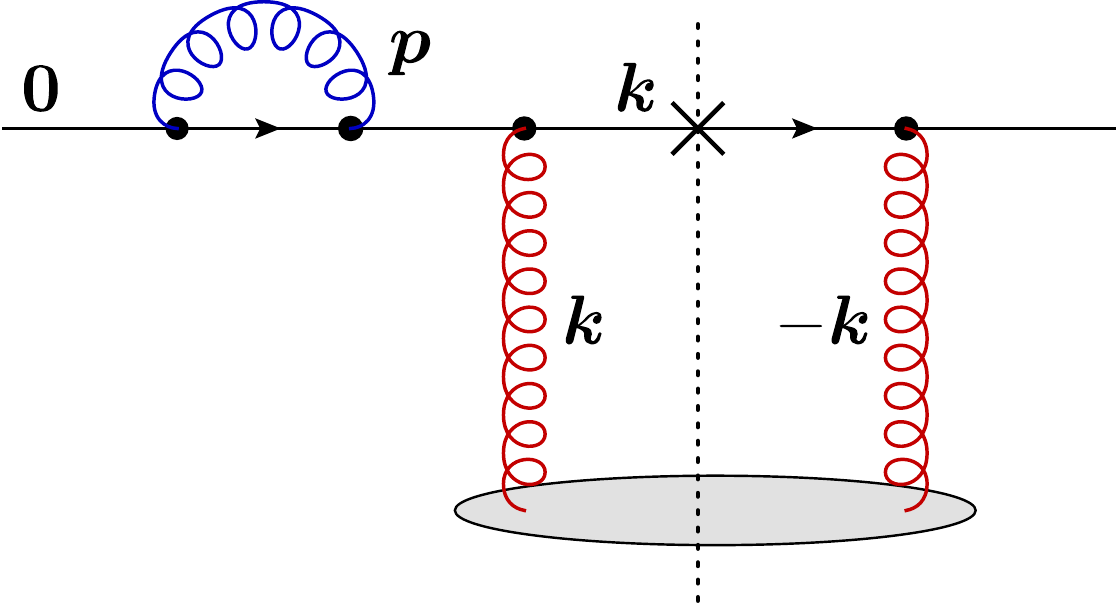}\\{\small (a)}
\end{center}
\end{minipage}
\begin{minipage}[b]{0.33\textwidth}
\begin{center}
\includegraphics[scale=0.45]{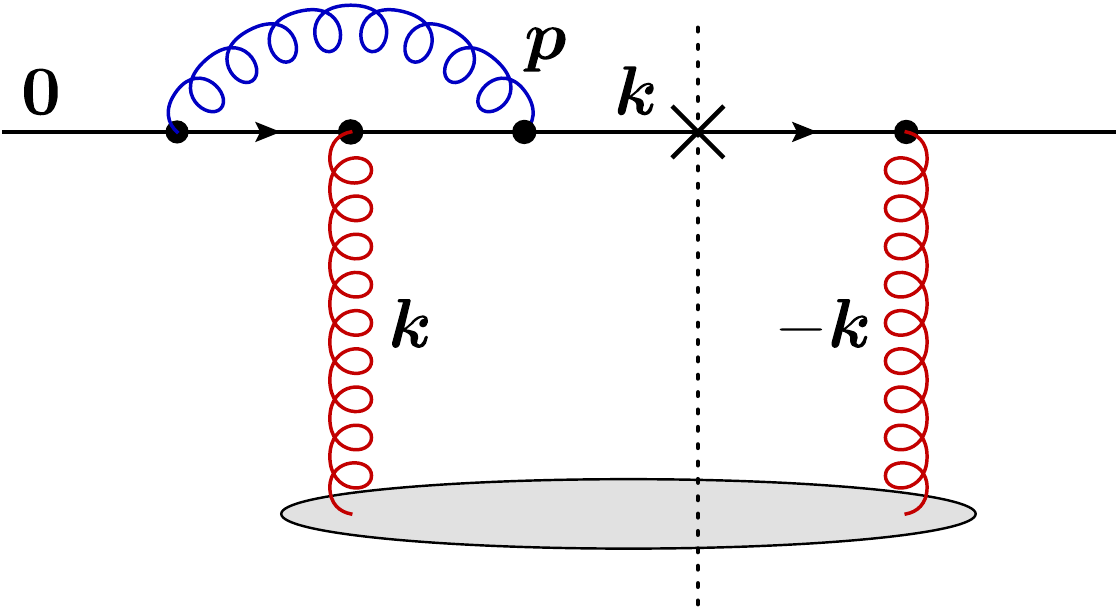}\\{\small (b)}
\end{center}
\end{minipage}
\begin{minipage}[b]{0.33\textwidth}
\begin{center}
\includegraphics[scale=0.45]{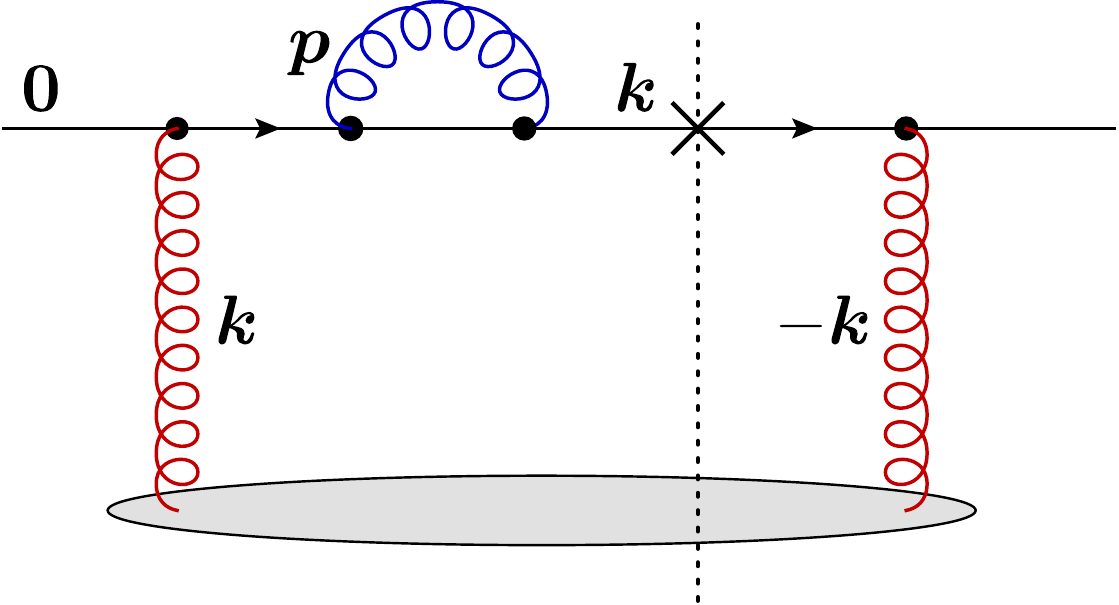}\\{\small (c)}
\end{center}
\end{minipage}
\caption{\small The virtual diagrams when $p_{\perp} \gg k_{\perp}$. In such a regime the shortly lived $p$-line is attached only to the projectile quark and the sum of the diagrams vanishes due to probability conservation. Transverse momenta are shown in the graphs.}
\label{fig:large}
\end{figure}

({\tt i}) $p_{\perp} \gg k_{\perp}$. This is the simplest case, as the virtual fluctuation cannot be
resolved by the scattering (the transverse separation $\Delta x_\perp \sim 1/\pt$ between the virtual gluon 
and its parent quark is much smaller than the transverse resolution $\sim 1/\kt$ of the exchanged gluon).
Accordingly, all the diagrams shown in Fig.~\ref{fig:large} are equally weighted by the $S$-matrix, so their
sum must vanish by probability conservation.
In practice, this occurs because the two self-energy graphs in figures (a) and (c) have a different sign as 
compared to the vertex correction in figure (b) and also an additional factor 1/2.

({\tt ii}) $p_{\perp} \lesssim k_{\perp}$. This case is much less trivial than ({\tt i}), since now one has to also consider the diagrams in which the virtual gluon with momentum $p$ is emitted and/or absorbed by the exchanged gluon. For the calculation that follows, it is convenient to change our apporach and use Light Cone Perturbation Theory (LCPT); see e.g. \cite{Kovchegov:2012mbw} for an introduction. 

\begin{figure}
\begin{center}
\begin{minipage}[b]{0.32\textwidth}
\begin{center}
\includegraphics[width=0.92\textwidth,angle=0]{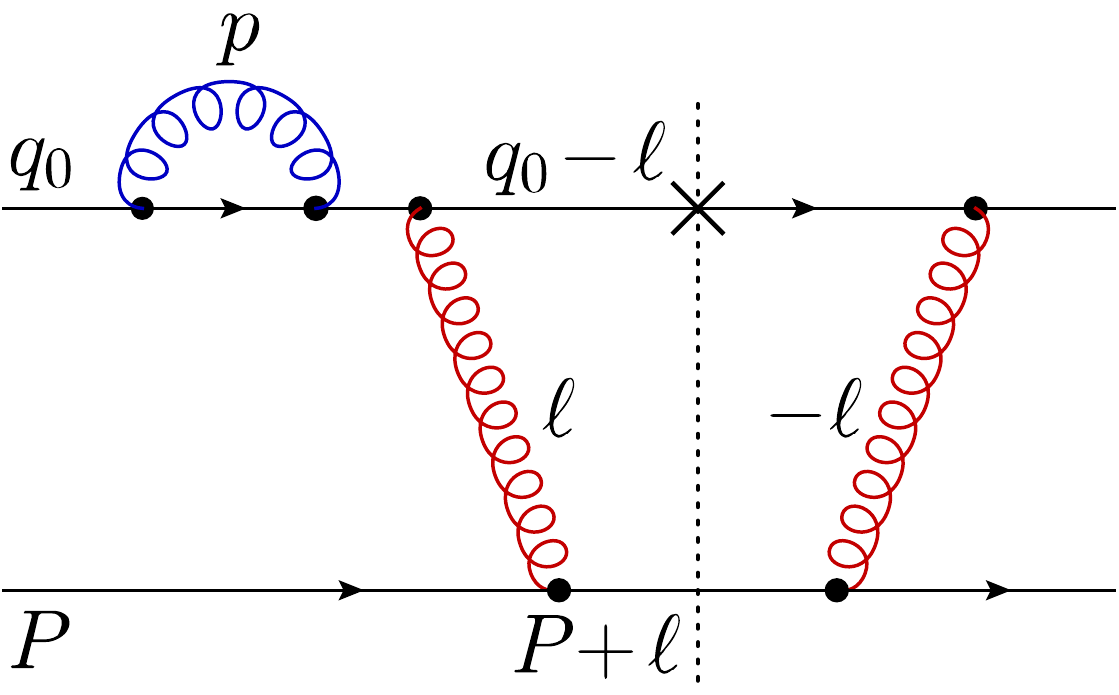}\\(A)\vspace{0.5cm}
\end{center}
\end{minipage}
\begin{minipage}[b]{0.32\textwidth}
\begin{center}
\includegraphics[width=0.92\textwidth,angle=0]{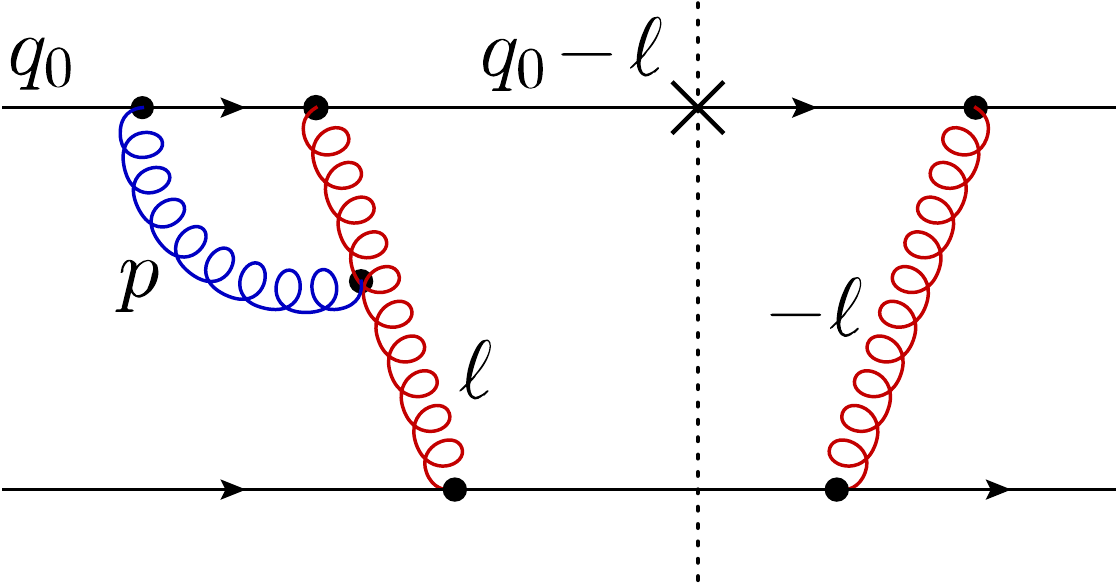}\\(B1)\vspace{0.5cm}
\end{center}
\end{minipage}
\begin{minipage}[b]{0.32\textwidth}
\begin{center}
\includegraphics[width=0.92\textwidth,angle=0]{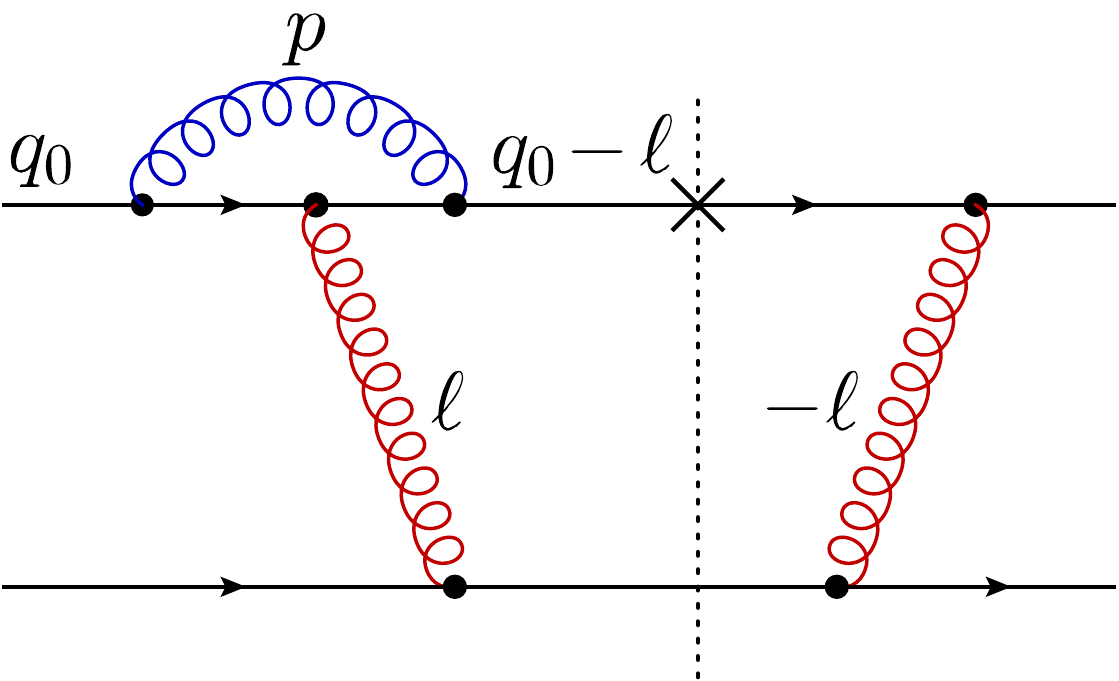}\\(B2)\vspace{0.5cm}
\end{center}
\end{minipage}
\begin{minipage}[b]{0.32\textwidth}
\begin{center}
\includegraphics[width=0.92\textwidth,angle=0]{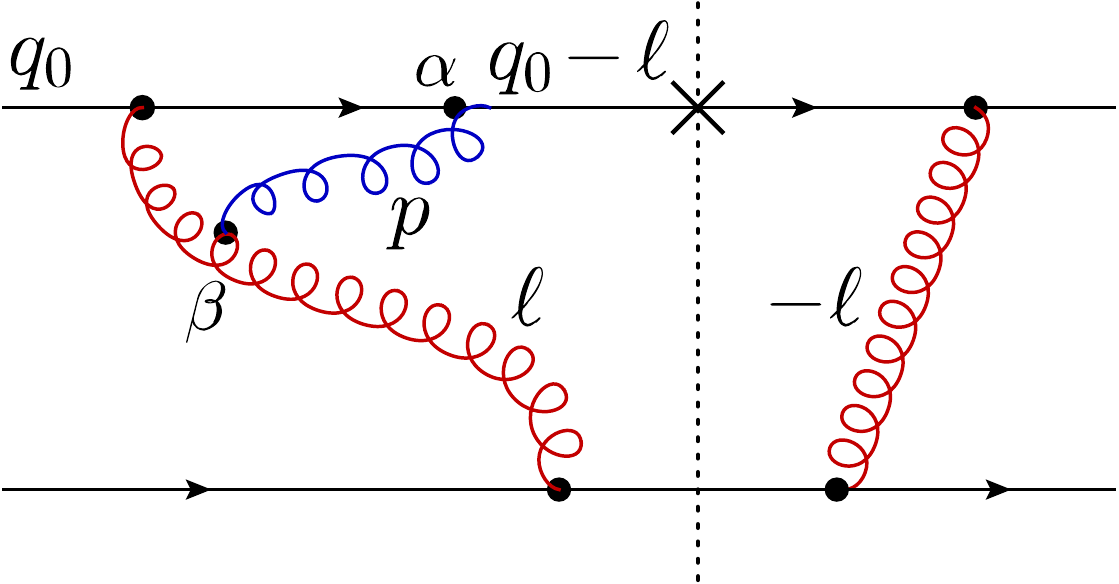}\\(C1)\vspace{0.5cm}
\end{center}
\end{minipage}
\begin{minipage}[b]{0.32\textwidth}
\begin{center}
\includegraphics[width=0.92\textwidth,angle=0]{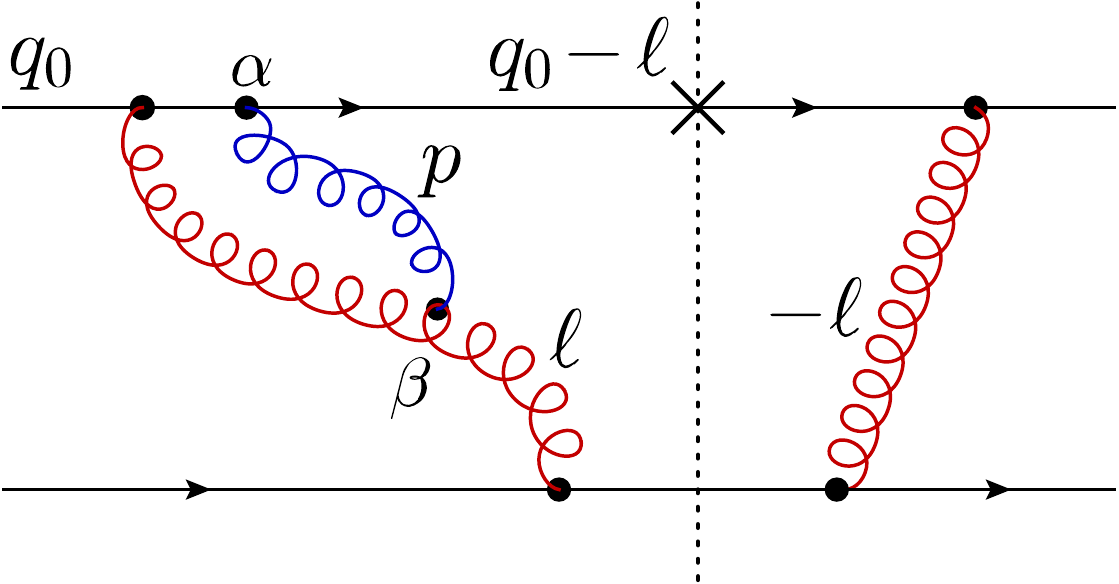}\\(C2)\vspace{0.5cm}
\end{center}
\end{minipage}
\begin{minipage}[b]{0.32\textwidth}
\begin{center}
\includegraphics[width=0.92\textwidth,angle=0]{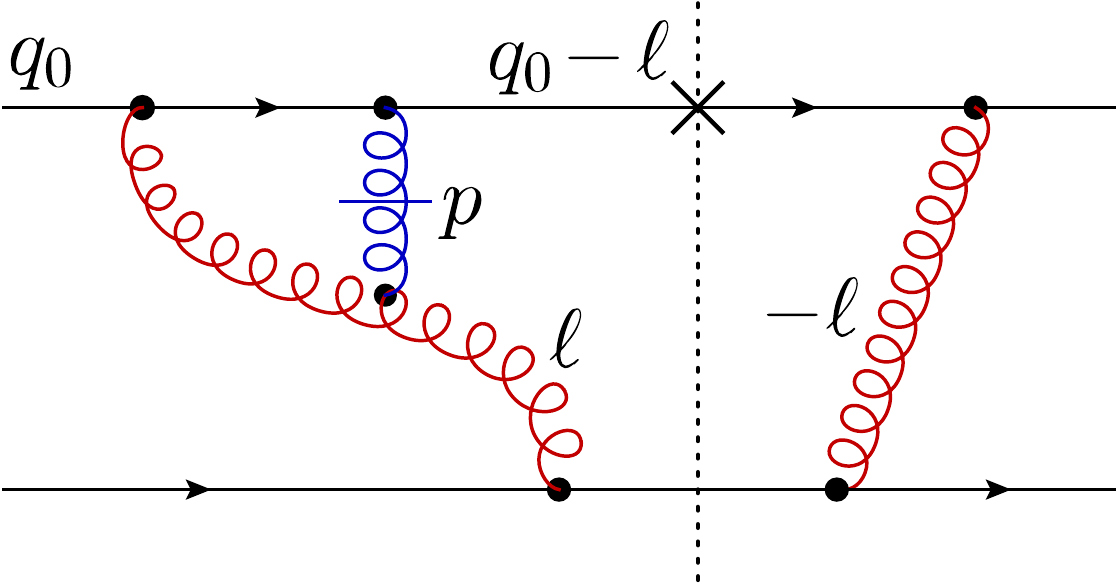}\\(C3)\vspace{0.5cm}
\end{center}
\end{minipage}
\begin{minipage}[b]{0.32\textwidth}
\begin{center}
\includegraphics[width=0.92\textwidth,angle=0]{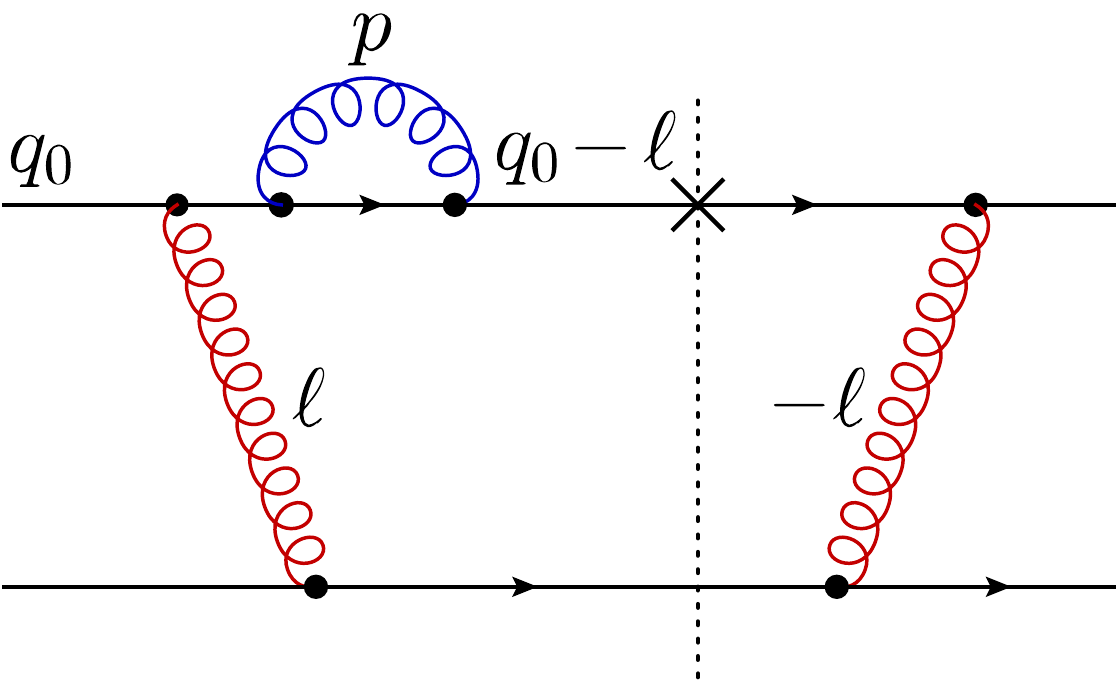}\\(D)\vspace{0cm}
\end{center}
\end{minipage}
\begin{minipage}[b]{0.32\textwidth}
\begin{center}
\includegraphics[width=0.92\textwidth,angle=0]{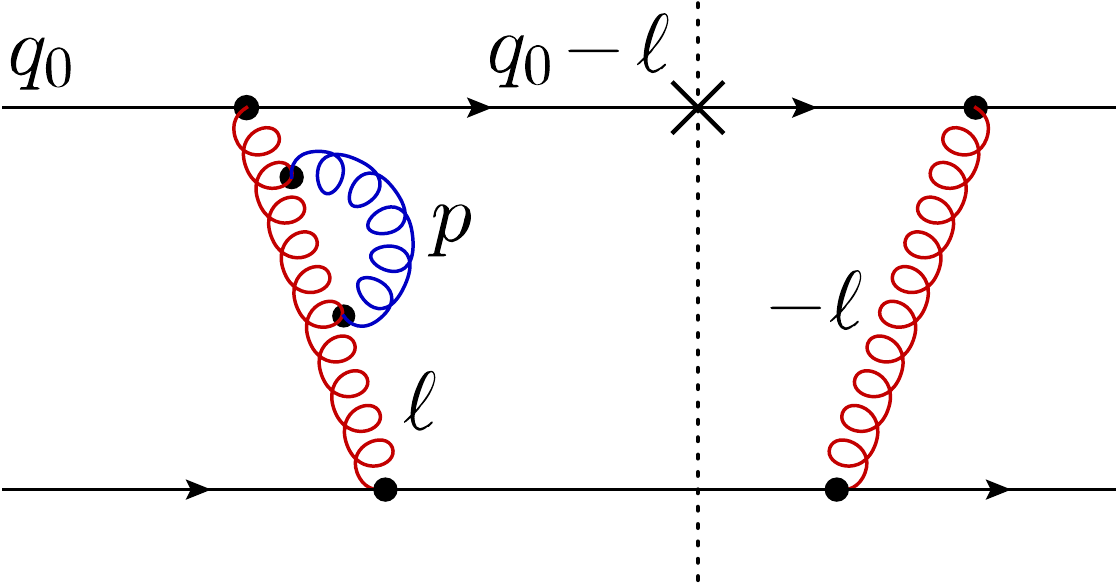}\\(E)\vspace{0cm}
\end{center}
\end{minipage}
\end{center}
\caption{\label{fig:same} \small The virtual diagrams when $p_{\perp} \lesssim k_{\perp}$. In contrast to those in Fig.~\ref{fig:large}, the shortly lived $p$-line can be now attached to the gluon ``exchange'' $\ell$-line. Four-momenta are shown in the graphs and flow from the left to the right. As in the text, $q_0 = (q_0^+,0,\bm{0})$, $P = (0,P^-,\bm{0})$, $\ell=(\ell^+,\ell^-,-\bm{k})$ and $p=(p^+,p^-,\bm{p})$, with $\ell^-=k_{\perp}^2/2 \ell^+$ and $p^-=p_{\perp}^2/2 p^+$. Just for C1, and for our convenience in the calculation, we let $\bp \to -\bp$.}
\end{figure}

To this end, both the virtual and the `exchanged' gluon are viewed as part of the projectile wavefunction and all the respective graphs are shown in Fig.~\ref{fig:same}. For simplicity, but without any loss of generality, we have taken the target to be a single quark. The four-momenta involved in the process are $q_0 = (q_0^+,0,\bm{0})$ for the incoming projectile quark, $P = (0,P^-,\bm{0})$ for the incoming quark from the target, 
$\ell=(\ell^+,\ell^-,-\bm{k})$ for the `exchanged' gluon and $p=(p^+,p^-,\bm{p})$ for the fluctuation. 
According to the rules of LCPT, all internal lines carry a positive plus longitudinal momentum and all
particles are on-shell, meaning that the respective minus components are fixed by the on-shell condition;
e.g. $\ell^-=k_{\perp}^2/2 \ell^+$ and $p^-=p_{\perp}^2/2 p^+$. 
Our convention in drawing Fig.~\ref{fig:same} is that momenta flow from the left to the right (since
this is the natural flow direction for the plus components).
In particular, the four-momentum of the $\ell$-gluon has the opposite flow from the one in the main text, 
but we have set $\bm{\ell}_{\perp} =- \bk_{\perp}$ so that the respective transverse momenta are
still the same.

The target quark must remain on-shell after the scattering, i.e. in the final state
crossing the cut. This condition fixes its minus longitudinal momentum, that is,
 \beq
 \label{onshell}
 (P + \ell)^2 = 0 
 \,\Rightarrow\,
 (P + \ell)^- = \frac{k_{\perp}^2}{2 \ell^+} = \ell^-.
 \eeq 
Now we invoke light-cone energy conservation (between the initial and final states), which implies
 \beq
 \label{lcecons}
 P^- = (P+ \ell)^- + (q_0-\ell)^- = 
 \ell^- + \frac{k_{\perp}^2}{2 (q_0^+ - \ell^+)}
 \simeq \ell^- + \frac{k_{\perp}^2}{2 q_0^+}.
 \eeq
Combining Eqs.~\eqref{onshell} and \eqref{lcecons} and using once more $\ell^+ \ll q_0^+$, we see that 
$(P+\ell)^- \simeq P^- \simeq \ell^-$ and more importantly we can determine the plus component of the $\ell$-gluon which reads
 \beq
 \label{ellplus}
 \ell^+ = \frac{k_{\perp}^2}{2 P^-}.
 \eeq
Looking deeply into the regime $2 p^+/p_{\perp}^2 \ll 1/P^-$, which means that the $p$-gluon is a fluctuation which has by far the shortest lifetime, and using $p_{\perp} \lesssim k_{\perp}$, we arrive at the strong ordering condition $p^+ \ll \ell^+$.

\begin{figure}[t]
\begin{minipage}[b]{0.45\textwidth}
\begin{center}
\includegraphics[scale=0.5]{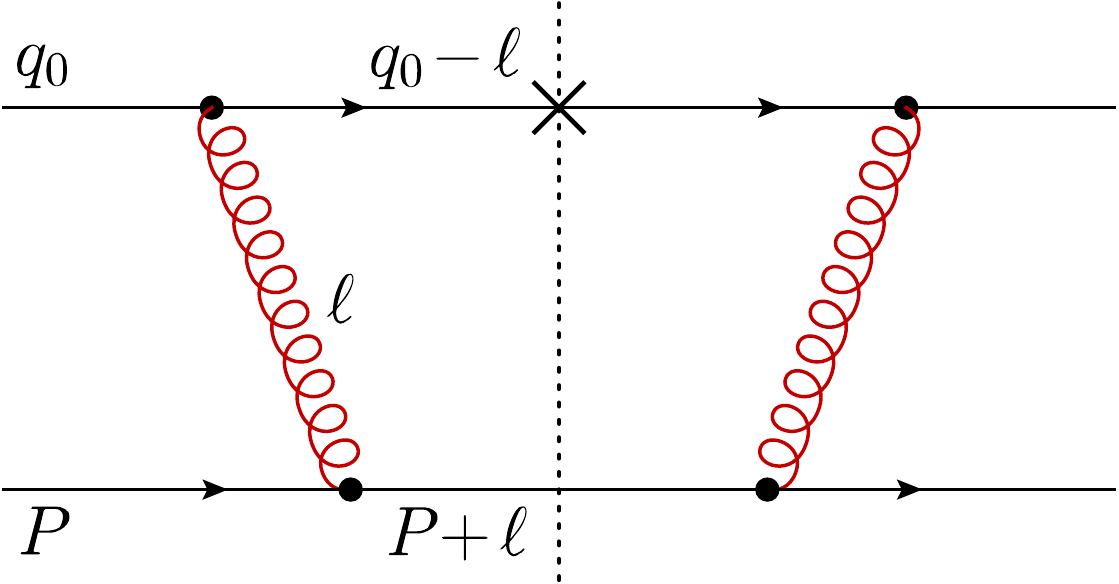}\\{\small (a)}
\end{center}
\end{minipage}
\begin{minipage}[b]{0.45\textwidth}
\begin{center}
\includegraphics[scale=0.55]{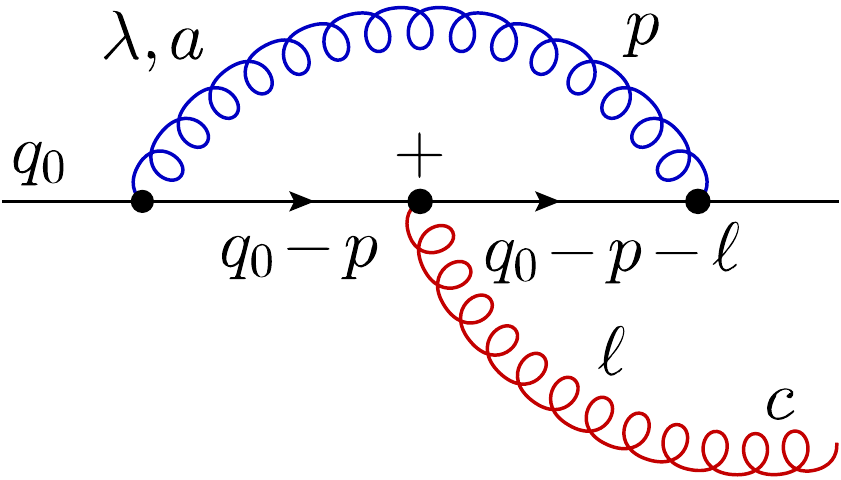}\\{\small (b)}
\end{center}
\end{minipage}
\caption{\small (a) Tree level diagram and (b) loop correction for the diagram B2 in Fig.~\ref{fig:same}. Four-momenta are shown in the graphs and flow from the left to the right.}
\label{fig:b2dec}
\end{figure}

We are not going to give a detailed calculation for all the diagrams appearing in Fig.~\ref{fig:same}, but only deal with some representative cases. Therefore, let us start by considering the diagram B2, which is decomposed in terms of the basic tree-level graph and the loop correction as shown in Fig.~\ref{fig:b2dec}. In LCPT, we can write this loop correction as
 \begin{align}
  {\rm B2}|_{\rm loop} \to (\rmi g)^2 
  t^a t^c t^a
  \int& \frac{\dif^2 \bp\, \dif p^+}{(2\pi)^3 2 p^+}\,
  \frac{1}{\frac{\bp^2}{2p^+}
  +\frac{(\bq_0-\bp)^2}{2(q^+_0-p^+)}}\,
  \frac{1}{\frac{\bp^2}{2p^+}
  +\frac{(\bq_0-\bp+\bk)^2}{2(q_0^+-p^+-\ell^+)}
  +\frac{\bk^2}{2\ell^+}}
  \nn
  &\times
  \gamma \!\cdot\! \epsilon^{\lambda}(p)\, 
  \gamma \!\cdot\! (q_0-p-\ell)\,
  \gamma^{+} 
  \gamma \!\cdot\! (q_0-p)\, 
  \gamma \!\cdot\! \epsilon^{\lambda}(p)\,
  \frac{1}{(2q_0^+)^2}, 
 \end{align}
where $t^a$ are the SU(3) generators in the fundamental representation and $\epsilon^{\lambda}(p)$ are the polarization vectors in the projectile light-cone gauge $A^+=0$, given by  
 \beq
 \epsilon^{\lambda}(p) = 
 \Big(0,\epsilon^{\lambda -} = \frac{\bm{\epsilon}^{\lambda} 
 \!\cdot\! \bp}{p^+},\bm{\epsilon}^{\lambda}\Big),
 \eeq
with $\bm{\epsilon}^{\lambda}$ two complex orthonormal two-dimensional vectors. Notice that since $p^+$ is very small, $\epsilon^{\lambda -}$ is the dominant component of the polarization vector. Using $\bq_0=0$, $q_0^+ \gg \ell^+\gg p^+$ and the condition in \eqn{ellplus} which fixes $\ell^+$, one readily finds that the dominant
term within both energy denominators and in the regime of interest (i.e.~for $p_{\perp} \lesssim k_{\perp}$ and $2 p^+/p_{\perp}^2 \ll 1/P^-$) is the light-cone energy $\bp^2/2p^+$ of the virtual gluon.
This was to be expected since, as already said, the $p$-line is by far the one with the  shortest lifetime. 
Regarding the inner products, we have $\gamma \cdot (q_0-p-\ell) \simeq \gamma \cdot (q_0-p) \simeq \gamma^- q_0^+$ and $\gamma \cdot \epsilon^{\lambda}(p) \simeq \gamma^+ \bm{\epsilon}^{\lambda}\cdot \bp/p^+$. Then, by using $t^a t^c t^a = -(C_{\rm F} - N_{\rm c}/2)t^c$ and putting aside a factor $t^c \gamma^+$ which is part of the tree-level graph, we eventually arrive at
 \beq
 \label{b2loop}
 {\rm B2}|_{\rm loop} = 
 \frac{\alpha_s}{\pi}
 \left(C_{\rm F} - N_{\rm c}/2 \right)
 \int \frac{\dif p^+}{p^+}
 \frac{\dif p_{\perp}^2}{p_{\perp}^2}.
 \eeq
One can calculate in a similar fashion the diagrams A, B1, D and E to find
 \beq
 \label{ab1de}
 {\rm A}|_{\rm loop} = {\rm D}|_{\rm loop} = 
 - \frac{\alpha_s C_{\rm F}}{2\pi}
 \int \frac{\dif p^+}{p^+}
 \frac{\dif p_{\perp}^2}{p_{\perp}^2}\,,
 \qquad
 {\rm B1}|_{\rm loop} = -{\rm E}|_{\rm loop} = 
 \frac{\alpha_s N_{\rm c}}{2\pi}
 \int \frac{\dif p^+}{p^+}
 \frac{\dif p_{\perp}^2}{p_{\perp}^2}.
 \eeq
The factor $-1/2$ for diagrams A and D emerges from the calculation and it is the usual factor associated with the wavefunction renormalization. A similar argument applies to diagram E, however in that case we have put the factor 1/2 by hand, since half of the full result should be attached to the lower vertex of the $\ell$-line.

Now let us move to the C diagrams, for which it is not hard to see that they will have the same color factor $N_{\rm c}/2$ with diagram B1. We have
 \begin{align}
 \hspace*{-1cm}
 \label{cloop1}
  {\rm C}|_{\rm loop} \to
  -(\rmi g)^2 \frac{N_c}{2} 
  \int \frac{\dif^2 \bp\, \dif p^+}{(2\pi)^3 2 p^+}\,
  \frac{1}{\frac{(\bk+\bp)^2}{2\ell^+}}
  \left[
  \frac{4 n_\alpha n_{\beta}}{p^+}
  -\frac{\epsilon^{\lambda}_{\alpha}(p) 
  \epsilon^{\lambda}_{\beta}(p)}{\frac{\bp^2}{2p^+} 
  + \frac{\bk^2}{2\ell^+}}
  -\frac{\epsilon^{\lambda}_{\alpha}(p) 
  \epsilon^{\lambda}_{\beta}(p)}{\frac{\bp^2}{2p^+} 
  + \frac{(\bk+\bp)^2}{2\ell^+}}
  \right]
 \frac{\gamma^{\alpha}\,\gamma \!\cdot\!q_0\, \gamma^+}{2 q_0^+}\,
 \frac{2\ell^{\beta}}{2\ell^+},
 \end{align}
where each term in the square bracket represents the respective contribution from C1, C2 and C3. The 4-vector $n$, appearing in the instantaneous term, is such that $n \cdot \upsilon = \upsilon^+$ for any vector $\upsilon^{\alpha}$. In \eqn{cloop1} we have already simplified the energy denominators (in particular we have neglected the terms suppressed by $1/q_0^+$) and kept only the terms that will contribute to the final result (notice also that in C1 we have let $\bp \to -\bp$ in order to combine them in an elegant way). Since $\epsilon^{\lambda -}$ is the large component of the polarization vector, we need to keep only the $--$ component of the tensor structure in the square bracket (notice that the indices here have been raised), which becomes
 \beq
 \label{bracketmm1}
 \Big[\cdots \Big]^{--} =
 \frac{4}{p^+} 
 - \frac{\frac{\bp^2}{(p^+)^2}}{\frac{\bp^2}{2p^+} 
  + \frac{\bk^2}{2\ell^+}} 
  -\frac{\frac{\bp^2}{(p^+)^2}}{\frac{\bp^2}{2p^+} 
  + \frac{(\bk+\bp)^2}{2\ell^+}}.
 \eeq 
Here it becomes clear that we cannot make the immediate approximation to keep only the $\bp^2/2p^+$ in the energy denominators (as we did for the previous class of diagrams), since the leading term cancels. It is a matter of straightforward algebra to find the dominant surviving terms:
 \beq
 \label{bracketmm}
 \Big[\cdots \Big]^{--} \simeq
 \frac{4}{\bp^2}\,
 \left[ \frac{(\bk+\bp)^2}{2\ell^+} + \frac{\bk^2}{2\ell^+}\right].
 \eeq
Since the transverse momentum of the exchange gluon is $\bk+\bp$ when attached to the upper vertex and $\bk$ when attached to the lower one, only the first term in \eqn{bracketmm} should be kept for our purposes, and \eqn{cloop1} leads to
 \beq
 \label{cloop}
 {\rm C}|_{\rm loop} = 
 \frac{\alpha_s N_{\rm c}}{2\pi}
 \int \frac{\dif p^+}{p^+}
 \frac{\dif p_{\perp}^2}{p_{\perp}^2}.
 \eeq
 
Putting Eqs.~\eqref{b2loop}, \eqref{ab1de} and \eqref{cloop} together, it is obvious that
 \beq
 {\rm A + B + C +D + E} =0,
 \eeq
which is the aforementioned cancellation of virtual corrections in the regime under consideration. Perhaps the most intuitive way to view this result is to realize that in diagrams C, D and E the shortly-lived gluon with momentum $p$ is emitted after the `exchanged' gluon. The gluon in the `exchange' line has a transverse velocity $|\bk+\bp|/\ell^+ \sim k_{\perp}/\ell^+$, so that over the small lifetime $2p^+/p_{\perp}^2$ of the $p$-fluctuation it separates from the quark $q_0$ only a very small distance
 \beq
 \label{separation}
 \Delta x_{\perp} \simeq \frac{k_{\perp}}{\ell^+}\,\frac{2 p^+}{p_{\perp}^2} 
 \sim\frac{2}{k_{\perp}}\,
 \frac{k_{\perp}^2}{2\ell^+}\,
 \frac{2 p^+}{p_{\perp}^2} 
 = \frac{2}{k_{\perp}}\,
 \frac{2 p^+ P^-}{p_{\perp}^2}
 \ll \frac{2}{k_{\perp}}.
 \eeq
To arrive at the above we have used \eqn{ellplus} and the fact that $2 p^+/p_{\perp}^2 \ll 1/P^-$. Therefore, the system composed of the quark and the `exchanged' gluon looks like a quark during 
the time interval defined by the emission and the reabsorption of the virtual, $p$-gluon. Indeed, one has
 \beq
 \label{cde}
 \left({\rm C + D + E}\right)|_{\rm loop} =
 - \frac{\alpha_s C_{\rm F}}{2\pi}
 \int \frac{\dif p^+}{p^+}
 \frac{\dif p_{\perp}^2}{p_{\perp}^2},
 \eeq
which can be associated with the quark wavefunction renormalization factor, more precisely with
 \beq
 \label{cdez2}
 \sqrt{Z_2}-1 = \sqrt{1 +\left(Z_2-1\right)} -1
 \simeq \frac{1}{2}\left(Z_2 -1\right),
 \eeq
 where the r.h.s. follows since $Z_2-1\sim \alpha_s\ll 1$.
Similarly graph A gives another factor of $1/2(Z_2-1)$, while graphs B can be identified with the 
quark-gluon vertex renormalization factor $Z_1$ to order $\alpha_s$ :
 \beq
 \label{bz1}
 ({\rm B_1 +B_2})|_{\rm loop} = \frac{1}{Z_1} -1\simeq 1- Z_1.
 \eeq 
Therefore, it is instructive to rewrite the sum of the one-loop virtual corrections as follows
 \beq
 \label{zprod}
 \left[1+\frac{1}{2}\left(Z_2 -1\right)\right]
 \left[1+\left(\frac{1}{Z_1}-1\right)\right]
 \left[1+\frac{1}{2}\left(Z_2 -1\right)\right]
 -1   \simeq \sqrt{Z_2}\, \frac{1}{Z_1}\, \sqrt{Z_2} -1 ,
 \eeq
so that the first factor in the square bracket can be identified with $1+{\rm A|_{loop}}$, the second with $1+({\rm B_1 +B_2})|_{\rm loop}$ and the last with ${\rm 1+(C+D+E)|_{loop}}$. The fact that our calculation gives $Z_2 - 1 =$ $- (1/Z_1 -1)$ at order $\alpha_s$, making the right hand side in \eqn{zprod} equal to zero, can be interpreted as the condition $Z_1 = Z_2$ valid in light cone gauge. At a first glance, it may seem strange to identify
 our calculation with an evaluation of renormalization constants when $p_{\perp}$ is not large, however one can check that the Slavnov-Taylor-Ward identities indeed require ${\rm 2A+B}=0$, as explicitly shown above.
 
 \section{Target versus projectile evolution beyond LO}
\label{sec:proj}

The NLO expressions for the quark multiplicity in Eqs.~\eqref{NLOkt} or \eqref{NLOX0} 
involve dipole $S$-matrices like $\calS\big(\bq,  X(x)\big)$, which in the main text have been 
assumed to follow the target evolution with decreasing $X=q^-/P^-$, 
from $X=X_0$ down to $ X(x)\ll 1$. 
The viewpoint of target evolution was indeed more convenient at a conceptual level, 
for developing a physical picture and also for formulating kinematical constraints 
like the energy conservation \eqref{Xx}. This is however less convenient in practice, because
the high-energy evolution of a dense nucleus is not known beyond LO (i.e. beyond rcBK).
Yet, as we shall now explain, this unknown target evolution can be replaced for the present 
purposes with the evolution of the dilute projectile, which is indeed known to the accuracy of interest.

More precisely, what is known is the NLO version of the BK equation in coordinate
space  \cite{Balitsky:2008zza,Balitsky:2013fea,Kovner:2013ona} together with its
 `collinear improvement' \cite{Beuf:2014uia,Iancu:2015vea,Iancu:2015joa,Hatta:2016ujq}.
When adapting these equations to the problem at hand, the only subtlety refers to the
proper identification of the longitudinal phase-space for the evolution of the projectile
--- that is, the interval in $p^+$ which is spanned by the evolution gluons.
This identification was straightforward at LO, where the `plus' and `minus' rapidities can be identified
with each other, as discussed in Sect.~\ref{sec:TvsP}, but it is less trivial beyond LO,
where the exact kinematics becomes important (recall also the discussion towards the end
of Sect.~\ref{sec:evol}).

When discussing projectile evolution in what follows, one should keep in mind that the 
relevant  `projectile' is the right-moving partonic pair made with the incoming quark
and its primary gluon. The kinematics of that pair, in particular the 3-momentum
$(p^+=xq_0^+,\bp)$ of the primary gluon, must be considered as fixed
for the purposes of the evolution --- this matter only for the boundaries of the 
evolution phase-space. The evolution rather refers to the three dipoles
$S$-matrices $\calS(\bq)$, $\calS(\bl)$, and $\calS(\bk)$ which appear
in Eqs.~\eqref{Jbare} and \eqref{Jvbare} --- and hence implicitly in
equations like \eqref{NLO} or \eqref{NLOkt} --- and describe the scattering 
between the primary quark-gluon pair (the `projectile') and the nuclear target
(whose wavefunction is not evolving anymore, as we now work in the infinite 
momentum frame of the projectile). These dipoles, that we referred to as 
`daughter dipoles' when discussing the primary emission in Sect.~\ref{sec:CXY},
will now act as {\em parent} dipoles for the high-energy evolution.
That is, the `evolution gluons' will be soft gluons which belong to the wavefunctions
of these three dipoles and whose longitudinal momenta $p^{+}_i = x_i q_0^+$
are necessarily smaller than the corresponding momentum $p^+=xq_0^+$
of the primary gluon. In what follows, we shall pick one of them, say $\calS(\bq)$, 
and study its high-energy evolution beyond LO.  Previously, we have (formally)
expressed the result in terms of the evolution of the target, like $\calS\big(\bk,  X(x)\big)$.
In what follows, we would like to compute the same quantity from the evolution
of the dipole $\calS(\bq)$ itself, which is known beyond NLO.

Let us first recall, from the discussion in Sect.~\ref{sec:TvsP}, that at LO
one simply has $\calS_\sT\big(\bq,  X(x)\big) = \calS_\sP\big(\bq, x_g/x\big)$,
where we have temporarily introduced the subscripts T (`target') and P (`projectile'),
to make the discussion more transparent. The rapidity argument $x_g/x$ of $\calS_\sP$
can be understood as follows:  the evolution variable is the
longitudinal fraction $z_i\equiv x_i/x$ of a generic evolution gluon w.r.t. the parent
dipole. At LO, this is bounded by $x_g/x < z_i <1$, where the lower limit comes from
the kinematical limit $X(x_i)=x_g/x_i\le 1$.
In other terms, the function $\calS_\sP\big(\bq, x_g/x\big)$ encodes 
the probability to find a gluon with longitudinal fraction $x_g$ within the quark-gluon projectile
 (see also Fig.~\ref{QtoQG}).

In what follows, we shall argue that beyond LO, the above relation should be extended to
\beq\label{TvsP}
\calS_\sT\big(\bq,  X(x)\big)\,=\,\calS_\sP\big(\bq,  x_\sT/x_\sP\big)\,,\quad\mbox{where}\quad 
x_\sT\equiv \frac{(1-x)Q_0^2}{\hat s}\,,\quad x_\sP
\equiv x(1-x)\mbox{min}\bigg\{1,\,\frac{\qt^2}{\kt^2}\bigg\} \,,
\eeq
where $Q_0$ is the `infrared cutoff' introduced by the initial condition at $x_0\sim 1$, 
e.g. the target saturation momentum at low energy (and the lowest 
transverse momentum scale in the problem). As before, $x_\sT$ refers to the softest
gluon in the wavefunction of the projectile which is involved in the collision, whereas 
$x_\sP$ to the hardest gluon that can be emitted by the primary quark-gluon pair.
The differences between $x_\sT$ and $x_g$ and, respectively, between $x_\sP$ and $x$,
are consequences of {\em time-ordering}, as we shall shortly explain. It is understood that
the function $\calS_P$ obeys a suitable `beyond LO' version of the BK equation,
which includes `collinear  improvement' 
\cite{Beuf:2014uia,Iancu:2015vea,Iancu:2015joa,Albacete:2015xza,Lappi:2016fmu,Hatta:2016ujq}
--- that is, which includes an all-order resummation of the radiative corrections enhanced
by double collinear logarithms. This equation, conveniently written as a differential equation
in the {\em plus} rapidity $Y\equiv \ln(x/x_\sT)$, 
must be integrated from $Y_0\simeq 0$ (where one can use the initial condition
$\calS_0$ from the MV model) up to $Y_\sP\equiv \ln ({x_\sP}/{x_\sT})$.
Full NLO accuracy can be achieved by using the
NLO version of the BK equation  \cite{Balitsky:2008zza}
amended by collinear improvement \cite{Iancu:2015vea,Iancu:2015joa} (the
feasibility of such a calculation has been demonstrated in \cite{Lappi:2016fmu}).

\begin{figure}[t] \centerline{
\includegraphics[width=.50\textwidth]{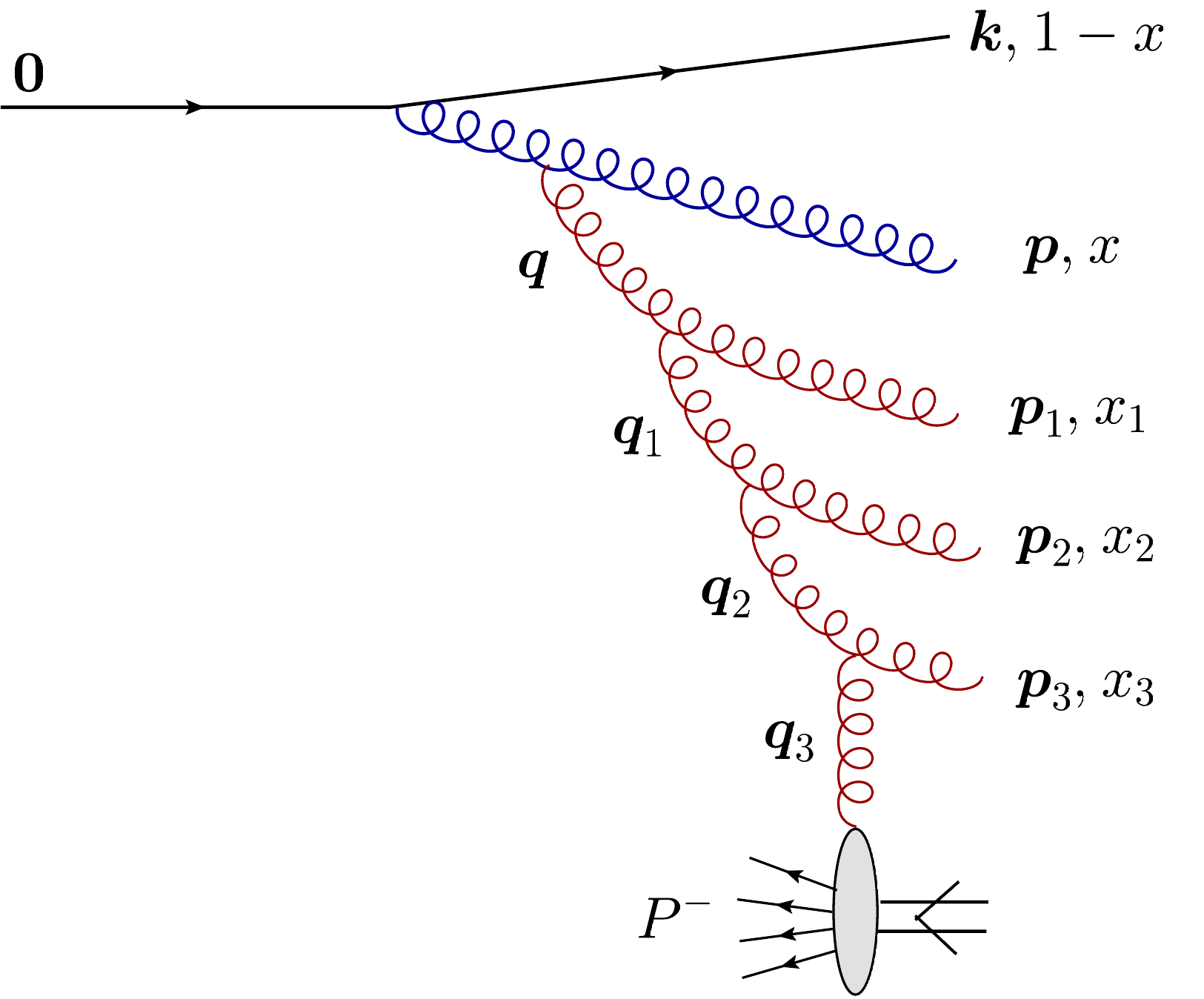}\quad
}
 \caption{\small A sequence of emissions contributing to the evolution of the quark-gluon projectile
 in the collinear regime. Both longitudinal and transverse momenta are strongly ordered down
 the cascade. The transferred momenta $q_{\perp i}$ are relatively soft, $q_{\perp i}
 \ll p_{\perp i}$, hence we also have $q_{\perp}\simeq p_{\perp 1}$, $q_{\perp 1}
 \simeq p_{\perp 2}$, etc. }
 \label{fig:TO}
\end{figure}

To justify \eqn{TvsP}, consider the evolution of the original quark-gluon pair via successive
emissions of soft gluons (see Fig.~\ref{fig:TO}). 
We focus on the `collinear regime' at $\kt \gg Q_0$, where the collinear resummations
are actually needed.  Then the projectile evolution looks rather similar as at LO,
in the sense that it is dominated by gluon emissions which are strongly ordered in both  
longitudinal ($x\gg x_1\gg x_2 \dots \gg x_\sT$) and transverse 
($\pt\gg p_{\perp 1}\gg p_{\perp 2} \dots \gg Q_0$) momenta, 
but which beyond LO must be also ordered in {\em time}~:
$\Delta x^+ > \Delta x^+_1> \Delta x^+_2 \dots > 1/P^-$. 

In the above, $x=p^+/q^+_0$ and $\pt=|\bp|$
refer to the kinematics of the primary gluon, as before, while 
$x_i=p^+_i/q^+_0$ and $p_{\perp i}=|\bp_i|$, with $i=1,\,2,\,\dots n$,
refer to the subsequent gluons in the cascade, which are softer.
The `lifetime'\footnote{This is truly the gluon {\em formation time}, 
i.e. the time during which a soft gluon fluctuation remains coherent with its 
parent partons at larger values of $x$.}
 of the $i$th generation, as given by the uncertainty principle or by the
respective energy denominator, is roughly $\Delta x^+_i \simeq 2p^+_i/p_{\perp i}^2$.
The `lifetime' $\Delta x^+$ of the primary gluon can be similarly estimated as
$\Delta x^+\simeq 2p^+/\pt^2$ or, more precisely (after using the complete energy
denominator for the quark-gluon fluctuation),
\beq
\frac{1}{\Delta x^+}\simeq \frac{\kt^2}{2(1-x)q^+_0}+\frac{\pt^2}{2xq^+_0}
\sim \frac{\kt^2}{2x(1-x)q^+_0}
\,,\eeq
where in the final equality we have used $\pt\sim\kt$, in line with the arguments leading 
to \eqn{NLOkt}.

Notice that, since both longitudinal ($p^+_i$) and transverse ($p_{\perp i}$) momenta
are simultaneously decreasing during the evolution, the condition of time-ordering, 
$\Delta x^+_i > \Delta x^+_{i+1}$, is not automatically satisfied: the LLA
includes unphysical configurations for which this condition is in fact violated. This is corrected
when going beyond LO, but the respective corrections are large (since they have to
`subtract' for double-logarithmic regions in phase-space) and hence must be resummed
to all orders. This is achieved by the `collinear resummations' aforementioned 
\cite{Beuf:2014uia,Iancu:2015vea,Iancu:2015joa,Hatta:2016ujq}, which essentially amount
to enforcing time-ordering within the projectile evolution.

Specifically, for the first emission, the condition $\Delta x^+_1 < \Delta x^+$ 
immediately implies $x_1\lesssim x_\sP$ with $x_\sP$ as defined in \eqn{TvsP}. 
(To obtain this, we have also used the fact
that $p_{\perp 1}\simeq \qt\ll \kt$, cf. Fig.~\ref{fig:TO}.)
That is, the longitudinal momentum of the first
gluon which counts for the evolution of $\calS_\sP(\bq)$ is not limited (from the above) by 
the respective momentum $p^+=xq_0^+$ of the primary quark, but by the generally smaller 
value $x_\sP q_0^+$. Similarly, the condition that the lifetime $\Delta x^+_n$ of the {\em last} (i.e.
softest) emitted gluon be larger than the longitudinal extent $1/P^-$ of the target, 
implies
\beq
x_n\,\gtrsim \,\frac{p_{\perp n}^2}{2q_0^+ P^-} =(1-x)\frac{p_{\perp n}^2}{\hat s} \,
 > \,(1-x)\frac{Q_0^2}{\hat s}=x_\sT \,,
\eeq
where we have also used  $q_0^+/Q^+ = x_p/(1-x)$ for the `real' terms. These considerations 
confirm that the longitudinal phase-space for the high-energy evolution of the quark-gluon `projectile' 
is given by $x_\sT < x_i < x_\sP$, in agreement with \eqn{TvsP}.

Note that  rapidity interval  $Y_\sP=\ln(x_\sP/x_\sT)$ available for the  evolution of the
projectile is larger (when $ {\qt}\gg {Q_0}$) than the corresponding interval 
$Y=\ln\big({1}/{X(x)}\big)$ for the evolution of the target:
\beq\label{yY}
Y_\sP= \,\ln \frac{x_\sP}{x_\sT}\,=\,\ln\frac{x\hat s}{\kt^2} + \ln\frac  {\qt^2}{Q_0^2}\,
=Y + \ln\frac  {\qt^2}{Q_0^2}\,.\eeq
This difference is compensated by time-ordering, which effectively
reduces the phase-space for the evolution of the projectile.

\section{NLO BK evolution and its collinear improvement}
\label{sec:NLOBK}

The BK equation describes the rapidity evolution of the $S$-matrix $S_{\bx\by} = 1 -T_{\bx\by}$ for the scattering of a color dipole with transverse coordinates ($\bx,\by$) off a hadronic target.
This equation is currently known to NLO accuracy \cite{Balitsky:2008zza}. Neglecting the terms suppressed in the multi-color limit $\Nc\gg 1$, one finds a closed equation for $S_{\bx\by}$ which, in the spirit of \eqn{BKint}, can be written in the integral form
 \begin{align}
 \label{nlobkint}
 \hspace*{-0.7cm}
 \delta S_{\bx\by}(x_{\rm \scriptscriptstyle T }) = \,&
 \frac{\abar}{2 \pi}
 \int_{x_{\rm \scriptscriptstyle T }}^1 \frac{\dif x}{x}
 \int \dif^2 \bz \,
 \frac{(\bx\minus\by)^2}{(\bx \minus\bz)^2 (\by \minus \bz)^2}\,
 \bigg\{ 1 + \abar
 \bigg[\bar{b}\, \ln (\bx \minus \by)^2 \mu^2 
 -\bar{b}\,\frac{(\bx \minus\bz)^2 - (\by \minus\bz)^2}{(\bx \minus \by)^2}
 \ln \frac{(\bx \minus\bz)^2}{(\by \minus\bz)^2}
 \nn
 & \hspace*{2.5cm}
 +\frac{67}{36} - \frac{\pi^2}{12} - \frac{5 \Nf}{18 \Nc}- 
 \frac{1}{2}\ln \frac{(\bx \minus\bz)^2}{(\bx \minus\by)^2} \ln \frac{(\by \minus\bz)^2}{(\bx \minus\by)^2}\bigg] 
 \bigg\}
 \left[S_{\bx\bz}(x) S_{\bz\by}(x) - S_{\bx\by}(x) \right]
 \nn
  & \hspace*{-1.8cm} + 
  \frac{\abar^2}{8\pi^2}
  \int_{x_{\rm \scriptscriptstyle T }}^1 \frac{\dif x}{x}
 \int \frac{\dif^2 \bu \,\dif^2 \bz}{(\bu \minus \bz)^4}
 \bigg\{-2
 + \frac{(\bx \minus\bu)^2 (\by \minus\bz)^2 + 
 (\bx \minus \bz)^2 (\by \minus \bu)^2
 - 4 (\bx \minus \by)^2 (\bu \minus \bz)^2}{(\bx \minus \bu)^2 (\by  \minus \bz)^2 - (\bx \minus \bz)^2 (\by \minus \bu)^2}
 \ln \frac{(\bx \minus \bu)^2 (\by  \minus \bz)^2}{(\bx \minus \bz)^2 (\by \minus \bu)^2}
 \nn
 & \hspace*{2.6cm} +
 \frac{(\bx \minus \by)^2 (\bu \minus \bz)^2}{(\bx \minus \bu)^2 (\by  \minus \bz)^2}
 \left[1 + \frac{(\bx \minus \by)^2 (\bu \minus \bz)^2}{(\bx \minus \bu)^2 (\by  \minus \bz)^2 - (\bx \minus \bz)^2 (\by \minus \bu)^2} \right]
 \ln \frac{(\bx \minus \bu)^2 (\by  \minus \bz)^2}{(\bx \minus \bz)^2 (\by \minus \bu)^2}\bigg\}
 \nn
 & \hspace*{2.6cm} \left[S_{\bx\bu}(x) S_{\bu\bz}(x) S_{\bz\by}(x) - S_{\bx \bu}(x) S_{\bu \by}(x)\right]
 \nn
  & \hspace*{-1.8cm} + 
 \frac{\abar^2}{8\pi^2}\,
 \frac{\Nf}{\Nc}
 \int_{x_{\rm \scriptscriptstyle T }}^1 \frac{\dif x}{x}
 \int \frac{\dif^2 \bu \,\dif^2 \bz}{(\bu \minus \bz)^4}
 \bigg[2
 - \frac{(\bx \minus\bu)^2 (\by \minus\bz)^2 + 
 (\bx \minus \bz)^2 (\by \minus \bu)^2
 - (\bx \minus \by)^2 (\bu \minus \bz)^2}{(\bx \minus \bu)^2 (\by  \minus \bz)^2 - (\bx \minus \bz)^2 (\by \minus \bu)^2}
 \ln \frac{(\bx \minus \bu)^2 (\by  \minus \bz)^2}{(\bx \minus \bz)^2 (\by \minus \bu)^2} \bigg]
 \nn
 & \hspace*{2.6cm} \left[ S_{\bx\bz}(x) S_{\bu\by}(x)- S_{\bx\bu}(x) S_{\bu\by}(x) \right],
 \end{align}
where $\Nf$ is the number of flavors, the coupling $\abar$ is evaluated at the renormalization scale $\mu$, while $\abar$ and $\bar{b}$ have been defined in the main text. Notice also that, for economy with respect to the notation used in Eq.~\eqref{Scoll} and afterwards, we have let $x \to x_{\rm \scriptscriptstyle T}/x$ and put the dependence on the transverse coordinates as subscripts. The $\delta$ in front $S_{\bx\by}(x_{\rm \scriptscriptstyle T })$ stands for the change from the initial condition, while the quantity $\Delta S$ that we have used in the main text stands for the $\abar^2$ terms on the r.h.s.~of the above equation, except for those which are proportional to $\bar{b}$.

The term with a single integration (SI) over the transverse coordinate $\bz$ keeps the same structure
as the LO equation, but receives a correction of order $\mathcal{O}(\abar^2)$ to the kernel. In particular, it contains the running coupling corrections proportional to $\bar{b}$. The terms of order $\mathcal{O}(\abar^2)$ with a double integration (DI) over the coordinates $\bu$ and $\bz$ arise from partonic fluctuations involving two additional partons at the time of scattering. The first of such terms is independent of $\Nf$ and clearly represents the case where both daughter partons are gluons. The $S$-matrix structure $S_{\bx\bu} S_{\bu\bz} S_{\bz\by} - S_{\bx \bu} S_{\bu \by}$ corresponds to a sequence of emissions in which the original dipole $(\bx,\by)$ emits a gluon at $\bu$ giving rise to the two dipoles  $(\bx,\bu)$ and  $(\bu,\by)$, and then the dipole $(\bu,\by)$ emits a gluon at $\bz$ leading to the dipoles  $(\bu,\bz)$ and  $(\bz,\by)$. The `real' term 
$S_{\bx\bu} S_{\bu\bz} S_{\bz\by}$ describes the situation where both gluons interact with the target, while the `virtual' term $ - S_{\bx \bu} S_{\bu \by}$ stands for the the case where the gluon at $\bz$ has been emitted and reabsorbed either before, or after, the scattering. This `virtual' term  ensures that the potential `ultraviolet' singularity due to the $1/(\bu - \bz)^4$ factor in the kernel is in fact harmless. The same discussion applies to the second DI term, proportional to $\Nf$,  which represents the case that the additional partons at the time of scattering are a quark and an antiquark, and thus the number of dipoles involved in the scattering is just two.

The issue with the NLO BK equation given in \eqref{nlobkint} is that there are terms in the kernels  which can get large in certain kinematic regimes, thus invalidating the strict $\abar$-expansion. The first class of such terms contain the corrections proportional to $\bar b$ in the SI term in \eqn{nlobkint}, and has already been discussed in Sect.~\ref{sec:evol}. The scale $\mu$ should be chosen in such way, that the logarithms of transverse dipole sizes proportional to $\bar{b}$ (and only those) become innocuous in any kinematic regime. It is trivial to see that both choices suggested in Sect.~\ref{sec:evol} will cancel any potentially large logarithmic contribution. In fact, our {\it fac} prescription, cf.
\eqn{azero}, is by definition the one in which the sum of all terms proportional to $\bar{b}$ vanishes identically.  

Now we wish to discuss NLO corrections enhanced by logarithms associated with large separations in transverse sizes (or momenta) between successive emissions. These `collinear' corrections become large only in the weak-scattering regime where all the dipoles are small compared to the target saturation scale $1/Q_s$, so that we can linearize w.r.t.~to the scattering amplitude $T$. At this level, we can drop the term proportional to $\Nf/\Nc$ in \eqn{nlobkint} since it vanishes after linearization, as one can check by using the symmetry of the kernel under  the interchange $\bu \leftrightarrow \bz$. More precisely, we consider the strongly ordered regime
\beq
\label{stor}
 1/Q_{s} \gg |\bz-\bx|\simeq |\bz-\by| \simeq |\bz-\bu| \gg |\bu-\bx| \simeq |\bu-\by| \gg |\bx-\by|,
\eeq   
which means that the parent dipole is the smallest one, a gluon is emitted far away at $\bu$, a second one even further at $\bz$,
but all the possible dipole sizes remain smaller than the inverse saturation momentum. Denoting by $r$, $\bar{u}$ and $\bar{z}$ the size of the parent dipole, the size of the dipoles involving $\bu$
and the size of the dipoles involving $\bz$, respectively, we have $r^2\ll \bar{u}^2\ll \bar{z}^2$. 

Inspecting the SI piece in the NLO BK equation \eqref{nlobkint}, it is obvious that the term which involves the double transverse logarithm (DTL) is the dominant one in the kinematic region defined in \eqn{stor}. In this regime we are also allowed to write $S_{\bx\bz} S_{\bz\by} - S_{\bx\by} \simeq -T_{\bx\bz} -T_{\bz\by} + T_{\bx\by} \simeq -2 T(\bar{z})$, with the second approximate equality arriving from the fact that the dipole amplitude for a small dipole is roughly proportional to the dipole size squared. Thus, the net result comes from the `real' term, i.e.~the one which involves the large daughter dipoles.

A single transverse logarithm (STL) is hidden in the DI term. To see that, let us isolate the kernel
\begin{align}
 \label{mstl}
 \mathcal{M}_{\slog} \equiv
 \frac{\abar^2}{8\pi^2(\bu \minus \bz)^4}
 \bigg[-2
 + &\,\frac{(\bx \minus\bu)^2 (\by \minus\bz)^2 + 
 (\bx \minus \bz)^2 (\by \minus \bu)^2
 - 4 (\bx \minus \by)^2 (\bu \minus \bz)^2}{(\bx \minus \bu)^2 (\by  \minus \bz)^2 - (\bx \minus \bz)^2 (\by \minus \bu)^2}
 \nn
 &\, \times \ln \frac{(\bx \minus \bu)^2 (\by  \minus \bz)^2}{(\bx \minus \bz)^2 (\by \minus \bu)^2} \bigg],
 \end{align} 
which in the collinear regime can be successively written as
\begin{align}
\label{mstllim}
\mathcal{M}_{\slog} \simeq &\,
\frac{\abar^2}{8\pi^2\bar{z}^4}
 \bigg[-2 + \frac{2 \bar{u}^2 - 2 \bar{u} r \cos\phi -3 r^2}{r^2 - 2 \bar{u} r \cos\phi}
 \ln \left(1+ \frac{r^2 - 2 \bar{u} r \cos\phi}{\bar{u}^2} \right)
  \bigg]
  \nn
  \simeq &\,
 -\frac{\abar^2}{\pi^2}\,\frac{6 -\cos^2\phi}{12}\,
 \frac{r^2}{\bar{u}^2\bar{z}^4}
 \to
 -\frac{11 \abar^2}{24\pi^2}\,
 \frac{r^2}{\bar{u}^2\bar{z}^4},
\end{align} 
where $\phi$ is the angle between $\br$ and any of the two dipoles involving $\bu$. To arrive at \eqref{mstllim}, we first set all dipole sizes which
include $\bz$ equal to each other, since any subleading term would be highly suppressed by inverse powers of $\bar{z}$. This simplifies significantly the expansion in the regime of interest, since the only $z$ dependence left is the one explicit in the prefactor. Then we have taken the limit $r \ll \bar{u}$ and finally we have performed an average over the angle $\phi$ between the parent dipole and those involving $\bu$. As evident in \eqn{mstllim},  the would-be leading term of order $1/\bar{z}^4$ has cancelled and the first non-vanishing term is suppressed by the power factor $r^2/\bar{u}^2$. Now by linearizing the $S$-matrices multiplying  $\mathcal{M}_{\slog}$ and realizing, as in the DTL case, that the `real' term dominates, we can approximate  $S_{\bx\bu} S_{\bu\bz} S_{\bz\by} - S_{\bx\bu}S_{\bu\by} \simeq
-T_{\bu\bz} - T_{\bz\by} + T_{\bu\by} \simeq -2T(\bar{z})$. Since this is independent of the intermediate dipole size $\bar{u}$, the integration over the latter, within the range limited by $r$ and $\bar{z}$, leads to the anticipated STL. 

Thus, putting together the LO term and the two NLO ones enhanced by the STL and the DTL, one finds that the NLO BK equation in the collinear regime \eqref{stor}
reduces to
 \begin{equation}
 \label{deltatlog}
         \delta T(r;x_{\rm \scriptscriptstyle T})
         = \abar 
         \int_{x_{\rm \scriptscriptstyle T}}^1
         \frac{\dif x}{x}
         \int_{r^2}^{1/Q_s^2}
         \dif\bar{z}^2\, \frac{r^2}{\bar{z}^4}
         \left(1 
         -\frac{1}{2}\,\abar
         \ln^2 \frac{\bar{z}^2}{r^2} - \frac{11}{12}\,\abar
         \ln \frac{\bar{z}^2}{r^2} \right) T(\bar{z}; x).
 \end{equation}
 Clearly, for sufficiently large daughter dipoles, the NLO contributions are enhanced by large transverse logarithms and become comparable to, or larger than, the LO one. Then, the present perturbative expansion of the kernel, which is of fixed order in $\abar$, cannot be trusted anymore. Eventually this problematic behavior is transmitted to the solution of equations like \eqref{deltatlog} and \eqref{nlobkint} which becomes unstable, as indeed seen in \cite{Avsar:2011ds,Lappi:2015fma,Iancu:2015vea}.
 
Therefore, in order to have a meaningful evolution equation, we need to identify the physical origin of the large transverse logarithms and subsequently resum them to all orders in the coupling $\abar$. It should be obvious by now, that such higher order terms become more important than the pure $\abar^2$ NLO terms (that is, the $\mcal{O}(\abar^2)$ corrections not enhanced by any transverse logarithm), so that the latter will be discarded in what follows. This is the approach taken in \cite{Iancu:2015vea,Iancu:2015joa} leading to the collinearly improved BK evolution equation which admits a stable solution. Before giving this equation, we shall first write its collinear limit, i.e.~we shall first resum the kernel appearing in \eqn{deltatlog}. 

The origin of the DTLs is in the kinematics \cite{Beuf:2014uia,Iancu:2015vea}. Our main observation in \cite{Iancu:2015vea} was that these large corrections arise from LCPT Feynman graphs in which the successive gluon emissions are not only strongly ordered in both longitudinal momenta and transverse momenta (or `dipole sizes'), but are also ordered in {\em lifetimes} (or, equivalently, in light-cone energies \cite{Beuf:2014uia}). The all-order resummation of the double collinear logarithms $[\abar\ln^2(\bar{z}^2/r^2)]^{n}$, with $n=1,2,3,
\dots$, eventually leads to a modification of the kernel by a multiplicative factor given in terms of 
the Bessel function $\rmJ_1$ (see below).

The STLs find their origin in the DGLAP dynamics \cite{Iancu:2015joa}. Since each power of $\abar$ is accompanied by a collinear logarithm, it is intuitively clear that such terms must represent DGLAP corrections to the BFKL dynamics. This is further confirmed by the fact that the coefficient $-11/12$ in front of the single logarithm in \eqn{deltatlog} can be recognized as the second-order term in the small-$\omega$ expansion of the largest eigenvalue $\mathcal{P}(\omega)$ of the DGLAP anomalous dimension matrix in the large-$\Nc$ limit, more precisely
\begin{equation}
    \label{pomega}
    \mathcal{P}(\omega) \simeq
         \int_0^1 \dif z\, z^\omega 
         \left[ P_{\rm GG}(z) + \frac{\CF}{\Nc}\, P_{\rm qG}(z) \right]
         = \frac{1}{\omega} - A_1
         +\mathcal{O}\left(\omega\right)
         \quad \mbox{with} \quad A_1 = \frac{11}{12} +\frac{\Nf}{6\Nc^3},
 \end{equation} 
where the second term of $A_1$ can be dropped when working in the large-$\Nc$ limit. 
Like the factor $1/\omega$ generates a small-$x$ gluon, the piece $-A_1$ corresponds to the most dominant sub-leading gluon emission (the term linear in $\omega$ would correspond to the next-to-most dominant one and so on). It is not hard to follow the combinatorics for an arbitrary number of such gluon emissions associated with $A_1$, and one finds that the resummation of these $[\abar \ln(\bar{z}^2/r^2)]^{n}$ terms leads to the exponentiation of the NLO correction. Of course this is only a partial resummation of STLs, but it is well-defined and in the spirit of the $\omega$-expansion introduced and developed in \cite{Salam:1998tj,Ciafaloni:1999yw,Ciafaloni:2003rd}. Higher orders in $\omega$ would be related to next-to-most dominant gluons and beyond, and the corresponding STLs would start at higher orders in perturbation theory (NNLO or higher).

Putting together the two types of resummation, one finds that the collinear kernel in \eqn{deltatlog} should be replaced by \cite{Iancu:2015vea,Iancu:2015joa}
 \beq
 \frac{r^2}{\bar{z}^4}\,
 \,\frac{\rmJ_1\Big(2 \sqrt{\abar \ln^2(\bar{z}^2/r^2)} \Big)}{\sqrt{\abar \ln^2(\bar{z}^2/r^2)}}\,
 \left(\frac{r^2}{\bar{z}^2}\right)^{\abar A_1}.
 \eeq
The last step in our construction consists of matching  the above kernel with the LO BFKL/BK one in a consistent way and reinserting the virtual and non-linear terms in the respective evolution equation. Then one arrives at the local equation
 \beq
 \delta S_{\bx\by}(x_{\rm \scriptscriptstyle T }) =
 \frac{\abar}{2 \pi}
 \int_{x_{\rm \scriptscriptstyle T }}^1 \frac{\dif x}{x}
 \int \dif^2 \bz \,
 \frac{(\bx -\by)^2}{(\bx- \bz)^2 (\by - \bz)^2}\,
 \mathcal{K}_{\rm DLA} \mathcal{K}_{\rm SL}
 \left[S_{\bx\bz}(x) S_{\bz\by}(x) - S_{\bx\by}(x) \right],
 \eeq
where the precise form of the kernels $\mathcal{K}_{\rm DLA}$ and $\mathcal{K}_{\rm SL}$ has been given in  Eqs.~\eqref{kdla} and \eqref{ksl}.

 \providecommand{\href}[2]{#2}\begingroup\raggedright\endgroup

\end{document}